# Optimizing Global Genomic Surveillance for Early Detection of Emerging SARS-CoV-2 Variants

Haogao Gu[#,1], Jifan Li[#,2], Wanying Sun[1], Mengting Li[1], Kathy Leung[1,3,4], Joseph T. Wu[1,3,4], Hsiang-Yu Yuan[5,6], Maggie H. Wang[7], Bingyi Yang[1], Matthew R. McKay[8,9,10,11], Ning Ning[2], Leo L.M. Poon[*,1,4,12,13]

## Affiliations

[1]School of Public Health, LKS Faculty of Medicine, The University of Hong Kong, Hong Kong SAR, China.

[2]Department of Statistics, Texas A&M University, United States.

[3]Laboratory of Data Discovery for Health (D²4H), Hong Kong Science and Technology Park, Hong Kong SAR, China.

[4]HKJC Global Health Institute, LKS Faculty of Medicine, The University of Hong Kong, Hong Kong SAR, China.

[5]Department of Biomedical Sciences, Jockey Club College of Veterinary Medicine and Life Sciences, City University of Hong Kong, Hong Kong, China.

[6]Centre for Applied One Health Research and Policy Advice, Jockey Club College of Veterinary Medicine and Life Sciences, City University of Hong Kong, Hong Kong, China.

[7]JC School of Public Health and Primary Care, Faculty of Medicine, the Chinese University of Hong Kong, Hong Kong SAR, China

[8]Department of Electrical and Electronic Engineering, University of Melbourne, Parkville, VIC, 3010, Australia.

[9]Department of Microbiology and Immunology, The Peter Doherty Institute for Infection and Immunity, University of Melbourne, Melbourne, VIC, 3000, Australia.

[10]Victorian Infectious Diseases Reference Laboratory, Royal Melbourne Hospital, at the Peter Doherty Institute for Infection and Immunity, Victoria, Australia.

[11]Department of Electronic and Computer Engineering, Hong Kong University of Science and Technology, Hong Kong SAR, China.

[12]Centre for Immunology & Infection, Hong Kong Science and Technology Park, Hong Kong SAR, China.

[13]HKU-Pasteur Research Pole, School of Public Health, LKS Faculty of Medicine, The University of Hong Kong, Hong Kong SAR, China.

[#]Contributed equally.
*Correspondence: llmpoon@hku.hk.

# Abstract

**Background**
Global viral threats underscore the need for effective genomic surveillance, but high costs and uneven resource distribution hamper its implementation. Targeting surveillance to international travelers in major travel hubs may offer a more efficient strategy for the early detection of SARS-CoV-2 variants.

**Methods**
We developed and calibrated a multiple-strain metapopulation model of global SARS-CoV-2 transmission using extensive epidemiological, phylogenetic, and high-resolution air travel data. We then compared baseline surveillance with various resource-allocation approaches that prioritize travelers, focusing on Omicron BA.1/BA.2 retrospectively and on hypothetical future variants under different emergence, transmission and vaccine effectiveness scenarios.

**Findings**
Focusing existing surveillance resources on travelers at key global hubs significantly shortened detection delays without increasing total surveillance efforts. In retrospective analyses of Omicron BA.1/BA.2, traveler-targeted approaches consistently outperformed baseline strategies, even when overall resources were reduced. Simulations indicate that focusing surveillance on key travel hubs outperform baseline practices in detecting future variants, across different possible origins, even with reduced resources. This approach also remains effective in future pandemic scenarios with varying reproductive numbers and vaccine effectiveness.

**Interpretation**
These findings provide a quantitative, cost-effective framework for strengthening global genomic surveillance. By reallocating resources toward international travelers in select travel hubs, early detection of emerging variants can be enhanced, informing rapid public health interventions and bolstering preparedness for future pandemics.

**Funding**
US National Institutes of Health, Research Grants Council of Hong Kong, InnoHK, Hong Kong Jockey Club, and Individual Research Grant at Texas A&M University.

# Introduction

The COVID-19 pandemic highlights the importance of genomic surveillance in identifying and monitoring emerging viral variants[1,2]. Variants with increased transmissibility[3], immune escape[8], or treatments resistance[4] pose substantial challenges to public health responses. Genomic surveillance via systematic collection, sequencing, and analysis of viral genomes, is essential for understanding viral evolution, guiding vaccine updates[5], and shaping public health and treatment strategies[6,7]. Beyond SARS-CoV-2, genomic surveillance helps track pathogens such as influenza, HIV, Ebola, Zika, and dengue[8-11], offering insights into transmission patterns, drug resistance, and outbreak dynamics.

Despite its value, global genomic surveillance faces high financial costs and uneven resource distribution. By 2023, each SARS-CoV-2 whole genome cost about £70 to sequence[12], and over 16.8 million samples have been processed—an unsustainable financial burden. Limited surveillance capacity in resource-poor areas[13] further amplifies global disparities and weakens international containment efforts. Addressing these inequities demands sustainable and efficient strategies that can operate across diverse settings. Existing initiatives (e.g., WHO's GISRS, PIP, and GISAID)[14], encourage data sharing but do not offer concrete guidance on how many cases to sequence, which populations to prioritize, or how to maximize cost-effectiveness within a global network.

Focusing diagnostic and sequencing efforts on groups of individuals with a high risk of propagating global spread, such as international travelers in major travel hubs which are critical conduits for viral transmission[15], may be a promising strategy. Our previous research in Hong Kong demonstrated that focusing surveillance efforts on major transit points improved early detection of emerging variants[16-18]. As an international transit hub, Hong Kong has mirrored global variant trends[17], underscoring the value of prioritizing high-travel-volume areas to enhance the timeliness and effectiveness of variant detection[19].

In this study, we evaluate the effectiveness of current genomic surveillance in tracking the global dissemination of SARS-CoV-2 variants and propose refined strategies for improvement. We demonstrate this in the context of Omicron BA.1/BA.2 emergence (when surveillance data was most abundant), and for hypothetical novel variants. Using a comprehensive metapopulation multiple-strain model calibrated with extensive data, we demonstrate that targeting existing genomic surveillance resources towards travelers and key international travel hubs significantly improves early detection. These findings provide evidence-based recommendations for building a more adaptive and targeted global genomic surveillance network. By reducing detection delays and improving public health responsiveness, reducing detection lags, informing rapid public health actions, and providing a stronger foundation for future pandemic preparedness.

## Methods

### *Data*

In addition to epidemiological surveillance data, we include integrates daily air travel data from high-resolution ADS-B sources, allowing detailed analyses of regional importation and exportation dynamics. Phylogenetic analyses inferring the probable origins of over five million sequences further illuminate transmission pathways and regional contributions to variant dissemination. Covariates like vaccination coverage, existing immunity, and travel policies further shape viral spread and were included in the model. Detailed data collection and processing are recorded in Supplementary Methods.

### *Models*

Evaluating the effectiveness of global genomic surveillance for early detection of SARS-CoV-2 variants requires understanding how international travel intersects with uneven regional capacity. We addressed this by developing a set of multiple-strain metapopulation models using the Spatiotemporal Partially-Observed Markov Process (SpatPOMP) framework[20-22] to simulate global SARS-CoV-2 spread and assess surveillance strategies. Each region (29 spatial units, aggregated from 136 countries/territories, Figure 1a and Figure S1) tracks susceptible, vaccinated, exposed, infectious, recovered, and deceased compartments, while distinguishing multiple viral strains. Infection can occur locally or be imported via travelers—both direct arrivals and those transiting through hubs (Figure S1b).

### Primary Model ($M_0$)

$M_0$ models co-circulation of four viral strains (Omicron BA.1, BA.2, Delta, others) from September 15, 2021, to March 15, 2022. It incorporates daily air travel data, vaccination coverage, and varying levels of local surveillance.

In $M_0$ and its extensions, genomic surveillance follows a two-step observation process (Figure S2), reflecting disparities in testing and sequencing: first identifying a fraction of infected individuals through diagnostic testing (via the infection-to-diagnostic ratio, IDR), then sequencing a subset of those cases (via the diagnostic-to-sequencing ratio, DSR).

We calibrate $M_0$ with recently optimized likelihood-based methods[20,22], we used daily cases, deaths, variant frequencies, and phylogenetic inferences of sequence origins. leveraging over 300,000 CPU hours through accelerated C++ coding.

### Comparisons and Extensions

Alternative or extension models were summarized in Table S1:

$M_1$, $M_2$: Simpler or purely statistical models, used to benchmark the mechanistic realism of $M_0$. Comparisons with simpler benchmark models $M_1$ and $M_2$ confirmed $M_0$'s superior capacity to capture complex spatial-temporal transmission dynamics (Supplementary Methods).

$M_3$: Extends $M_0$ with adjustable parameters for testing (IDR), sequencing (DSR), and traveler-focused surveillance strategies.

$M_4$: Evaluates hypothetical variants in a future pandemic scenario without travel restrictions, using average 2019 air travel data.

# Results

## *Modeling the Global Spread Dynamics of Omicron Variants*

We applied our primary model ($M_0$) to examine how Omicron BA.1 and BA.2 disseminated worldwide from September 15, 2021, to March 15, 2022. Built on the daily air travel data (Methods) and calibrated with surveillance indicators (daily cases, deaths, variant frequencies, inferred cases origins). Some smaller regions emerged as key travel hubs because of high passenger volume (Figure 1a, columns 1 and 2), while places with stronger diagnostic and sequencing capacities (e.g., the US, UK) maintained more robust genomic surveillance (Figure 1a).

Simulations using $M_0$ emphasize the influence of international travel on Omicron's rapid spread. Although South Africa seeded BA.1 and BA.2 early on, the UAE and Europe soon contributed, and by weeks three and four, North America and Asia became major sources (Figure 1b). With 49 calibrated parameters (Table S3), $M_0$ closely captured real-world fluctuations in cases, deaths, and sequenced variant counts across the 29 regions over 182 days (Figure 1c). These findings validate the model's capacity to capture complex transmission patterns and demonstrate its utility for evaluating genomic surveillance strategies for emerging SARS-CoV-2 variants.

## *Community Outbreaks Predominate in Early Variant Detection*

Simulations using model $M_0$ provide quantitative insights into how Omicron BA.1 and BA.2 spread globally and were first detected (Figure 2). Overall daily infections peaked near January 25, 2022, dominated by BA.1 (Figure 2a). The global IDR and DSR averaged 17.1% and 6.79%, respectively, but showed significant regional and temporal variability (Figure S3). Diagnostic cases peaked at 5.5 million daily, while sequencing peaks appeared earlier and were more uneven, reflecting disparities in sequencing capacity (Figure 2a).

Despite imported cases sparking introductions, early variant detection—both diagnosis and sequencing—primarily occurred among communities (Figure 2b). For BA.1, only five regions (United Kingdom, Hong Kong, Japan, New Zealand, and Papua New Guinea) diagnosed the variant first in travelers, and just four for BA.2 (Germany, Japan, South Korea, New Zealand). Similarly, earliest sequencing among travelers was rare: five regions detected BA.1 (Hong Kong, Brazil, Germany, Japan, and New Zealand) and only two detected BA.2 (Japan, New Zealand) that way.

Traveler-related infections for BA.1 peaked on January 18, 2022—119 days post-emergence—averaging 153,668 daily cases, about a week before the global infection peak. BA.2 traveler-related infections peaked on February 27, 2022, at 18,750 daily, 110 days post-emergence (Figure 2c). During the *seeding period* (<200 daily global travel-related infections), South Africa was the dominant source of travel-related infections, contributing 64.1% for BA.1 and 98.6% for BA.2. BA.1 spent 60 days in this phase (ending November 21, 2021), while BA.2 spent 44 days (ending December 22, 2021) (Figure 2d). Afterwards, transmission dynamics shifted significantly. For BA.1, Europe emerged as the leading source, with Turkey, Spain, the United Kingdom, and the United States contributing substantially (Figure 2c). South Africa, while dominant during the seeding period (Figure 2d), contributed only 0.41% of BA.1 and 1.79% of BA.2 travel-related infections afterward. By February 2022, Germany became a notable source of BA.2 infections (Figure 2c), corresponding to its peak in confirmed cases during that time[23].

Time lags underscore how variants moved globally. On average, BA.1 and BA.2 took 40.7 and 30.1 days, respectively, from emergence (defined as reaching 2,000 community cases) to first arrival in new regions, longer than arrival-to-diagnostic (11.0 days for BA.1; 9.0 days for BA.2) or diagnosis-to-sequencing (16.7 days for BA.1; 11.7 days for BA.2) intervals. Regions with high traveler volume (e.g., Europe and North America) exhibited shorter emergence-to-arrival lags and detected variants sooner (Figure S4).

Our results demonstrate that under baseline surveillance, early detection of BA.1/BA.2 is largely driven by community outbreaks. Enhancing surveillance capacities at travel hubs could shorten arrival-to-diagnostic and diagnostic-to-sequencing lags, thereby improving early detection and global public health responses.

### *Evaluating Traveler-Targeted Approaches for Early Detection of Omicron Variants*

We used model $M_3$ (Table S1) to assess how reallocating surveillance resources toward travelers could accelerate Omicron BA.1/BA.2 detection. We focused on diagnostic efforts (IDR and diagnostic capacity), sequencing efforts (DSR and sequencing capacity), a "traveler weight" parameter defining the proportion of surveillance resources allocated to traveler cases in selected regions, and the travel hub configurations for targeted surveillance.

Under the baseline condition—where resources are distributed proportionally to traveler vs. community infections—median identification lags (community-based) were 11 days for BA.1 and 18 days for BA.2. Reducing diagnostic capacity alone had minimal impact on identification lags for either variant (Figure 3a, left panel), suggesting that diagnostic testing was not the primary bottleneck. By contrast, sequencing efforts was critical: doubling it significantly shortened identification lags for both community and traveler cases (Figure 3a, middle panel), while reducing it to 0.1% of baseline levels made detecting BA.2 among travelers nearly impossible due to their smaller infection pool.

Increasing traveler weight globally shortened identification lags for traveler cases but could prolong them among community cases. For BA.2, increasing traveler weight to 30% detected it sooner among travelers than baseline community detections (Figure 3a, right panel). A two-parameter interaction analysis showed that for BA.2, increasing traveler weight to 70% globally reduced identification lags by 9 days (median estimates), even when diagnostic capacity was reduced to 1% or sequencing efforts were halved (Figure 3c,d). However, BA.1 exhibited higher prevalence and diagnostic rates, so community detection generally dominated unless traveler weight exceeded 90% (Figure 3c,d). These differences reflect distinct epidemiological profiles during variants' emergence: BA.1 had higher community prevalence (6.0%) and IDR (6.5%) but lower DSR (1.6%), favoring earlier community detection, whereas BA.2 had lower community prevalence (4.1%) and IDR (0.2%) but higher DSR (6.1%), making traveler-focused surveillance more effective outside its origin region.

We next tested traveler-targeted strategies (TTS) that direct available genomic surveillance resources toward a small number of high-travel-volume hubs. Each TTS includes two elements: selecting which hubs to prioritize and deciding how much, if any, of the global resources should be reallocated from other regions. For selecting travel hubs, we ranked the regions by their total (T-ranked) or per-capita (P-ranked) travel volume, and selected the top multiple regions. For resource redistribution, in some standard approaches (R-type) no resources are diverted from non-hub areas; in others (H-type or M-type), 50% or 90% of resources from unselected regions are reassigned to selected hubs (Table S4). For instance, T3-R focuses surveillance on the top three total-volume (T-ranked) hubs without reducing resources elsewhere (R-type), whereas P3-H targets the top three per-capital (P-ranked) hubs with H-type redistribution. We also tested whether adding continental representative regions (denoted with "a", e.g. in P3a-H) can enhance surveillance more efficiently. We found allocating 50%–70% traveler weight to the top 11 per-capita hubs under an R-type scheme (P11-R) reduced median identification lags for both BA.1 and BA.2 by one day (Figure 3e), with 13%–30% of early detections through traveler surveillance. Notably, this modest adjustment enhanced early detection but did not diminish surveillance in other regions.

More aggressive redistribution schemes (M-type, redistributing 90% of surveillance resources from non-hub to hub regions) reduced detection delays further but risk weakening non-hub coverage; T3a-M, which focuses on the top three total-volume hubs plus four continental representative (Table S4), reduced BA.2 identification lags by three days when traveler weights exceeded 90% (Figure 3f), but left regions like South Africa with fewer resources.

Overall, these TTS matched or outperformed baseline approaches for early BA.1/BA.2 detection. R-type schemes, which do not pull resources from non-hub regions, proved practical and cost-effective. Effectiveness also depends on the variant's origin and global travel patterns, as illustrated by differing source-destination dynamics (with primary sources of imported cases differing from their main destinations for exported cases) in regions such as Europe and the United States (Figure 2d). These dynamics underscore the importance of adapting surveillance strategies to context-specific transmission pathways. If Omicron variants had originated elsewhere, spread patterns might have varied substantially. Building on these findings, we used model $M_4$ to evaluate TTS under diverse epidemiological conditions, aiming to develop robust strategies for future variant monitoring.

### *Optimizing Global Surveillance Strategies for Future Variant Detection*

Using model $M_4$ (Table S1), we examined how traveler weight and different TTS influence global identification lags (time until a novel variant is first detected anywhere in the world). We modeled a hypothetical scenario where a novel variant could emerge in any of the 29 spatial units, using 2019 average air travel data without travel restrictions. Maximal diagnostic and sequencing capacities were set to the 95th percentile of daily resources during the Omicron period for each region (Table S2), avoiding outliers of short-term high-resource conditions.

Analysis categorized each potential origin as a community-detected or traveler-detected source (Table S2). Community-detected origins typically have higher baseline IDR and DSR (Figure S5), enabling local surveillance to detect variants early with minimal benefit from extra traveler-focused measures (grids with predominantly red cells, Figure 4a). In contrast, traveler-detected origins (grids with cells shifting from blue to red as traveler weight increases, Figure 4a) depend on infected individuals being identified abroad in more interconnected hubs. Increasing traveler weight substantially lowered global detection lags for these regions.

We tested TTS under a random emergence context (variant emergence probability proportional to population) using bootstrap resampling. Globally, higher traveler weights generally reduced identification lags (Figure 4b-d). Pairing H-type and M-type resource redistribution with P-ranked travel hubs produced further reductions (Figure 4c). Two TTS stood out as both effective and practical: P2-H, focusing and enhancing surveillance on the United Arab Emirates and Hong Kong at 0.1% traveler weight, while halving surveillance elsewhere, and T2-R, prioritizing the United States and United Kingdom at 70% traveler weight without redistributing resources (Table S4). P2-H (0.1% weight) and T2-R (70% weight) lowered the mean global identification lag from 14.13 days to 10.85 and 11.47 days, respectively (P < 0.001, two-sided t-test). Variations of the P2-H travel hub configuration with any extra region did not significantly outperform its core configuration (Figure S6), so we selected P2-H and T2-R for further analysis. These findings demonstrate that concentrating surveillance on a small number of highly connected travel hubs can markedly speed up detection while maintaining overall resource efficiency.

### *Enhancing Surveillance Efficiency Under Resource Constraints*

We next tested the two selected TTS—P2-H (0.1% traveler weight) and T2-R (70% traveler weight)—under reduced diagnostic and sequencing resources, compared to a baseline scenario of 1% traveler weight with no hub-focused strategy. Variants emerging from resource-limited regions (e.g., "Africa (others)" and "Asia (others)") benefited most, achiving earlier global detection despite surveillance resources were scaled down (Figure 5a). Established surveillance regions like Canada, the United States, and the United Kingdom maintained short identification lags under TTS, even at 70% traveler weight. This indicates that full surveillance capacity (Table S2) is not always necessary for effective detection and that selected TTS can optimize resource utilization substantially.

Substantially cutting global diagnostic capacity still shortened detection times under P2-H and T2-R. For example, at 10% and 1% of baseline diagnostic efforts, detection improved by 3.29 and 2.61 days, for P2-H (0.1% weight) and T2-R (70% weight) respectively (P < 0.001, two-sided t-test, Figure 5b). Similarly, reducing global sequencing efforts to 40% or 60% of baseline for P2-H and T2-R, respectively, continued to improve detection lags (P < 0.001, two-sided t-test). These findings illustrate the robustness of the tested TTS, enabling significant resource savings without compromising global detection timelines.

We further examined these TTS across varying transmissibility (relative scaling of the effective reproduction number $R_{eff}$, against the pre-existing variant) and different vaccine effectiveness (VE) against infection. Across most scenarios, the selected TTS consistently outperformed or matched the baseline with reduced global resources (Figure 5d), especially when variants are transmitted slowly (high VE and low relative $R_{eff}$). Under strong vaccine immunity (68% VE) and a less transmissible variant (relative $R_{eff}$ = 1.5), P2-H at 0.1% traveler weight reduced the median identification lag from 32 days to 17 days while using only half the baseline diagnostic and sequencing resources. This adaptability underscores that the selected TTS can function effectively under diverse epidemiological conditions.

Overall, redistributing existing surveillance resources toward international travelers and high-impact hubs significantly reduce global detection delays while containing costs. Focusing on P2-H or T2-R scenarios sustains early variant detection, bolsters preparedness for future pandemics, and offers a scalable blueprint for optimizing global genomic surveillance.

# Discussion

We present a quantitative framework for optimizing global genomic surveillance to detect emerging SARS-CoV-2 variants rapidly. By integrating extensive epidemiological data, high-resolution air travel patterns, and phylogenetic analyses into a multiple-strain metapopulation model, we capture complex transmission dynamics and informs targeted surveillance strategies, offering model-based insights for promoting global genomic surveillance initiatives[24,25].

A key component of our approach is the strategic allocation of genomic surveillance resources towards international travelers at major travel hubs. We demonstrate that such targeting significantly enhances early variant detection without additional costs. Focusing on a few high-impact regions, such as those identified in the P2-H (0.1% traveler weight) and T2-R (70% traveler weight) strategies, addresses global disparities in sequencing capacities and allows earlier detection of variants emerging in under-resource settings, improving global health security.

Our findings underscore the importance of international travel in spreading SARS-CoV-2 variants[19]. Building on local evidence from Hong Kong[17,18], where enhanced surveillance among travelers proved effective for early variant detection, we show that similar strategies can be applied globally. By targeting surveillance efforts on travelers in key hubs, our approach also addresses concerns about cryptic transmission preceding recognized outbreaks[26], potentially reducing undetected community spread and enabling more timely interventions. These results align with prior studies that highlight global inequities in sequencing capacities[13]. Our study proposes traveler-targeted strategies that mitigate these disparities by concentrating efforts on travelers in major hubs. By detecting variants more rapidly in hub regions, we enhance the overall effectiveness of global surveillance networks.

While increased diagnostic testing aids overall case detection[27], our analysis indicates that increasing diagnostic efforts alone has a limited impact when sequencing resources are constrained. Combining strong traveler-focused sequencing with strategic resource allocation yields marked improvements, pointing to the need for balanced investment in both testing and sequencing, adapted to local conditions.

We also explored the trade-offs posed by resource redistribution schemes. While H-type (half redistribution) and M-type (maximal redistribution) schemes can capture variants early, especially when the source region has lower surveillance capacity but is well-connected to the selected hubs, they risk weakening community surveillance in non-hub regions. Therefore, careful consideration is needed to balance resource allocation to maximize global surveillance efficiency without compromising local detection capabilities. In contrast, T2-R avoids redistributing resources altogether and still outperforms the baseline, making it operationally practical.

Our analysis also examined the robustness of the TTS under various epidemiological scenarios, including different levels of transmissibility (relative $R_{eff}$) and vaccine effectiveness (VE). As expected, stronger immunity (higher VE) and lower relative $R_{eff}$ corresponded to longer identification lags due to slower spread of the variant. P2-H (0.1% weight) remained robust even under reduced diagnostic (10% of baseline) and sequencing efforts (40% of baseline), and significantly outperforming the baseline with 50% of the global total resources. This strategy would only require a maximum diagnostic capacity of 1,303 (HK) and 519 (UAE) positive cases for travelers—effectively aiming for diagnostic testing among 130,300 (HK) and 51,900 (UAE) incoming travelers max (assuming a 1% infection rate)—and reserving a sequencing capacity of only 14 (HK) and 6 (UAE) samples for travelers per day. Such settings are operationally achievable and cost-effective.

Some limitations merit discussion. Our model depends on available data, which can be unreliable or incomplete, especially in areas with limited resources. We also assume that increasing traveler weight and shifting resources can be accomplished without major logistical or ethical hurdles. In reality, scaling traveler testing may face staffing, training, and compliance challenges. Furthermore, traveler behavior or mobility policies could shift, altering the success of traveler-focused strategies. Additionally, testing all possible traveler-targeted strategies is computationally infeasible; therefore, we focused on the most relevant travel hub configurations. It is worth noting that although our results highlight effective configurations for TTS, a broader range of other strategies could perform similarly or better.

Maintaining intensive sequencing levels achieved during the pandemic peak is unrealistic long-term. Governments have scaled back efforts amid shifting priorities and budget constraints. Yet our traveler-targeted strategies remain effective with reduced diagnostic or sequencing capacities, addressing global surveillance inequities by leveraging regions with greater capacity and connectivity, and strengthening preparedness for both current and future pandemics. By providing a quantitative, data-driven strategy, our study offers a valuable tool for policymakers seeking to optimize limited resources and bolster global health security in the face of current and future infectious disease threats.

# Figure Legends

**Figure 1: Global Dissemination of SARS-CoV-2 Variants and Genomic Surveillance Data Across 29 Spatial Units.**

**(a)** World map and ranked surveillance metrics: The illustrates 29 spatial units derived from 136 administrative regions, categorized into 24 natural units (representing individual regions) and five aggregated units (representing the remaining regions stratified by continent). Aggregated units are depicted with semi-transparent coloring. On the right, six columns rank these spatial units based on total travel volume during the study period, population size, number of infected cases, reported cases, community-sequenced cases, and traveler-sequenced cases. Ranks in each column are ordered from top to bottom, with rank numbers labeled on the far-right side.

**(b)** Weekly global transmission patterns of Omicron BA.1 (blue) and BA.2 (green): This panel depicts the spreading dynamics of the Omicron BA.1 and BA.2 during the first (top), second (middle), and the third-to-fourth (bottom) weeks after their respective emergences. Lines connecting spatial units represent directional transmission flows (total number of transmitted infections), with a circle at the head denoting the exporting region and a triangle at the tail representing the importing region. Line transparency reflects the number of transmitted cases aggregated over the specified period. Circles and triangles indicate the number of exporting and importing connections, respectively, for each spatial unit. The direction of BA.1 transmission flows is indicated by counterclockwise arrows, while BA.2 flows follow clockwise arrows. Results are averaged across 1024 simulations of model $M_0$.

**(c)** Temporal dynamics of cases and variant counts across 29 spatial units: Temporal dynamics of total reported cases, deaths, and sequenced cases are displayed for Delta, Omicron BA.1, Omicron BA.2, and other variants. Colored lines represent case counts for different metrics: total reported cases (grey), total deaths (brown), Delta (red), Omicron BA.1 (blue), Omicron BA.2 (green), and Other (pink) variants. Solid lines denote fitting data, while shaded bands represent the 5th and 95th percentiles of simulated data across 1024 model $M_0$ replications. Each grid corresponds to one of the 29 spatial units, positioned geographically on a world map using an equirectangular projection.

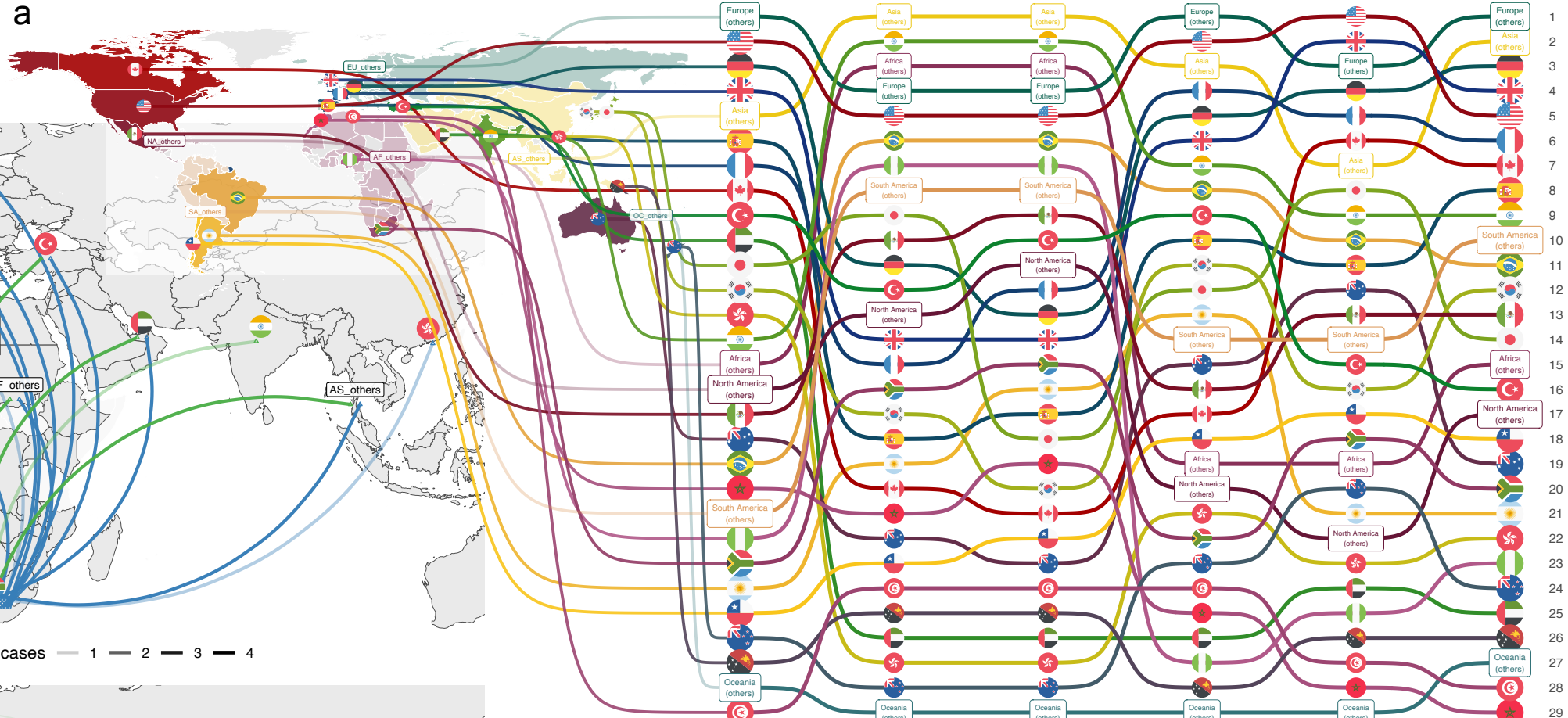

a

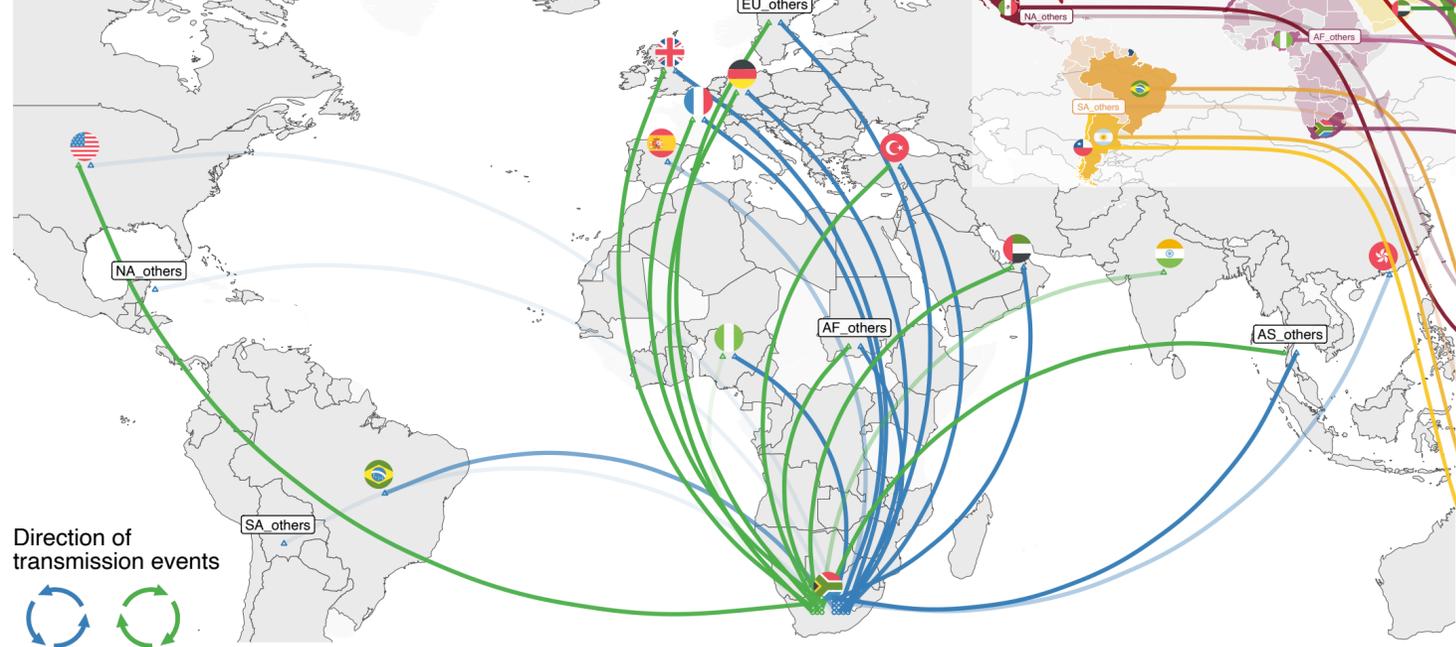

b

**The first week after emergence**

Direction of transmission events

Variant — Omicron BA.1 — Omicron BA.2    Number of cases — 1 — 2 — 3 — 4

**The second week after emergence**

Direction of transmission events

Number of cases — 2 — 4 — 6 — 8    Variant — Omicron BA.1 — Omicron BA.2

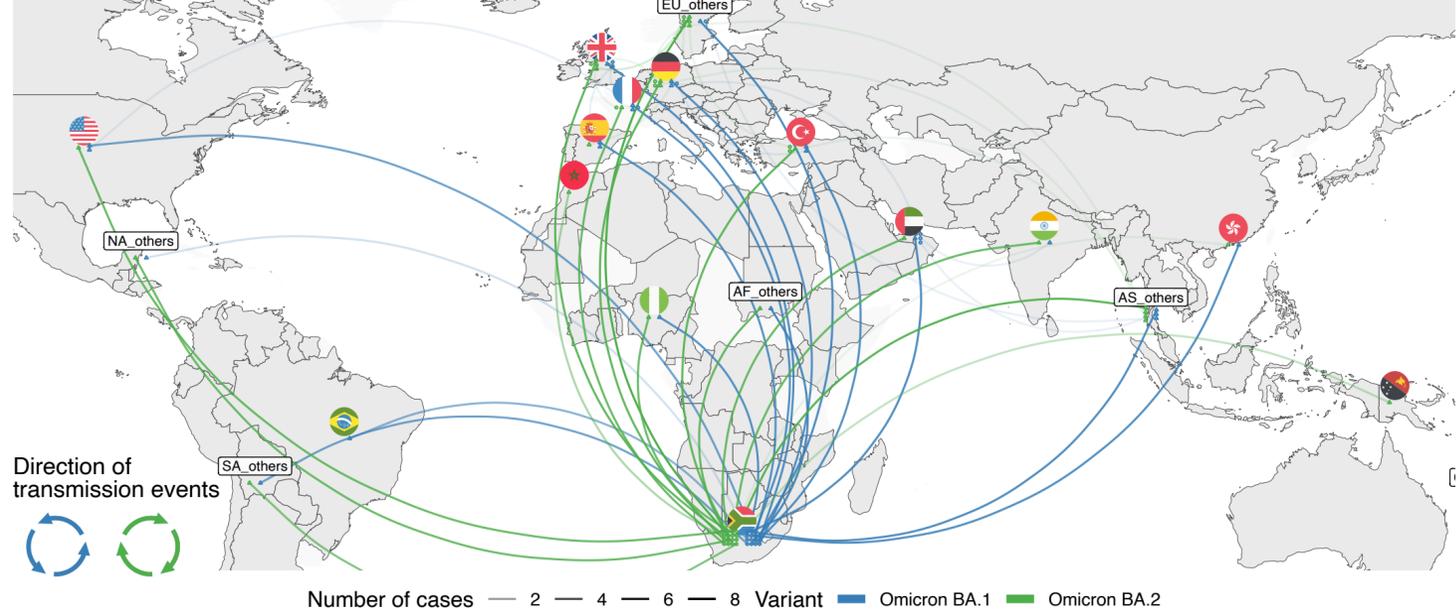

**The 3rd to 4th weeks after emergence**

Direction of transmission events

Variant — Omicron BA.1 — Omicron BA.2    Number of cases — 10 — 20 — 30 — 40

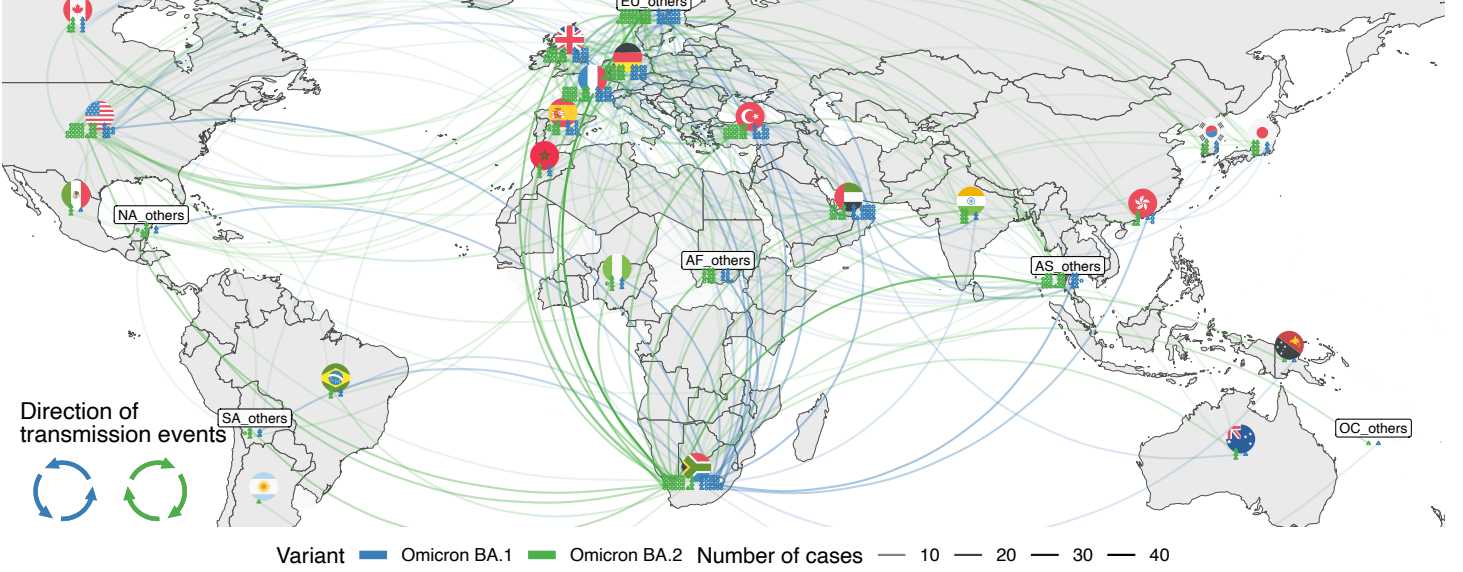

c

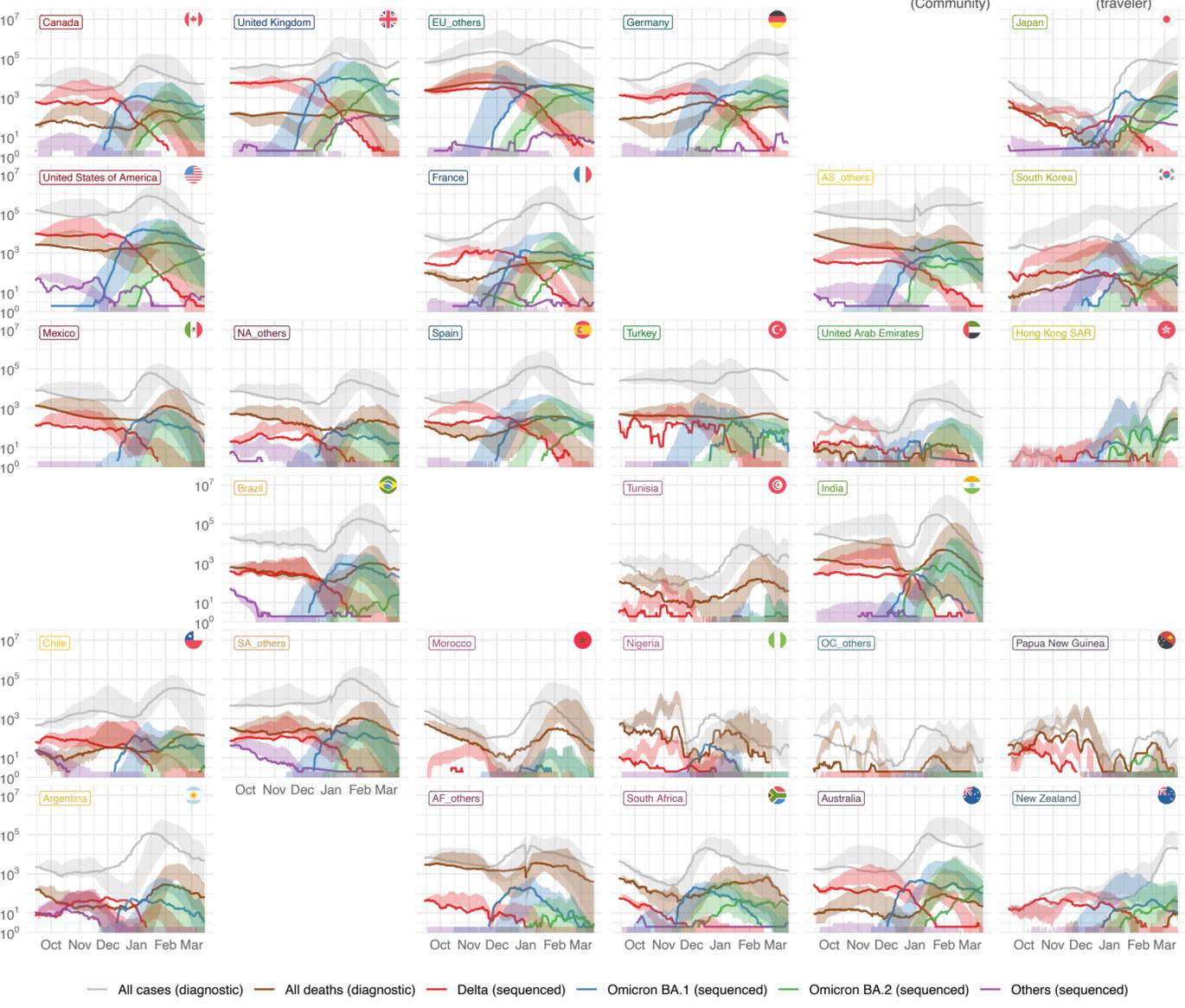

— All cases (diagnostic) — All deaths (diagnostic) — Delta (sequenced) — Omicron BA.1 (sequenced) — Omicron BA.2 (sequenced) — Others (sequenced)

**Figure 2: Infection, Diagnostic, Sequencing, and International Transmission Dynamics of Omicron BA.1 and BA.2 Variants.** This figure illustrates simulation results for infection, diagnostic, sequencing, and virus import/export dynamics of Omicron BA.1 and BA.2 variants, based on $M_0$ with 1024 simulation replications. Vertical solid and dashed lines indicate the presumed emergence dates of Omicron BA.1 and BA.2, respectively.

**(a)** Aggregate cases: The total number of infected, diagnostic, and sequenced cases across all spatial units are displayed, categorized by variant (Delta (red), Omicron BA.1 (blue), Omicron BA.2 (green), Others (pink)) and case type (community (light shade) or traveler (dark shade)). Median simulation values are summed across spatial units to represent the overall totals.

**(b)** Earliest cases across spatial units: Distribution of the earliest observed infected, diagnostic, and sequenced cases in each spatial unit, stratified by variant and case type, with color coding matching panel (a). Each horizontal boxplot represents the median (bold line), interquartile range (box edges), and whiskers (1.5 times the interquartile range). Regions are ordered by their median earliest arrival dates for Omicron BA.1.

**(c)** Temporal dynamics of travel-related cases: Temporal trends of the travel-related cases for Omicron BA.1 and BA.2, separated by their origins. The top 10 origin spatial units contributing the most exported cases are shown with individual colors, while all remaining origins are grouped as "others." Each area band indicates cumulative exported cases from each origin over time.

**(d)** Time-varying top origins and destinations: Horizontal bands depict the time-varying top origins and destinations for travel-related cases, shown separately for Omicron BA.1 and BA.2. For each spatial unit, the origin bands identify the unit contributing the most imported cases at any time point (top source), while destination bands indicate the unit receiving the most exported cases (top recipient). BA.1 origin (BA.1 Ori.), BA.1 destination (BA.1 Dest.), BA.2 origin (BA.2 Ori.), and BA.2 destination (BA.2 Dest.) are shown, with regions ordered by their median earliest arrival dates for Omicron BA.1.

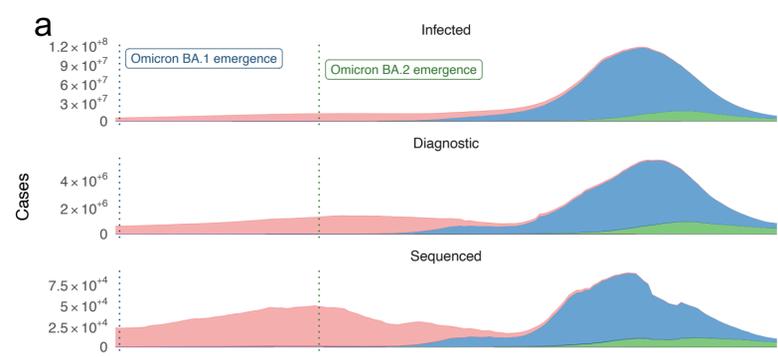

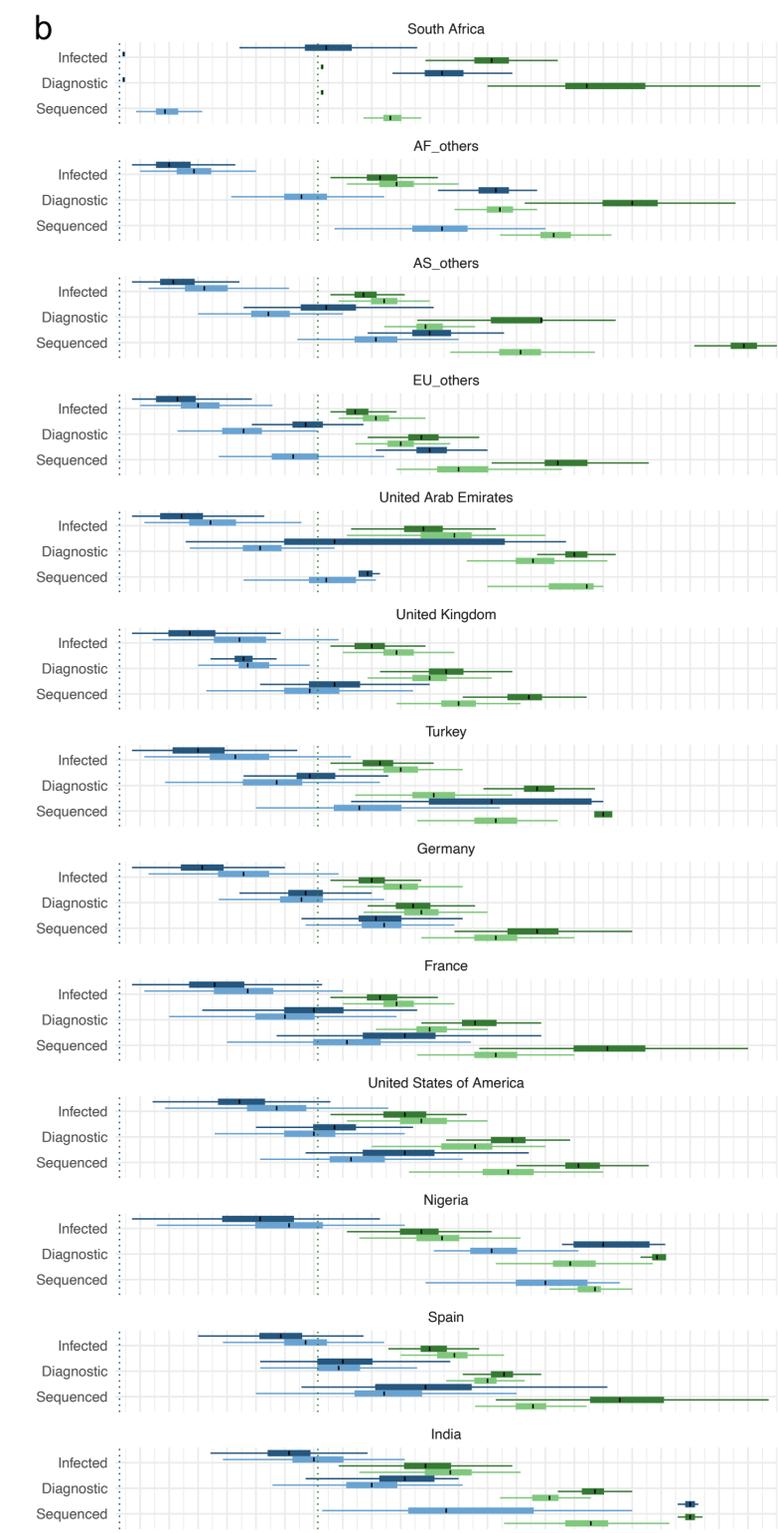

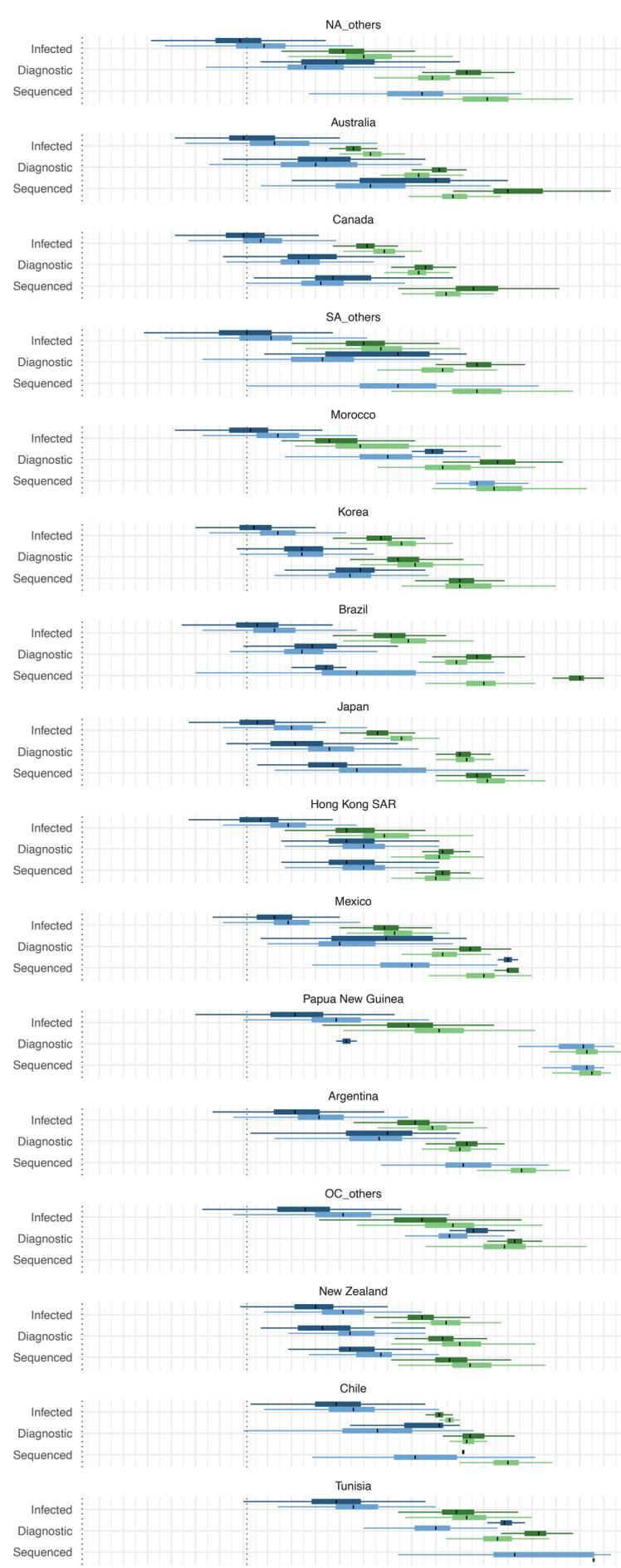

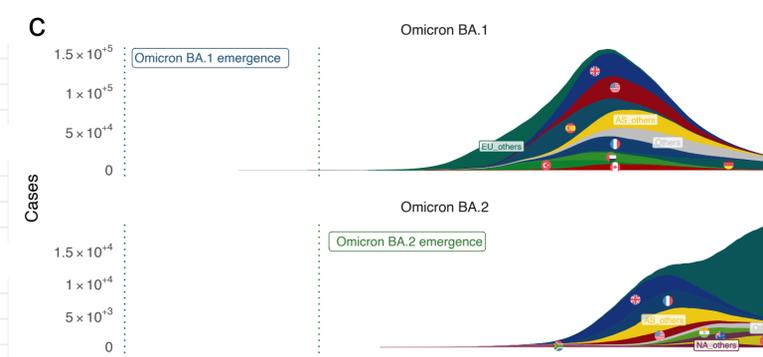

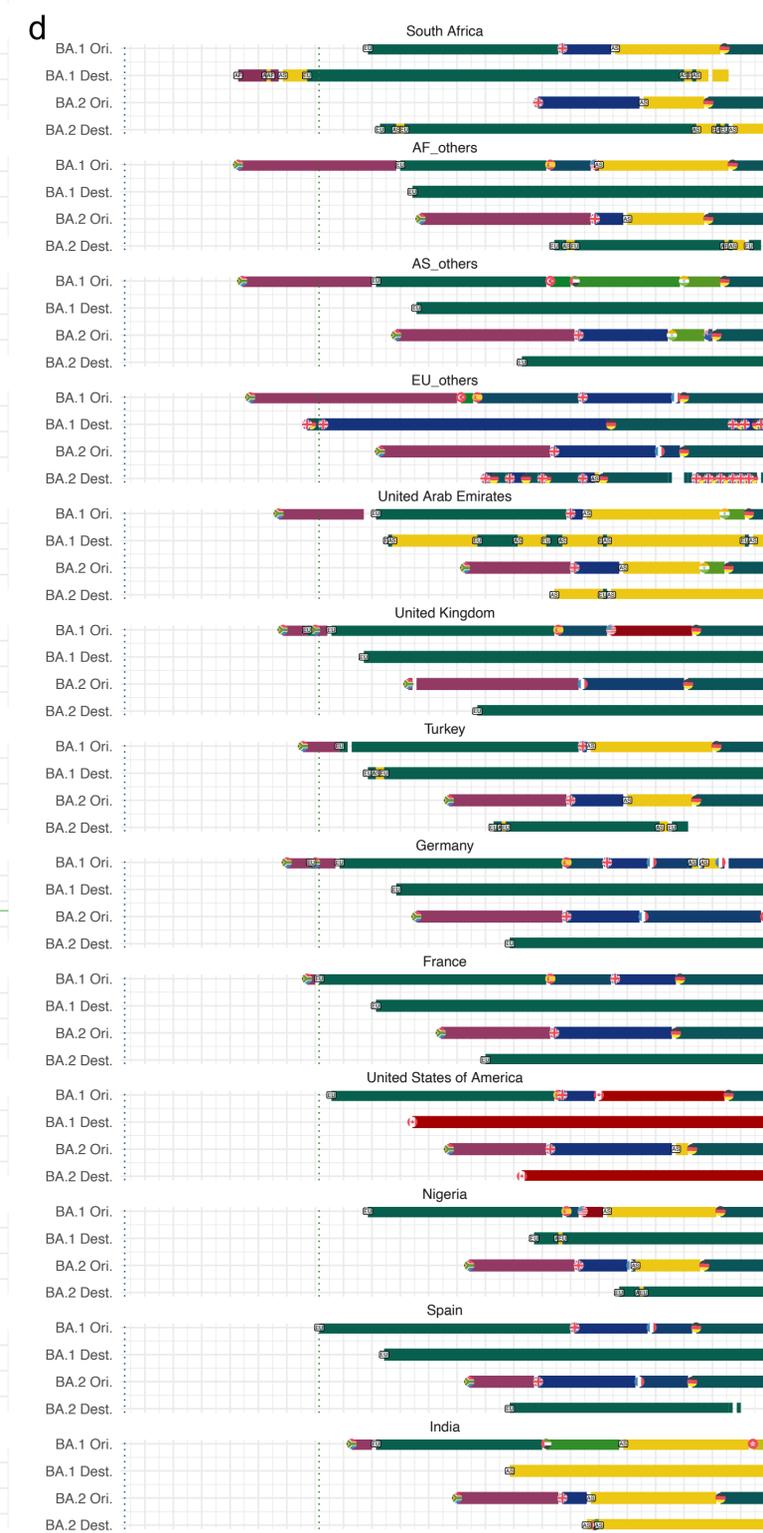

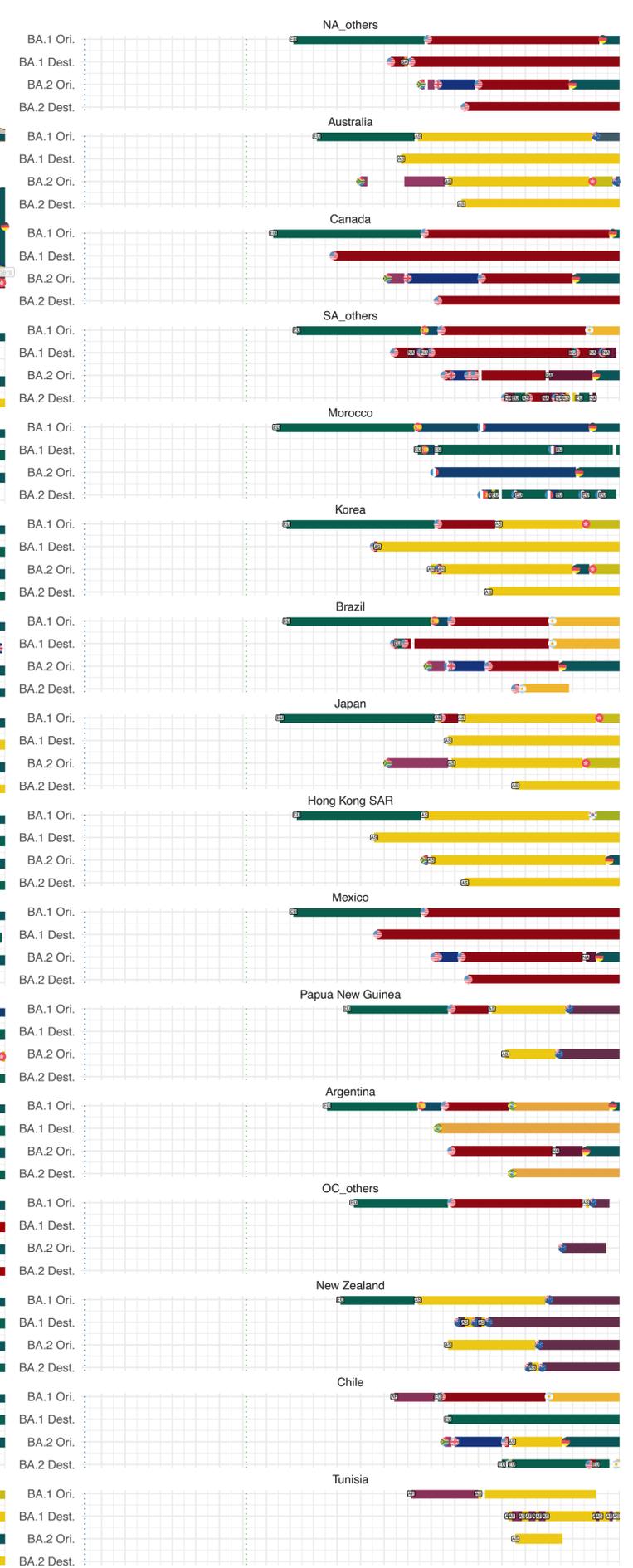

**Figure 3: Impact of Various Parameters on Identification Lags for Omicron BA.1 and BA.2 Variants.** This figure shows the effects of four key parameters—diagnostic efforts (identification-to-diagnostic ratio (IDR) and diagnostic capacity), sequencing efforts (diagnostic-to-sequencing ratio (DSR) and sequencing capacity), traveler weight (the proportion of surveillance resources allocated to travelers), and travel hub scenarios affect the identification lags of Omicron BA.1 and BA.2 variants under $M_3$. Results are based on 512 simulation replications.

**(a)** Boxplots of identification lags: Identification lag is defined as the period between the emergence of a variant and its first sequencing, regardless of spatial unit or case type (community or traveler). Boxplots illustrate the distribution of identification lags across simulations for changes in diagnostic efforts (left), sequencing efforts (middle), and traveler weight across all regions (right). Each boxplot displays the median (line within the box), first and third quartiles (box edges), and whiskers extending to 1.5 times the interquartile range. Red solid (traveler) and dashed (community) lines represent the baseline median identification lags from $M_0$ and are equivalent to the baseline in $M_3$ when diagnostic and sequencing efforts are set to 100%. The value of community ("C") and traveler ("T") baselines were also text-labeled in each panel. The median values were labeled above respective boxplots.

**(b-f)** Heatmaps of interaction effects: Heatmaps show the interaction effects between pairs of parameters on identification lags. Colors represent the median deviation from the baseline lag (11 days for BA.1, 18 days for BA.2), with white indicating selected baseline, red indicating earlier identification, and blue indicating later identification. Text within each cell denotes the percentage of simulations in which either community or traveler cases yielded the shortest identification lag; Plain text indicates community cases, while boxed text indicates traveler cases. In cases of tied outcomes, all contributing combinations are displayed.
**(b)** Interaction between diagnostics and sequencing efforts.
**(c)** Interaction between traveler weight (applied globally) and diagnostic efforts.
**(d)** Interaction between traveler weight (applied globally) and sequencing efforts.
**(e)** Interaction between traveler weight (allocated to selected travel hubs) and travel hub strategies based on per-capita travel volume (P-ranked hubs).
**(f)** Interaction between traveler weight (allocated to selected travel hubs) and travel hub strategies based on total travel volume (T-ranked hubs).

**Figure 4: Identification Lags for the Emergence of a Novel Variant Under Non-Pandemic Conditions.** This figure presents simulation results from $M_4$, examining hypothetical scenarios where a novel variant emerges in each of the 29 spatial units under non-pandemic conditions. Results are based on 512 simulation replications.

**(a)** Identification lags across spatial units for traveler-targeted scenarios: Heatmaps display the interaction between travel weight (the proportion of surveillance resources allocated to travelers in selected regions) and different travel hub strategies on identification lags. Each heatmap corresponds to a scenario where the novel variant emerges in a specific spatial unit. The color gradient represents median identification lags, with darker shades indicating longer lags and white representing the global baseline median lag of 15 days (obtained from bootstrap resampling under a 1% traveler weight setting for all regions). Text within cells indicates the proportion of simulations in which either community or traveler cases yielded the shortest identification lag: plain text denotes community cases, while boxed text denotes traveler cases. If tied outcomes occur, all contributing combinations are shown. Heatmaps are arranged according to the geographic layout of the spatial units for visual clarity.

**(b-d)** Global summative identification lags for traveler-targeted scenarios: These panels summarize the global identification lags across all spatial units and tested parameter combinations, simulating the variant's emergence in any of the 29 spatial units. Bootstrapped resampling (size: 1024 × 10) was used to account for population-weighted emergence probabilities, where regions with larger populations were more likely to be selected as the origin. **(b)** 25th percentile of identification lags. **(c)** Median identification lags. **(d)** 75th percentile of identification lags. Dashed cell outlines in (**b-d**) highlight the traveler-targeted scenarios chosen for further analysis. These include combinations of travel hub strategies and traveler weights that yielded globally optimal identification lags under specific settings.

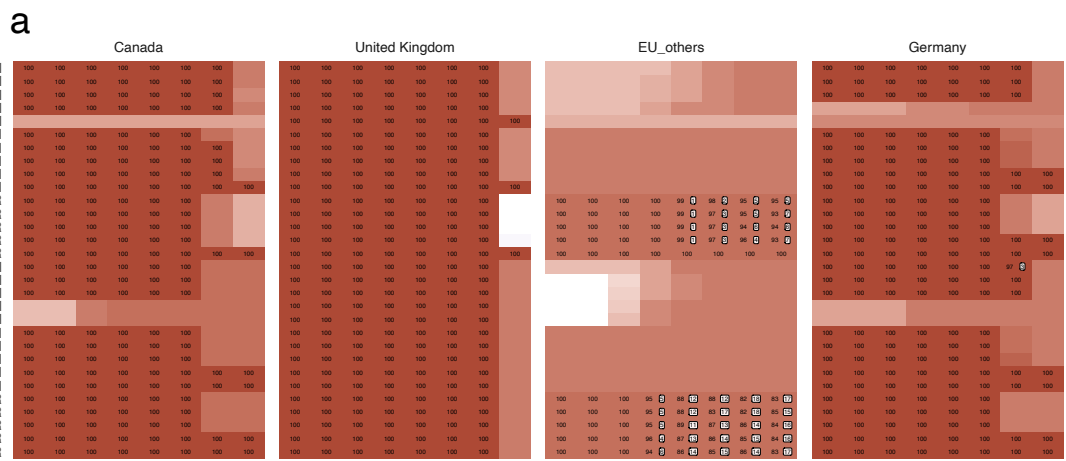

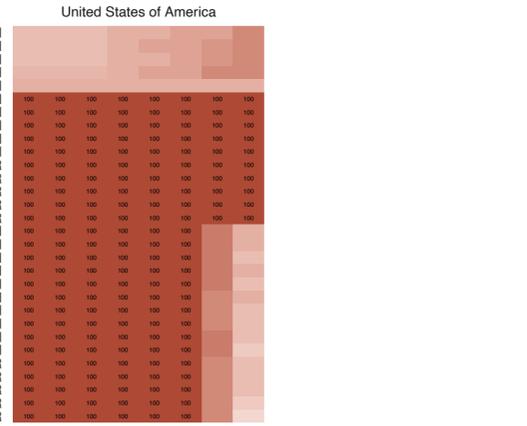
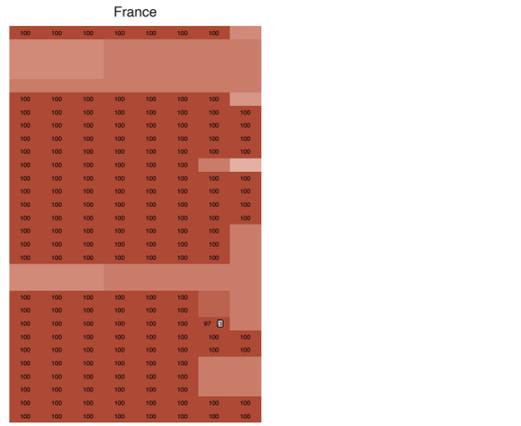
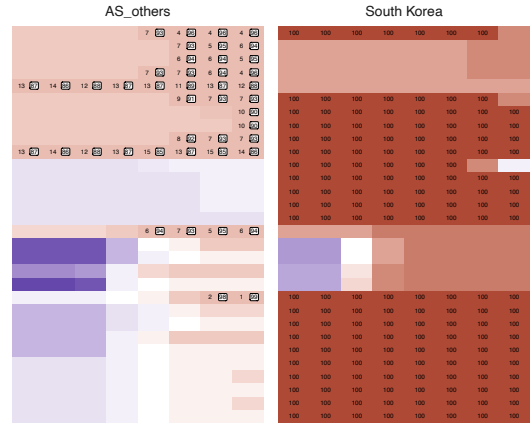
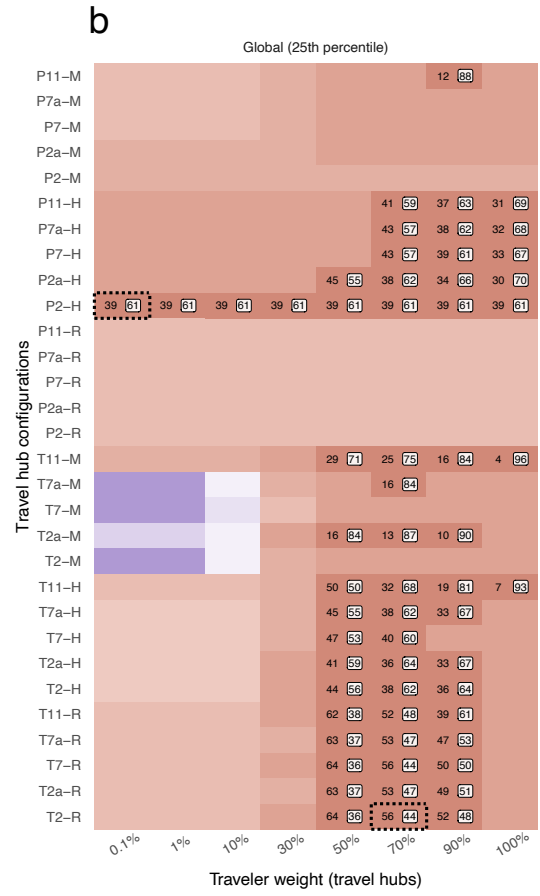

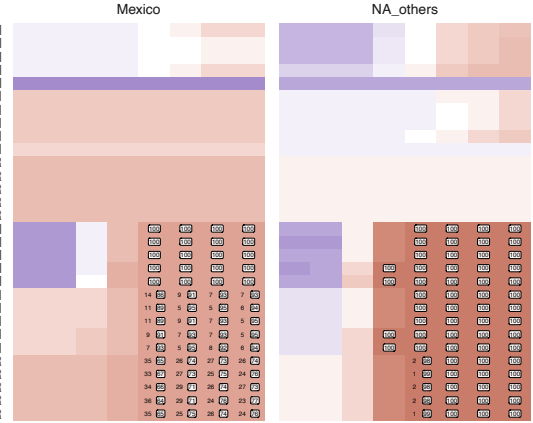
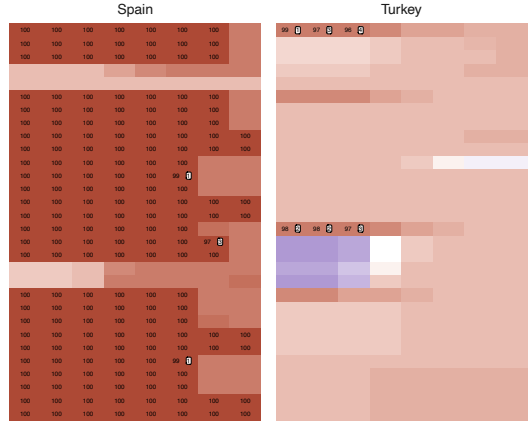
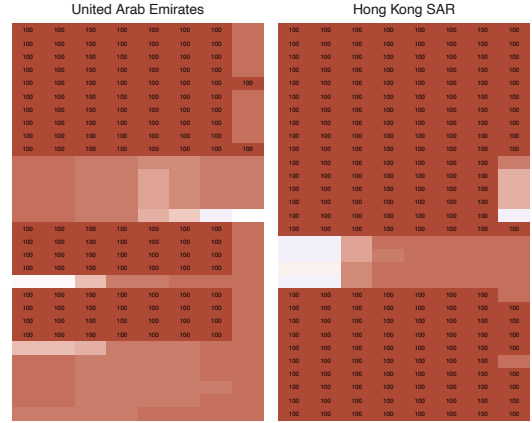
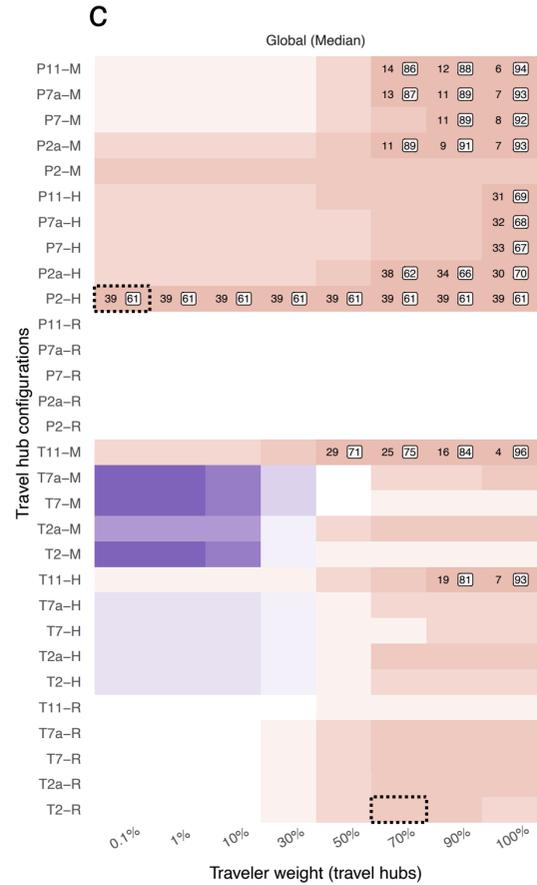

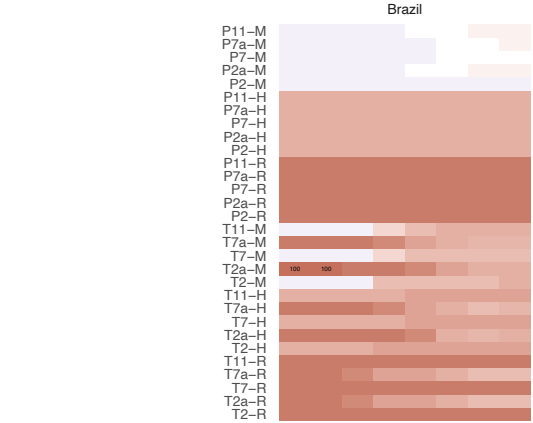
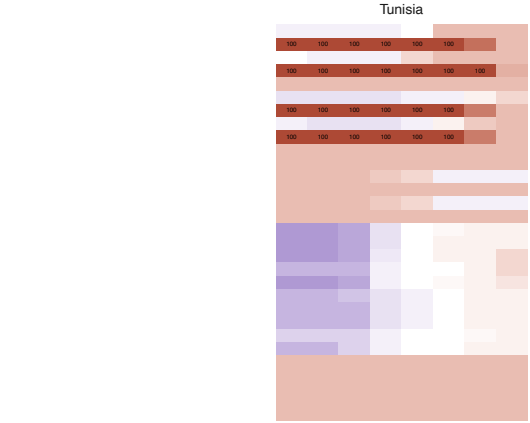
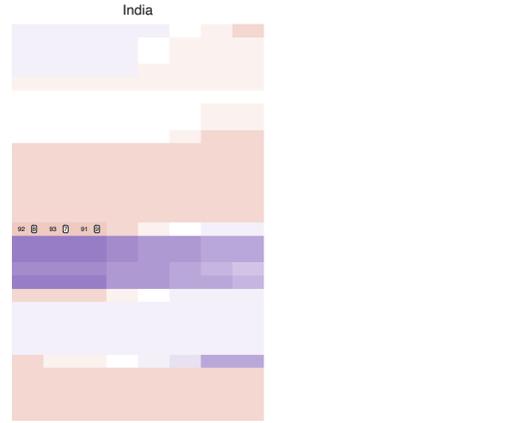

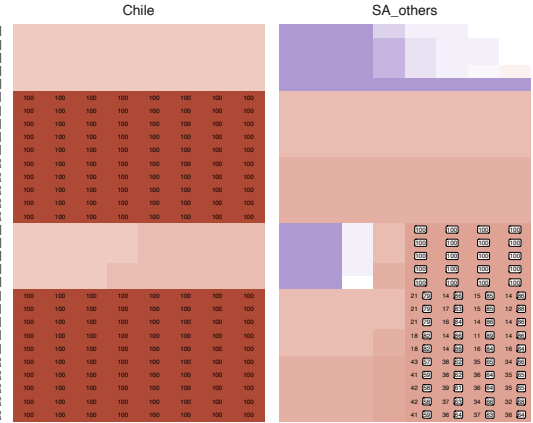
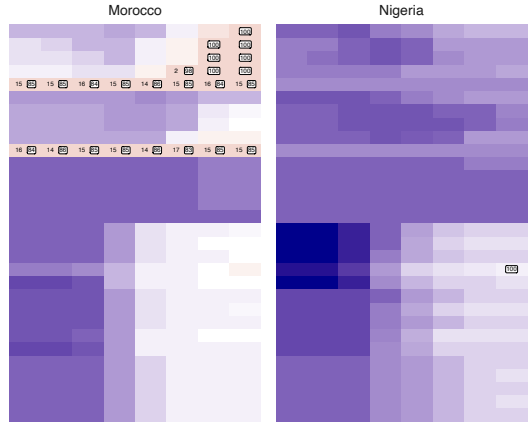
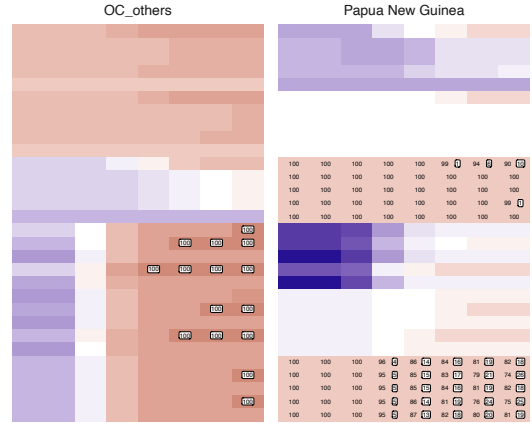
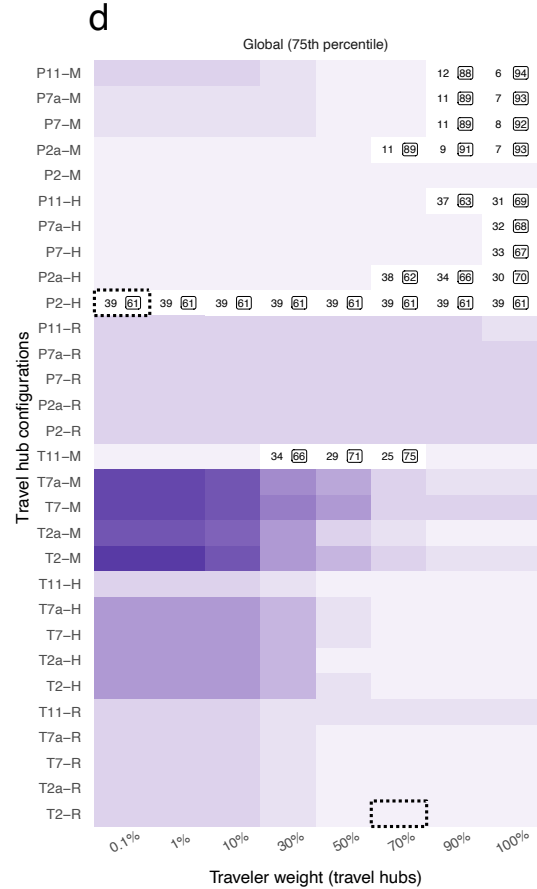

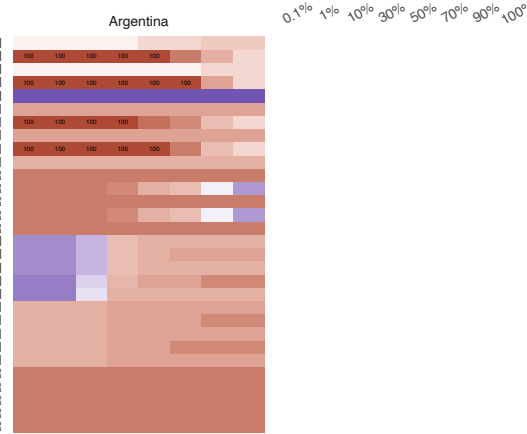
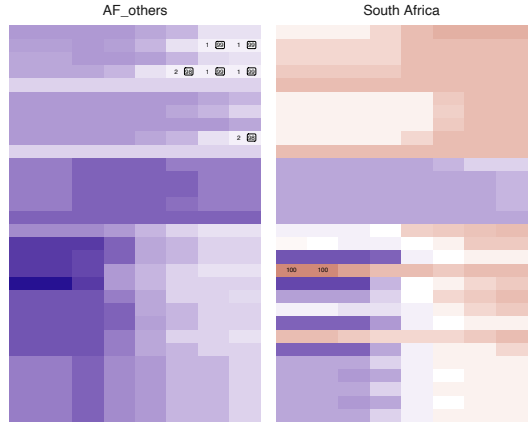
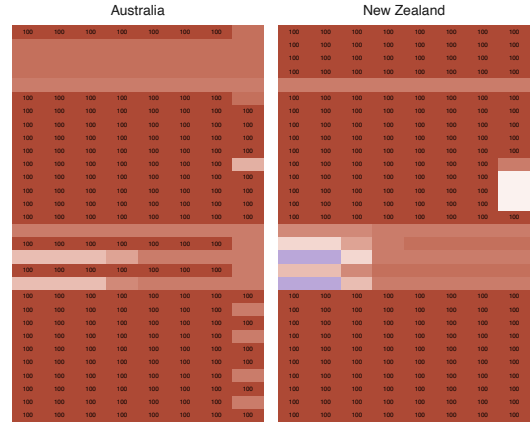

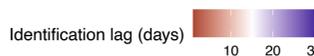

Identification lag (days)

**Figure 5: Impact of Diagnostics, Sequencing, and Immunity on Identification Lags in Traveler-Targeted Scenarios for Novel Variant Detection**

**(a)** Identification lags across spatial units under varying diagnostic and sequencing efforts: Boxplots illustrate the identification lags for traveler-targeted scenarios across 29 spatial units, where a novel variant emerges. Scenarios were tested with adjustments in diagnostic and sequencing efforts under two specific traveler-targeted scenarios: P2-H (half redistribution) and T2-R (regular redistribution). Each boxplot represents the median, first, and third quartiles, with whiskers extending up to 1.5 times the interquartile range. The horizontal solid and dashed red lines indicate the baseline identification lags for community (Com) and traveler (Tra) cases, respectively, under 1% traveler weight for all regions. Baseline values for community- and traveler- based detection lags are labeled within each subplot for reference.

**(b-c)** Global identification lags with changes in diagnostic and sequencing efforts: Summary boxplots aggregate the effects of changes in diagnostic (**b**) and sequencing (**c**) efforts on identification lags across all 29 spatial units. Simulations accounted for population-weighted probabilities of variant emergence, where regions with larger populations have a higher likelihood of being the origin. Bootstrapped resampling (resample size = $1024 \times 10$) was used to summarize results. Median identification lag values are labeled above the respective boxplots, with boxed values indicating parameter combinations selected for further exploration in panel (**d**). The baseline scenario is marked with horizontal solid (community) and dashed (traveler) red lines for comparison.

**(d)** Identification lags under varying vaccine effectiveness levels and relative $R_{eff}$: Violin and boxplots display the global identification lags for selected parameter combinations (from **b** and **c**) under different vaccine effectiveness levels against infection (low VE: 13%, moderate VE: 40.5%, high VE: 68%) and relative $R_{eff}$ of the novel variant. Median identification lag values are labeled above the boxplots, while violin plots depict the distribution of lags across resampled simulations. The tested scenarios include reductions in diagnostic or sequencing resources, as well as combinations of H-type and R-type schemes for P2-H and T2-R.

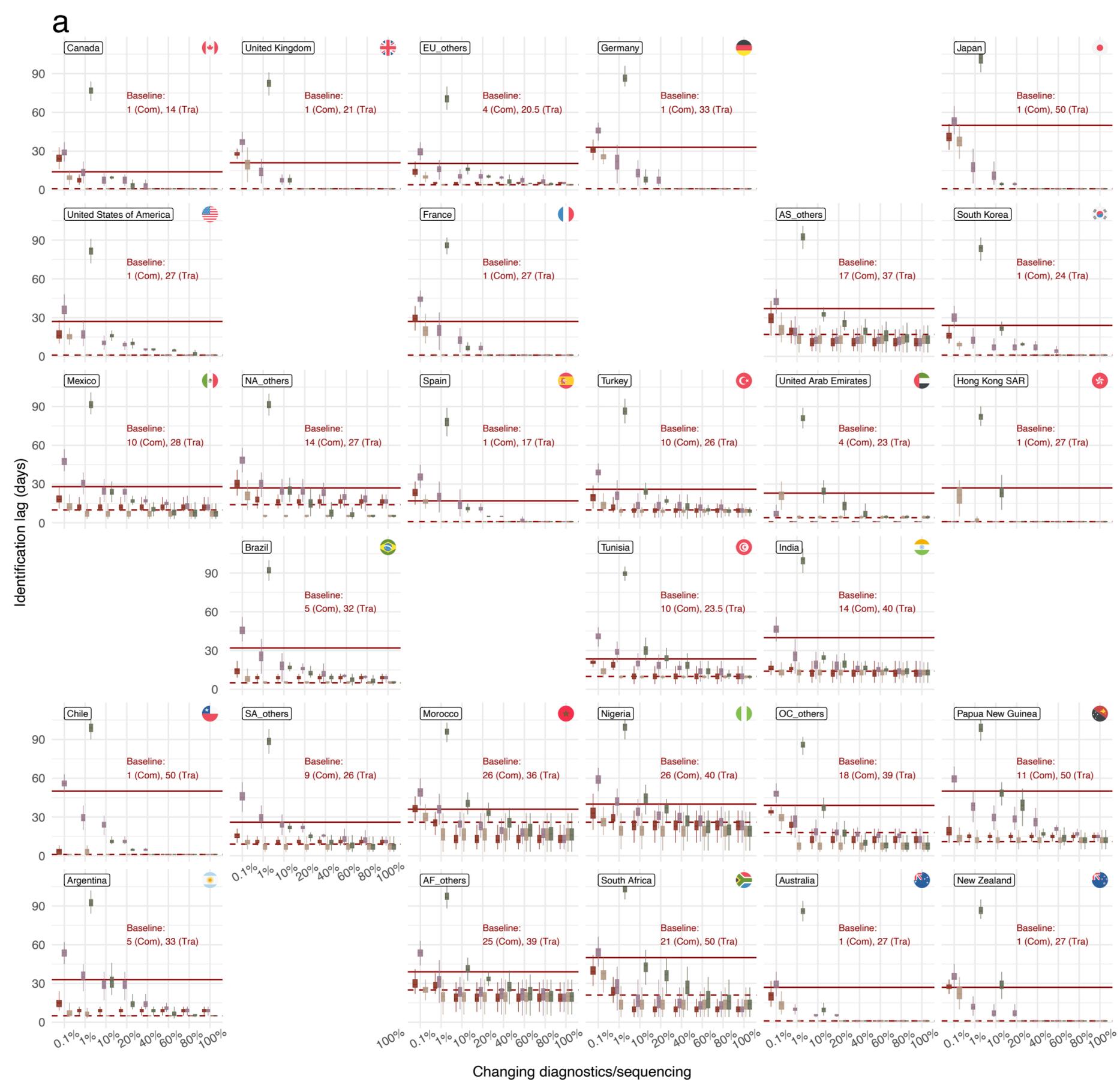

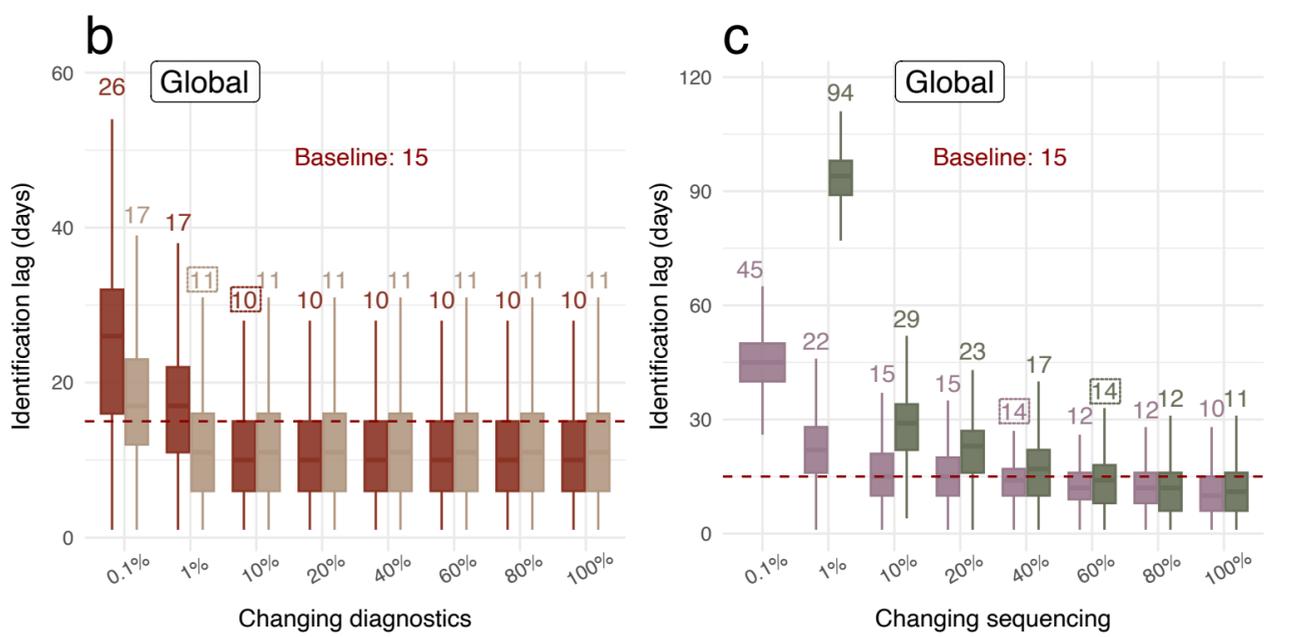

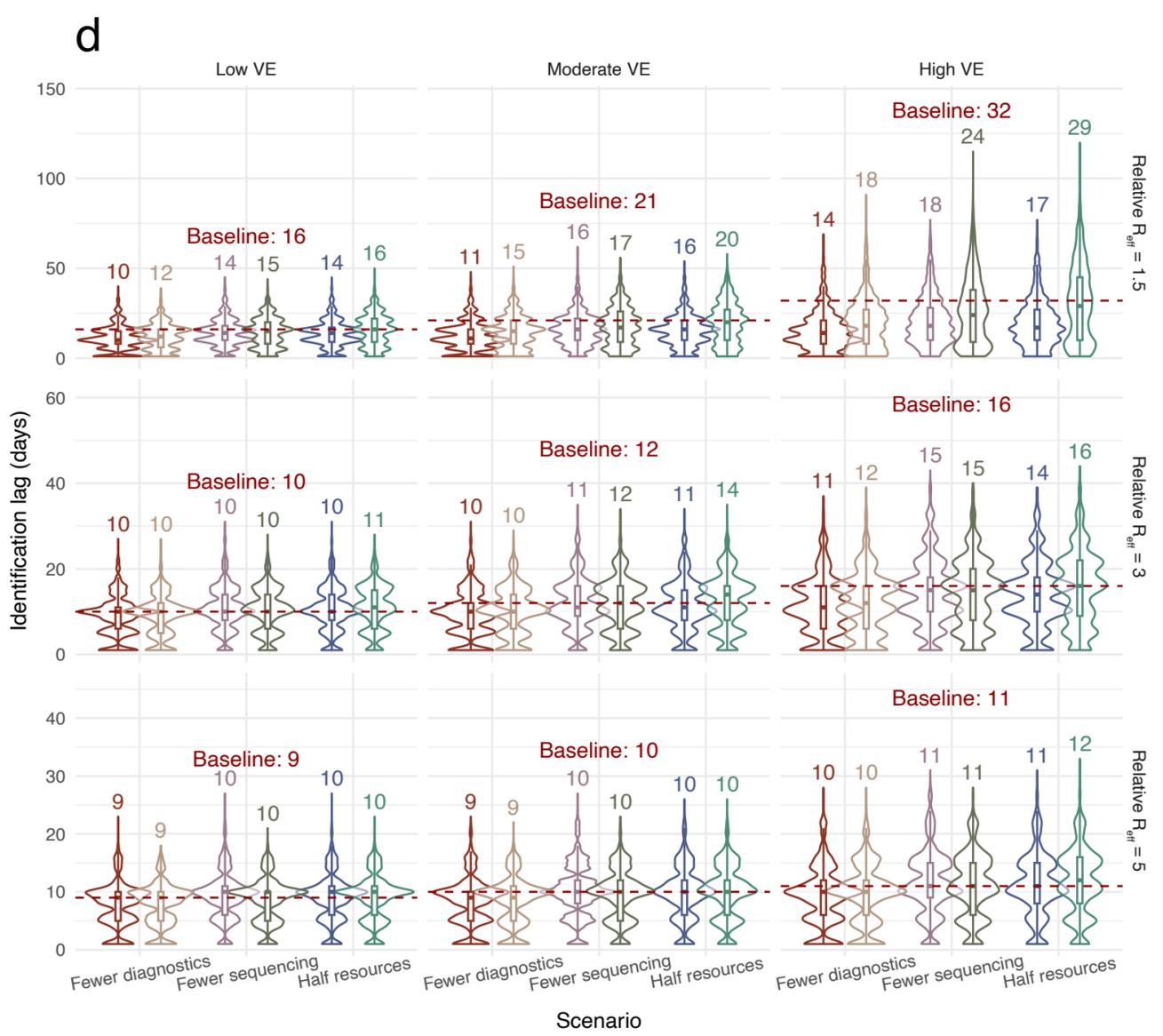

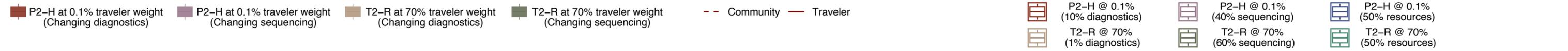

## Data Availability

Data sources are referenced in the Methods and Supplementary Methods sections. Intermediate analyzing data are available via the archived repository (Zenodo link to be provided).

## Code Availability

Analysing code are available via https://github.com/Leo-Poon-Lab/global_genomic_surveillance_spatpomp_article.

## Acknowledgements

The computations were performed using research computing facilities offered by Information Technology Services, the University of Hong Kong. Portions of this research were conducted with Texas A&M High Performance Research Computing. This work was supported by Theme-Based Research Scheme (T11-705/21-N: LLMP & MRM) from the Research Grants Council of Hong Kong, InnoHK from Hong Kong (C2i: LLMP), Hong Kong Jockey Club (HKJCGHI: KL, JTW & LLMP), NIH grant (1R21AI180492-01: NN; HHSN272201400006C: LLMP) and the Individual Research Grant at Texas A&M University (NN). ChatGPT 4o was used with prompt "polish to improve readability" to improve readability and language of the part of the work.

## Contributions

This study was designed by H.G. and L.L.M.P. Data curation was performed by H.G. and M.L. The phylogenetic analysis was conducted by W.S. and H.G. Coding and fitting the epidemiological model was done by H.G. and J.L. Data visualization was done by H.G. N.N. and L.L.M.P. were responsible for project supervision. Results were reviewed and discussed amongst all authors. The original draft of this manuscript was prepared by H.G. and was reviewed and edited by KL, JTW, HYY, MHW, MRM, BY, N.N. and L.L.M.P.

## Competing interests

None.

Supplementary Information is available for this paper.

# Supplemental Methods

## *Data collection*

### Incidence, death, and other covariate data

We compiled comprehensive data on COVID-19 cases, deaths, infections, population sizes, and vaccination coverage estimates from the Institute for Health Metrics and Evaluation (IHME) dataset titled "COVID-19 Mortality, Infection, Testing, Hospital Resource Use, and Social Distancing Projections". This dataset includes records from 177 countries and territories, covering a wide range of epidemiological variables essential for our analysis.

### Genomic surveillance data

Genomic sequencing data were retrieved from the GISAID hCoV database (accessed on October 16, 2023; GISAID Identifier: EPI_SET_240923dt). We extracted sequences with complete metadata, including collection date, Pango lineage, and submitting country or region. The remaining sequences were categorized into four main groups based on their Pango lineages:
1. **Omicron BA.1**: BA.1 or BA.1.*
2. **Omicron BA.2**: BA.2 or BA.2.*
3. **Delta**: B.1.617, B.1.617., or AY.
4. **Others**: All remaining lineages during the study period.

### Air travel data

Air travel patterns play a critical role in modeling the global spread of SARS-CoV-2 variants. We compiled daily air travel movement data between countries by integrating information from three primary sources:
1. **OpenSky COVID-19 Flight Dataset (v23.00)**[1]: Contains over 123 million individual flight records between 2019 and 2022, based on Automatic Dependent Surveillance-Broadcast (ADS-B) data.
2. **International Civil Aviation Organization (ICAO)**: Yearly air passenger statistics from the World Bank (Civil Aviation Statistics of the World and ICAO staff estimates; last updated date: 2023-07-25).
3. **World Tourism Organization (WTO)**: Annual tourism statistics from the World Bank (Yearbook of Tourism Statistics, Compendium of Tourism Statistics and data files; last updated date: 2023-07-25).

To estimate the number of daily air travelers, we assumed that 80% of flights carried passengers (excluding cargo flights) and that aircraft operated at 80% of their maximum passenger capacity. Passenger capacities were determined based on the aircraft's wake turbulence category (WTC). For instance, medium ("M") aircraft like the Boeing 737 were estimated to carry up to 250 passengers (Details in the next section).

Due to limitations in ADS-B coverage in certain regions (e.g., mainland China and parts of Africa), we employed a gravity model to estimate missing air travel data. This model utilized variables such as ICAO air passenger data, GDP per capita, urban population, and geographic distance to predict travel volumes. The gravity model was fitted separately for each year from 2019 to 2021 among regions with both ADS-B and ICAO data, excluding outliers like the United States (adjusted for high domestic flight volumes) and mainland China.

### Transit passenger data

Transit passenger volumes at major airports were obtained from previous estimates[2]. Where specific data were unavailable, a global average transit passenger share of 14% was applied. These figures were used as a set of fixed parameters ($q_u$) to adjust our models for the effect of transit passengers who do not contribute to local transmission dynamics (see Table S2).

### International travel control policy data

Data on international travel control measures were sourced from the Oxford COVID-19 Government Response Tracker[3], focusing on the "C8" indicator, which details policies on international travel controls. This information allowed us to account for the impact of travel restrictions and quarantine measures on virus transmission dynamics. The effects of travel control policy were modeled via $\zeta_u(t)$, as specified in the next section.

## Data Processing

To mitigate reporting inconsistencies and fluctuations, all surveillance data were smoothed using a 7-day moving average. Infection-to-diagnostic ratios (IDR) and infection-to-fatality ratios (IFR) were calculated based on IHME estimates. Diagnostic-to-sequencing ratios (DSR) were computed by dividing the number of daily sequenced cases by the number of daily reported cases. These ratios were used as covariates in our models to represent testing and sequencing capacities across different regions and times.

## Phylogenetic Analysis

We conducted phylogenetic analyses to infer the origins of sequenced cases. Due to the computational challenges of handling large datasets, we adopted a combined approach:

1. **Sequence Filtering and Subsampling**: After removing low-quality sequences and duplicates, we subsampled sequences for each strain group using a pipeline designed for phylogeographic analysis. This ensured proportional representation based on case counts per location and time, resulting in datasets of approximately 20,000 sequences per strain group.
2. **Phylogenetic Tree Construction**: Maximum likelihood trees were generated for each strain group using FastTree, with sequences aligned to the Wuhan-Hu-1 reference genome.
3. **Ancestral State Reconstruction**: We used the TreeTime mugration model to infer the geographic origins of internal nodes within the phylogenetic trees, applying a molecular clock model with a fixed clock rate.

The remaining sequences were sequentially placed into the background trees using UShER. Origins were inferred based on the locations of ancestral nodes. For sequences classified as imported-like, we constructed additional phylogenetic trees focusing on these sequences to refine origin assignments. This approach allowed us to classify sequences as either imported or locally transmitted, providing critical insights into transmission dynamics. As a result, a total of 9,583,188 sequences between December 24, 2019 and March 15 2022 were included (via DOI: 10.55876/gis8.240923dt) in phylogenetic inference, of which the inferred origins of 5,296,695 sequences were directly used for model fitting. Detailed methodologies are provided in the next section.

## *Detailed Data Processing Steps*

## Air travel data

The OpenSky COVID-19 Flight Dataset contains more than 123 million individual flight records between 2019 and 2022. The origin and destination airports determined based on the Automatic Dependent Surveillance-Broadcast (ADS-B) trajectories on approach/takeoff, and aircraft information was sourced from the OpenSky aircraft database. We summarized the daily number of departure and arrival flights between each of studied countries/regions. The total number of air travelers were estimated assuming 80% of the maximum passenger capacity for each flight and 80% of the flights carrying passengers (not cargo).

The passenger load capacity for each flight was determined by the registered wake turbulence category (WTC) based on their maximum certificated takeoff mass. We approximated light ("L"), small ("S"), medium ("M", e.g., Boeing 737), and heavy ("H", e.g., Boeing 777) aircrafts as having maximum passenger load of 8, 19, 250, 550 respectively. We call these estimated daily air travel statistics ADS-B estimates.

Yearly summary ADS-B estimates matched well with ICAO and WTO yearly census data from the World Bank. For example, strong positive correlations and high representativeness were found between ADS-B estimates and WTO data (International inbound and outbound tourists (overnight visitors) with indicator code "ST.INT.ARVL" and "ST.INT.DPRT") (R=0.71, P<0.001, Figure S7a) and ICAO data (Air passengers carried with indicator code "IS.AIR.PSGR") (R=0.59, P<0.001, Figure S7b) among Europe and North American countries in 2019. This suggests that the ADS-B estimates reliably reflect real-world air-travel movement patterns. The OpenSky Network coverage is strong in areas globally with high air traffic concentrations[1], excluding mainland China, making it a representative sample of the most significant routes and airspaces.

Due to limited flight data coverage in certain countries, such as those with few ADS-B feeding receivers, we filled the gaps by fitting a gravity model linking ICAO data with the ADS-B estimates, and using the trained model to predict daily air travel movements in countries without ADS-B coverage. We used ICAO data instead of WTO data due to the latter's extensive missing values post-2019. We adjusted the ICAO air travel passenger for US to 36.3%, cross-referencing ICAO and WTO data, to account for the high number of domestic flights included in the ICAO dataset. The gravity model is a generalized linear model with Poisson regression that accounts for several variables, including the ICAO flows, the GDP, the urban population, and distance between countries. The model was fitted separately for each year from 2019 to 2021, excluding mainland China which is a known outlier[1].

As the ICAO data for 2022 was unavailable for most countries, we use the 2021 data to fit the 2022 ADS-B data in the 2022 model. Data on GDP per capita (indicator: "NY.GDP.PCAP.CD") and urban population ("SP.URB.TOTL") were retrieved from World Development Indicators (World Bank). The distance between the origin and destination countries was determined using the shortest distance between two points, or "great-circle-distance", based on the "Vincenty (sphere)" approach, which assumes a spherical earth. As a result, the gravity model predicts yearly air travel flows between countries, and we assumed a uniform distribution throughout the year. Daily movements were calculated by dividing yearly statistics by 365.25. For air travel in and out of mainland China, we applied an additional 10% scalar to account for strict travel restrictions from 2020 to 2022.

The final movement matrix between countries was constructed by combining ADS-B estimates (used as primary data source) and gravity model estimates (used only when ADS-B data was unavailable).

## Phylogenetic Analysis

We obtained SARS-CoV-2 sequences from the GISAID hCoV database (accessed on October 16, 2023) for the period between December 1, 2019 and March 15, 2022. Sequences without sufficient metadata, such as ambiguous pangolin lineage and country/region code, were excluded. The remaining sequences were categorized into four distinct strain groups, 1. Omicron BA.1; 2. Omicron BA.2; Delta; and Others. We then filtered out sequences with low quality: After removing sequences with duplicated isolate names, we used NextClade (v 3.2.1)[4] to exclude sequences that: (i) had less than 90% sequence coverage, (ii) contained more than 30 ambiguous bases, (iii) deviated with the Wuhan01 reference genome (NCBI accession: MN908947) by more than 50 nucleotide mutations, or (iv) were flagged by NextClade for an excessive number of clustered or private mutations.

The resulting dataset comprises 60% of the unfiltered data, containing 4,258,762, 2,287,989, 654,916, and 2,493,154 sequences for the Delta, Omicron BA.1, Omicron BA.2, and Others groups, respectively. Constructing phylogenetic trees and making reliable inferences on such large datasets using traditional methods, such as maximum likelihood or Bayesian approaches, is computationally infeasible. Therefore, we employed a hybrid approach that combines phylogenetic reconstruction with sequence placement to infer and annotate the possible origins of the sequences. Our approach entails five major steps:
   1. Constructing Background Trees: We created background trees for each of the four strain groups using subsampled representative sequences.
   2. Ancestral Node Location Inference: We inferred the locations of all ancestral nodes within the background trees.
   3. Sequence Placement: Remaining sequences were placed into the background trees based on branch parsimony scores (BPS), in sequential time order.
   4. Origin Determination: We determined the origins of newly placed sequences using the location information of the ancestral nodes in the background trees.
   5. Origin Rectification for Imported-Like Sequences: We constructed separate phylogenetic trees for imported-like sequences to refine origin assignments.

For background tree construction, we selected representative sequences using a subsampling pipeline (DOI: 10.5281/zenodo.7065455), designed for phylogeographic analysis. The pipeline weighted sequences by case counts per geographic location and per unit of time (weekly in our case). We used global daily case count files from the CSSE at Johns Hopkins University (https://github.com/CSSEGISandData/COVID-19) and applied a weekly unit for analysis. Subsample sequences represented approximately 20,000 sequences for each strain group, yielding 22,575, 19,595, 23,888, and 24,223 sequences in the Delta, Omicron BA.1, Omicron BA.2, and Others groups, respectively. We aligned subsampled sequences with Nextalign (v2.14.0) against the Wuhan01 reference genome. Finally, we generated maximum likelihood trees for each strain group using FastTree (v2.1.11) with default parameters. We also used Tempest (v1.5.3) to assess the regression of root-to-tip genetic distance against sampling time and removed sequences with high residuals.

For ancestral state reconstruction, we used TreeTime and TreeTime mugration separately for each strain group. After constructing molecular clock phylogenetic trees with a clock rate of 0.0008 and a standard deviation of 0.0004, we employed TreeTime mugration with default parameters to infer migration dynamics between geographic regions. The resulting annotated trees, with the geographic states added to each node, were used in downstream analyses.

For sequence placement and origin inference, we utilized the UShER (v0.6.2) toolkit combined with a customized annotation strategy. Sequences were sequentially incorporated into the background trees in daily intervals. We inferred potential geographic origins for each sequence, classifying it as an imported-like sequence if the ancestral node's geographic state differed from its collection region. If multiple sequences shared the same ancestral nodes, we retained only those from foreign regions that were first-occurring within a consecutive 30-day period. This approach identified a set of imported-like sequences (N = 96,725 [2.3%], 60,966 [2.7%], 26,990 [4.1%], 61,637 [2.5%]).

To detect finer-scale geographic migration events, we constructed additional phylogenetic trees focusing on imported-like sequences. Using TreeTime mugration, we re-annotated the new imported-like-only trees. Consequently, some previously inferred imported-like sequences were reclassified as either (i) community origin if the collection region matched the ancestral node's geographic state, or (ii) a new origin if the previous inferred origin did not correspond with the ancestral node's geographic state. This final dataset included imported sequences with origin information (N = 58,112 [1.4%], 46,363 [2%], 22,000 [3.4%], 38,025 [1.5%]). All other unclassified sequences were considered to be of community origin.

## *Models*

We developed a series of spatiotemporal partially observed Markov process (SpatPOMP) models to simulate the global spread of SARS-CoV-2 variants and evaluate genomic surveillance strategies. The following part is a brief overview of the model settings; detailed implementation is available in the next section, and source code is provided.

### Primary Model ($M_0$)

Our primary model, $M_0$, is a multi-strain metapopulation model simulating the co-circulation of four viral strains across 29 spatial units from September 15, 2021, to March 15, 2022. The spatial units comprise 24 individual countries or regions and five aggregated units representing other regions within continents (Asia, Europe, North America, South America, and Oceania) (see Figure 1a).

### Model Structure

The model tracks individuals through compartments representing: **Susceptible ($S_u$), Vaccinated ($V_u$), Exposed ($E_{u,k}$), Infectious ($I_{u,k}$), Recovered ($R_u$), and Deceased ($D_u$). Here, $u$ denotes the spatial unit, and $k$ represents the viral strain. Exposed and infectious compartments are further divided into:**
        **Community Cases ($c$);**
        **Imported Cases ($i$), which include:**
            -  **Direct Arrivals**
            -  **Transit Passengers**

## Transmission Dynamics

**The force of infection for strain $k$ in unit $u$ at time t is defined as:**

$$\Psi_{uk}(t) = \eta_k \beta_u \frac{\bar{I}_{uk}(t)}{N_u(t)}$$

Where:

$\eta_k$: Strain-specific transmissibility adjustment.

$\beta_u$: Unit-specific transmission rate.

$\bar{I}_{uk}(t)$: Total infectious individuals in unit $u$ for strain $k$, accounting for both community and imported cases.

$N_u(t)$: Total population in unit $u$ at time $t$.

Travel movements are explicitly modeled using a time-varying movement matrix $M_{ju}(t)$, capturing the number of individuals traveling from unit $j$ to unit $u$ based on the air travel data. Transit passengers are accounted for by incorporating a transit fraction $q_u$ for each unit $u$.

## Emergence of Variants

The emergence of Omicron BA.1 and BA.2 was simulated by introducing 2,000 initial infections into the community-infected population of South Africa at specific time points ($\tau_{BA.1}$ and $\tau_{BA.2}$). This approach minimizes stochastic effects in early transmission dynamics.

## Additional Models

We developed supplementary models for comparative analyses:

$M_1$: Similar to $M_0$ but without origin-inference sequencing data, simplifying traveler case modeling.

$M_2$: An autoregressive statistical model serving as a benchmark to assess the benefits of our mechanistic modeling approach.

$M_3$: Extends $M_0$ with adjustable parameters affecting IDR, DSR, traveler surveillance weight ($\pi$), and different travel hub configurations (Table S4).

$M_4$: Based on $M_3$ but simulates a future pandemic scenario with neutral initial conditions, no travel restrictions, and average air travel data from 2019. The novel variant was modeled as Omicron BA.1 against an existing Delta variant.

## Observation Process

Our observation model links the latent states to observed data, accounting for under-reporting and variability in testing and sequencing efforts. Key components include:

**Infection-to-Diagnostic Ratio (IDR)** ($\delta_u(t)$): Proportion of infections detected through testing in unit $u$ at time $t$.

**Diagnostic-to-Sequencing Ratio (DSR)** ($\rho_u(t)$): Proportion of positive cases that are sequenced in unit $u$ at time $t$.

## Reporting Equations

**Reported Cases**:

$$Y_u^{cases}(t_n) \sim \text{Normal}\big(\delta_u(t_n)\big(\Delta N_{I_u R_u}(t_n) + \Delta N_{I_u D_u}(t_n)\big), \theta_{cases}\big) - Y_u^{deaths}(t_n),$$

**Reported Deaths**:

$$Y_u^{deaths}(t_n) \sim \text{Normal}\big(\Delta N_{I_u D_u}(t_n), \theta_{deaths}\big),$$

**Sequenced Cases**:

$$Y_{u,k}^{seq\_all}(t_n) \sim \text{Normal}\big(\rho_u(t_n)\delta_u(t_n)\big(\Delta N_{I_{u,k,c} R_u}(t_n) + \Delta N_{I_{u,k,c} D_u}(t_n)\big), \theta_{seq\_all}\big),$$

$$Y_{u,k,o}^{seq\_unit}(t_n) \sim \text{Normal}\big(\rho_u(t_n)\delta_u(t_n)(\Delta N_{I_{u,k,i,o} R_u}(t_n) + \Delta N_{I_{u,k,i,o} D_u}(t_n)), \theta_{seq\_unit}\big).$$

Where:

$Y_u^{cases}(t_n)$ and $Y_u^{deaths}(t_n)$: Reported cases and deaths at time $t_n$.

$\Delta N_{I_u R_u}(t_n)$ and $\Delta N_{I_u D_u}(t_n)$: Changes in the number of recoveries and deaths.

$\theta_{cases}$ and $\theta_{deaths}$: Variance parameters accounting for overdispersion.

$Y_{u,k}^{seqall}(t_n)$: Total sequenced cases for strain $k$ in unit $u$.

$Y_{u,k,o}^{seq\_unit}(t_n)$: Sequenced cases for strain $k$ in unit $u$ originating from unit $o$.

## Initial Conditions

Initial compartment sizes were set based on IHME data as of September 15, 2021:

**Vaccinated Individuals** ($V_u(t_0)$): Sum of effectively vaccinated individuals and cumulative infections among the unvaccinated.

**Exposed and Infectious Individuals**:

$E_{u,k,c}(t_0)$): Infections from days $-10$ to $-7$ prior to $t_0$.

$I_{u,k,c}(t_0)$: Infections from days $-6$ to $-1$ prior to $t_0$.

**Imported Cases**: Estimated based on travel volumes and the proportion of transit passengers ($q_u$).

## Model Fitting

We employed likelihood-based methods to estimate parameters:

**Iterated Filtering**: Used algorithms like the Iterated Unadapted Bagged Filter (IUBF)[5] and the Iterated Block Particle Filter (IBPF)[6] to maximize the likelihood function.

**Profiling**: Identified parameter values that individually maximized the likelihood.

Model fitting involved a burn-in period to reach reasonable starting points, followed by formal optimization. Likelihoods were estimated using particle filtering methods, ensuring that the models adequately captured the observed data trends in cases, deaths, and sequenced variant counts across spatial units. Details are provided in the next section.

## Model Validation and Comparison

We validated model $M_0$ by comparing its simulated outputs to real-world data, including daily cases, deaths, and sequenced variant counts. We compared $M_0$ to a simpler implementation in $M_1$, validating the robustness of the model fitting methods. We also compared $M_0$ to benchmark statistical model $M_2$, which uses autoregressive methods to predict case numbers. The higher log-likelihood values achieved by $M_0$ indicate that it more effectively captures the underlying mechanisms of disease transmission. Detailed comparisons are provided in the next section.

# *Detailed Model Specifications*

# General settings

Our tested models all comprise 29 spatial units, where 24 of them are natural units (representing 24 of the 136 administrative regions) and five aggregated units (representing the remaining administrative regions stratified by five continents). Specifically, the 24 natural units were selected as they ranked among the top three, in terms of either total reported cases or overall traffic flow, in their respective continents. The remaining units were grouped into aggregated continent-based units, labeled AS_others (Asia), EU_others (Europe), NA_others (North America), SA_others (South America), and OC_others (Oceania).

Our study includes five models, i.e., $M_0, M_1, M_2, M_3$ and $M_4$ (Table S1). The SpatPOMP model $M_0$ – a multi-strain metapopulation model with origin-inference sequencing data, is used as a base model, for calibration of essential parameters. For comparison, $M_1$ – a multi-strain metapopulation model without origin-inference sequencing data ($M_0$ excluding origin inference) and $M_2$ – an autoregressive normal model fitted to the time-series observations from $M_0$, were developed and calibrated. $M_3$ ($M_0$ with adjustable sequencing propensity) and $M_4$ ($M_3$ with a neutral initial condition) are variants of $M_0$, with different observational process or different covariates, designed for simulation under different conditions. The metapopulation models are partially observed Markov process (POMP) models with spatial structure, as known as SpatPOMP models[7]. A SpatPOMP model comprises two main processes, a latent dynamic process (which is a continuous-time Markov chain) and an observation process (which takes noisy and /or incomplete observational values from the latent process at different timepoints). The two process (the latent process and the observation process) models of $M_0, M_1, M_3$ and $M_4$ are described in the following sections.

In the $M_0$ model, we examined a scenario outlining the emergence of the Omicron variant, focusing on 29 spatial units from September 15, 2021, to March 15, 2022, and encompassing four viral variants: Delta, Omicron BA.1, Omicron BA.2, and Others. The same data was also used for $M_1, M_2$ and $M_3$. In $M_4$, we illustrate and simulate for a scenario with wider application, by applying a movement matrix in "peace time" (i.e., the traffic data in 2019, before COVID-19 pandemic), and by considering more spatial units (N=133) and only two viral variants ("existing" and "new"), with the aim of providing a thorough evaluation of genomic surveillance efficiency for future pandemics. Most parameters in $M_4$ are inherited from the fitted parameters of $M_0$. We utilize parameters estimated for the Delta and Omicron BA.2 variants to represent the hypothetical "existing" and "new" variants, respectively, in $M_4$. For spatial units directly incorporated in $M_0$, we apply the estimated unit-specific parameters (e.g., $\beta_{0,u}$ and $q_u$) to $M_4$. Meanwhile, for newly included spatial units, we integrate the estimates from the aggregated units within each continent into their respective units in $M_4$.

## Model Structure

### The latent process model of $M_0, M_3,$ and $M_4$

The latent state is represented as $\boldsymbol{X}(t) = (S_u(t), V_u(t), E_{uk}(t), I_{uk}(t), R_u(t), D_u(t), u \in 0 : U, k \in 1 : K)$, where $u$ denotes a spatial unit among the $U$ countries/regions under study and $k$ represents one of the $K$ circulating viral variants or strains. Each individual within a spatial unit is assigned to one of the following compartments: susceptible ($S_u$), immunized ($V_u$), exposed ($E_{u,k}$), infected and infectious ($I_{u,k}$), recovered ($R_u$) and deceased ($D_u$). The exposed ($E_{u,k}$) and infected ($I_{u,k}$) compartments are further divided into imported ($i$) and community ($c$) cases, indicated as $E_{u,k,i}, E_{u,k,c}, I_{u,k,i}$ and $I_{u,k,c}$ respectively. Both immigration and emigration are possible for the imported cases compartments ($E_{u,k,i}$ and $I_{u,k,i}$). However, for community cases ($E_{u,k,c}$ and $I_{u,k,c}$), only emigration is possible.

For some spatial units, a significant proportion of travelers can be classified as transit passengers, who enter and exit a spatial unit without remaining in the local population. To account for this, models $M_0, M_3$, and $M_4$ allow for travelers originating from a spatial unit to have a single transition at an intermediate spatial unit before arriving at their final destination. Further transitions are not considered due to increased model complexity.

The travelling population (represented in the imported compartments $E_{u,k,i}$ and $I_{u,k,i}$) is categorized into direct arrival (*direct*) and transit passengers (*transit*), resulting in the compartments $E_{u,k,direct}, E_{u,k,transit},$ $I_{u,k,direct},$ and $I_{u,k,transit}$. To track the different origins of the direct and transit travelers, we further subdivided these compartments by their origin spatial units ($o \in 1 : U$). Consequently, the total travelling population consists of $E_{u,k,direct,o \in 1:U}, E_{u,k,transit,o \in 1:U}, I_{u,k,direct,o \in 1:U},$ and $I_{u,k,transit,o \in 1:U}$.

For notational convenience, we introduce a transport compartment $T$, and a transiting passenger compartment $P$, to account for all travelling passengers. A flow diagram for the SEIRD model is shown in Figure S1.

The directional transitions between compartments $W$ to $M$ are represented by $N_{WM}(t)$, with the infinitesimal increment $dN_{WM} = N_{WM}(t + dt) - N_{WM}(t)$. We can express the state $\boldsymbol{X}(t) = (X_1(t), \ldots, X_u(t))$ in terms of its initial value $\boldsymbol{X}(t_0)$ and the flow equations:

$$dS_u = -dN_{S_u V_u} - \sum_{k=1}^{K} dN_{S_u E_{u,k,c}},$$

$$dV_u = dN_{S_u V_u} - \sum_{k=1}^{K} dN_{V_u E_{u,k,c}},$$

$$dE_{u,k,c} = dN_{S_u E_{u,k,c}} + dN_{V_u E_{u,k,c}} - dN_{E_{u,k,c} T} - dN_{E_{u,k,c} I_{u,k,c}},$$

$$dI_{u,k,c} = dN_{E_{u,k,c}I_{u,k,c}} - dN_{I_{u,k,c}T} - dN_{I_{u,k,c}R_u} - dN_{I_{u,k,c}D_u},$$

$$dE_{u,k,direct,o} = dN_{TE_{u,k,direct,o}} - dN_{E_{u,k,direct,o}P} - dN_{E_{u,k,direct,o}I_{u,k,direct,o}},$$

$$dI_{u,k,direct,o} = dN_{E_{u,k,direct,o}I_{u,k,direct,o}} + dN_{TI_{u,k,direct,o}} - dN_{I_{u,k,direct,o}P} - dN_{I_{u,k,direct,o}R_u} - dN_{I_{u,k,direct,o}D_u},$$

$$dE_{u,k,transit,o} = dN_{PE_{u,k,transit,o}} - dN_{E_{u,k,transit,o}I_{u,k,transit,o}},$$

$$dI_{u,k,transit,o} = dN_{E_{u,k,transit,o}I_{u,k,transit,o}} + dN_{PI_{u,k,transit,o}} - dN_{I_{u,k,transit,o}R_u} - dN_{I_{u,k,transit,o}D_u},$$

$$dR_u = \sum_{k=1}^{K}(dN_{I_{u,k,i}R_u} + dN_{I_{u,k,c}R_u}),$$

$$dD_u = \sum_{k=1}^{K}(dN_{I_{u,k,i}D_u} + dN_{I_{u,k,c}D_u}).$$

In this model, individuals travel between spatial units only when in the exposed ($E$) or infected and infectious ($I$) compartments. Note that there may be minor stochastic variations in the total population as we do not track individual entries and exits from the transport compartment $T$.

Each transition $dN_{WM}$ is associated with a rate, $\mu_{WM}$, which may depend on the state of other compartments, covariate processes, parameters, or time. For all compartments, except those related to travel ($T$ or $P$), rates are specified per capita. The non-zero transition rates are:

$$\mu_{S_uV_u} = v_u(t),$$

$$\mu_{S_uE_{u,k,c}} = \Psi_{u,k}(t)\Gamma(t),$$

$$\mu_{V_uE_{u,k,c}} = \lambda_k\Psi_{u,k}(t)\Gamma(t),$$

$$\mu_{E_{u,k,c}I_{u,k,c}} = \mu_{E_{u,k,direct,o}I_{u,k,direct,o}} = \mu_{E_{u,k,transit,o}I_{u,k,transit,o}} = \varepsilon_k,$$

$$\mu_{I_{u,k,c}D_u} = \mu_{I_{u,k,direct,o}D_u} = \mu_{I_{u,k,transit,o}D_u} = h_u(t)\gamma,$$

$$\mu_{I_{u,k,c}R_u} = \mu_{I_{u,k,direct,o}R_u} = \mu_{I_{u,k,transit,o}R_u} = (1 - h_u(t))\gamma.$$

The force of infection is given by:

$$\Psi_{uk}(t) = \eta_k\beta_u\frac{\tilde{I}_{uk}(t)}{N_u(t)}$$

where $\eta_k$ and $\beta_u$ together represent the strain-specific transmissibility adjustment and unit-specific transmission rate, respectively. $\tilde{I}_{uk}(t) = \zeta_u(t)I_{u,k,i}(t) + I_{u,k,c}(t)$ is the total number of infectious individuals in unit $u$ for strain $k$, accounting for both community and imported cases. $N_u(t)$ is the total population in unit $u$ at time $t$. $\varepsilon_k$ is the inverse of the incubation period for strain $k$. $h_u(t)$ is the infection-to-fatality ratio (IFR) which specifies the fraction of infected cases that are fatal at time $t$ in unit $u$, and is a pre-estimated covariate process. $\gamma$ is the rate at which infected individuals either recover or die. We define $\Gamma(t) = d\Gamma/dt$ as non-negative multiplicative gamma white noise with variance parameter $\sigma$. That is, $\Gamma(t)$ is a gamma process with stationary independent increments, such that $\Gamma(t) - \Gamma(s)$ is gamma distributed with mean $t - s$ and variance $\sigma(t - s)$[8,9].

*Travel-Related Transitions*

Transitions involving the transport ($T$) and transiting passenger ($P$) compartments (Figure S1) depend on the time-varying transport matrix, $M_{ju}(t)$, which describes the number of individuals moving from unit $j$ to unit $u$ at time $t$. The travel-related transitions are as follows:

$$dN_{E_{u,k,c}T} = Pois(\sum_{j}\frac{M_{uj}(t)E_{u,k,c}}{N_u}),$$

$$dN_{I_{u,k,c}T} = Pois(\sum_{j}\frac{M_{uj}(t)I_{u,k,c}}{N_u}),$$

$$dN_{TE_{u,k,direct,o}} = Pois\left(\frac{M_{ou}(t)E_{o,k,c}}{N_o}\right),$$

$$dN_{E_{u,k,direct,o}P} = Pois(q_u dN_{TE_{u,k,direct,o}}),$$

$$dN_{TI_{u,k,direct,o}} = Pois\left(\frac{M_{ou}(t)I_{o,k,c}}{N_o}\right),$$

$$dN_{I_{u,k,direct,o}P} = Pois(q_u dN_{TI_{u,k,direct,o}}),$$

$$dN_{PE_{u,k,transit,o}} = \sum_j dN_{E_{j,k,direct,o}P} \frac{M_{ju}(t)}{\sum_m M_{jm}(t)},$$

$$dN_{PI_{u,k,transit,o}} == \sum_j dN_{I_{j,k,direct,o}P} \frac{M_{ju}(t)}{\sum_m M_{jm}(t)}.$$

where $q_u$ specifies the fraction of arriving travelers that are transit passenger in unit $u$. $Pois(n)$ denotes a Poisson distribution with rate $n$. In $M_4$, we use the average $M_{ju}(t)$ from 2019, to represent a neutral travelling data during non-pandemic time (i.e., not under public health emergence conditions).

*Modeling Variant Emergence*

The emergence of BA.1 and BA.2 variants was simulated by introducing 2,000 initial infections into the community-infected population at the presumed origin region, as simulations with fewer cases resulted in inconsistent dynamics due to stochastic effects. Parameters $\tau_{BA.1}$ and $\tau_{BA.2}$ denote the time points when 2,000 BA.1 or BA.2 cases were introduced into the exposed and infected compartments of South Africa. Of note, 1000 infections approximate to <3% of the total infections in South Africa on September 9, 2021, based on the IHME estimates. In $M_3$ and $M_4$, we include an additional parameter $o_{in} \in 1:29$, representing the random origin of the novel variant. This parameter is sampled proportional to the population size of different spatial units.

*Latent process model of $M_1$*

Model $M_1$ inherits notations from $M_0$ but dose not track the number of incoming travelers based on inferred origins. Consequently, the transitions for imported cases are simplified:

$$dE_{u,k,direct} = dN_{TE_{u,k,direct}} - dN_{E_{u,k,direct}P} - dN_{E_{u,k,direct}I_{u,k,direct}},$$

$$dI_{u,k,direct} = dN_{E_{u,k,direct}I_{u,k,direct}} + dN_{TI_{u,k,direct}} - dN_{I_{u,k,direct}P} - dN_{I_{u,k,direct}R_u} - dN_{I_{u,k,direct}D_u},$$

$$dE_{u,k,transit} = dN_{PE_{u,k,transit}} - dN_{E_{u,k,transit}I_{u,k,transit}},$$

$$dI_{u,k,transit} = dN_{E_{u,k,transit}I_{u,k,transit}} + dN_{PI_{u,k,transit}} - dN_{I_{u,k,transit}R_u} - dN_{I_{u,k,transit}D_u}.$$

The transition rates are adjusted accordingly:

$$\mu_{E_{u,k,c}I_{u,k,c}} = \mu_{E_{u,k,direct}I_{u,k,direct}} = \mu_{E_{u,k,transit}I_{u,k,transit}} = \varepsilon_k,$$

$$\mu_{I_{u,k,c}D_u} = \mu_{I_{u,k,direct}D_u} = \mu_{I_{u,k,transit}D_u} = h_u(t)\gamma,$$

$$\mu_{I_{u,k,c}R_u} = \mu_{I_{u,k,direct}R_u} = \mu_{I_{u,k,transit}R_u} = (1 - h_u(t))\gamma.$$

The travel-related transitions are simplified as follows:

$$dN_{TE_{u,k,direct}} = Pois\left(\sum_j \frac{M_{ju}(t)E_{j,k,c}}{N_j}\right),$$

$$dN_{E_{u,k,direct}P} = Pois(q_u dN_{TE_{u,k,direct}}),$$

$$dN_{TI_{u,k,direct}} = Pois\left(\sum_j \frac{M_{ju}(t)I_{j,k,c}}{N_j}\right),$$

$$dN_{I_{u,k,direct}P} = Pois(q_u dN_{TI_{u,k,direct}}),$$

$$dN_{PE_{u,k,transit}} = \sum_j dN_{E_{j,k,direct}P} \frac{M_{ju}(t)}{\sum_m M_{jm}(t)},$$

$$dN_{PI_{u,k,transit}} == \sum_j dN_{I_{j,k,direct}P} \frac{M_{ju}(t)}{\sum_m M_{jm}(t)}.$$

Other transitions remain the same as in $M_0$.

*Observation process model of $M_0$*

For each spatial unit $u$ at time $t_n$, the observed data include daily reported cases and deaths, the cumulative number of sequences for four variants, as well as $K \times U = 116$ inferred numbers of sequenced cases for four variants across 29 different origins. Daily reported cases ($Y_u^{cases}(t_n)$) and deaths ($Y_u^{deaths}(t_n)$) are modelled as random variables reflecting the fraction of latent infections.

Both cumulative and the unit-specific sequenced cases for each variant are represented as $Y_{u,k}^{seq\_all}(t_n)$ and $Y_{u,k,o}^{seq\_unit}(t_n)$, respectively. These values account for community and imported cases categorized by origin ($o$). The models are defined as follows:

$$Y_u^{cases}(t_n) \sim \text{Normal}\big(\delta_u(t_n)\big(\Delta N_{I_u R_u} + \Delta N_{I_u D_u}\big), \theta_{cases}\big) - Y_u^{deaths}(t_n),$$

$$Y_u^{deaths}(t_n) \sim \text{Normal}\big(\Delta N_{I_u D_u}(t_n), \theta_{deaths}\big),$$

$$Y_{u,k}^{seq\_all}(t_n) \sim \text{Normal}\big(\rho_u(t_n)\delta_u(t_n)(\Delta N_{I_{u,k,c}R_u}(t_n) + \Delta N_{I_{u,k,c}D_u}(t_n)), \theta_{seq\_all}\big),$$

$$Y_{u,k,o}^{seq\_unit}(t_n) \sim \text{Normal}\big(\rho_u(t_n)\delta_u(t_n)(\Delta N_{I_{u,k,i,o}R_u}(t_n) + \Delta N_{I_{u,k,i,o}D_u}(t_n)), \theta_{seq\_unit}\big).$$

where $\Delta N_{I_u R_u}(t_n) = N_{I_u R_u}(t_n) - N_{I_u R_u}(t_{n-1})$ and $\Delta N_{I_u D_u}(t_n) = N_{I_u D_u}(t_n) - N_{I_u D_u}(t_{n-1})$ represent transitions from infected to recovered and deceased states, respectively. Similarly, $\Delta N_{I_{u,k,c}R_u}(t_n)$ and $\Delta N_{I_{u,k,c}D_u}(t_n)$ correspond to recovery and death transitions for community cases, and $N_{I_{u,k,i,o}R_u}(t_n)$ and $\Delta N_{I_{u,k,i,o}D_u}(t_n)$ represent these transitions for imported cases from different origins. Here, $\text{Normal}(\Delta N, \theta)$ denotes a normal distribution with mean $\Delta N$ and variance $\theta$. $\delta_u(t)$ and $\rho_u(t)$ are covariate processes specifying infection-to-diagnostic ratio (IDR) and diagnostic-to-sequencing ratio (DSR), respectively.

*Observation process model of $M_1$*

In $M_1$, community and imported sequences are not differentiated, so only cumulative sequenced cases for each variant ($Y_{u,k}^{seq\_all}$) are modeled, while $Y_{u,k,o}^{seq\_unit}$ is excluded. The observation processes for daily reported cases and deaths remain unchanged.

*Observation process model of $M_3$ and $M_4$*

Models $M_3$ and $M_4$ build on $M_0$ by incorporating additional parameters, $\kappa_{IDR}, \kappa_{DSR}, \pi$, and traveler-targeted strategies, to explore different genomic surveillance strategies. Specifically, $\kappa_{IDR}$ and $\kappa_{DSR}$ adjust IDR/diagnostic capacity and DSR/sequencing capacity relative to $M_0$ baseline. $\pi$ controls the focus on imported cases in diagnostic and sequencing efforts. Specifications of traveler-targeted strategies are listed in Table S4. These parameters can be adjusted for each unit or shared across units to simulate various scenarios and assess the effectiveness of surveillance strategies in detecting novel variants. For example, for figure 3e, the hypothetical strategy for concentrated resources in hubs was simulated by scaling up both diagnostics and sequencing using separate adjustment factors ($\kappa_{IDR}$ and $\kappa_{DSR}$). These scaling factors were calculated to ensure that the global total effort remained consistent before and after reallocation.

*Initial conditions for $M_0, M_1, M_3$, and $M_4$*

All models start from September 15, 2021, when the Delta variant was widespread. The initial vaccinated/immunized ($V_u(t_0)$) is estimated by combining the number of effectively vaccinated (one and two dose with efficacy) individuals and cumulative infections among the unvaccinated (mean estimate), based on IHME data. This setting assumes vaccine-induced immunity is the same as that conferred by natural infection. The effectiveness of immunity is measured by $1 - \lambda_k$, and the immunized population subject to exposed for strain $k$ is $\lambda_k V_u(t_0)$.

The exposed ($E$) and ($I$) compartments are initiated using infections from the past 7 to 10 days and 1 to 6 days, respectively:

$$E_{u,k,c}(t_0) = N_{u,k}^{\{i \in \mathbb{Z} | -10 \le i \le -7\}},$$

$$E_{u,k,i}(t_0) = N_{u,k}^{\{i \in \mathbb{Z} | -10 \leq i \leq -7\}} \frac{[-7-(-10)+1]E(\sum_m M_{mu}(t))}{N_u},$$

$$E_{u,k,direct,o}(t_0) = 0.86 \frac{E_{u,k,i}(t_0)}{U},$$

$$E_{u,k,transit,o}(t_0) = 0.14 \frac{E_{u,k,i}(t_0)}{U},$$

$$I_{u,k,c}(t_0) = N_{u,k}^{\{i \in \mathbb{Z} | -6 \leq i \leq -1\}},$$

$$I_{u,k,i}(t_0) = N_{u,k}^{\{i \in \mathbb{Z} | -6 \leq i \leq -1\}} \frac{[-1-(-6)+1]E(\sum_m M_{mu}(t))}{N_u},$$

$$I_{u,k,direct,o}(t_0) = 0.86 \frac{I_{u,k,i}(t_0)}{U},$$

$$I_{u,k,transit,o}(t_0) = 0.14 \frac{I_{u,k,i}(t_0)}{U}.$$

Where 0.14 represents the global average of transition travelers ($q_u$). $R(t_0)$ is set to 0, and the initial susceptible population is calculated as follows:

$$S_u(t_0) = N_u - V_u(t_0) - \sum_k E_{u,k,c}(t_0) - \sum_k E_{u,k,i}(t_0).$$

*Traveler-Targeted Surveillance Strategies ($M_3$ and $M_4$)*

To evaluate the impact of traveler-targeted surveillance on early variant detection, we tested multiple resource allocation strategies focused on selected travel hubs. Travel hubs were prioritized based on rankings in total (T-ranked) or per-capita (P-ranked) travel volume, as detailed in Table S2. To ensure geographic diversity, continental representative hubs were optionally included in some configurations (denoted with "a").

Resource allocation across selected travel hubs was adjusted using three redistribution scenarios, designed to reflect varying levels of resource concentration: (1) R-type (Regular Redistribution): Traveler weights were adjusted within selected hubs without drawing additional resources from non-hub regions. This minimal adjustment maintained overall community surveillance while prioritizing selected travel hubs for traveler-targeted efforts; (2) H-type (Half Reallocation): Fifty percent of the surveillance resources from non-hub regions were reallocated to selected hubs. This moderate adjustment enhanced diagnostic and sequencing capacities in selected hubs in proportion to their traveler volumes, while moderately reducing resources for community surveillance in non-hub regions; (3) M-type (Maximal Reallocation): Ninety percent of the surveillance resources from non-hub regions were reallocated to selected hubs. This aggressive strategy maximized surveillance capacities in high-travel-volume hubs, prioritizing traveler-focused efforts at the expense of community surveillance in non-hub regions, including low-travel areas.

These resource allocation strategies were applied to various sets of travel hubs. The configurations included subsets ranked by travel volume (e.g., top 3, top 7, and top 11 hubs) and combinations with or without continental representatives (e.g., T3, T3a, P3, P3a). The resulting strategies provided a systematic framework for examining the trade-offs between concentrating resources in high-travel regions and maintaining broad community-based genomic surveillance.

Changing the relative $R_{eff}$ is achieved by changing the parameter $\eta_2$ in the model, and modeling vaccine effectiveness against infection is by changing the parameter $\lambda_2$.

This approach enabled us to evaluate the efficiency of targeted strategies under varying conditions of resource availability, travel volume, and geographic coverage. Details of the tested configurations are summarized in Table S4.

*Assumptions for the novel-variant scenario under non-pandemic conditions ($M_4$)*

Model $M_4$ assumes no travel restrictions and uses 2019 air travel data (average). For each region, IDR and DSR values are set to average levels observed during the study period. The model simulates the emergence of a novel variant (with parameters adopted from Omicron BA.1 in $M_0$) alongside a circulating existing variant (with parameters based on Delta in $M_0$), introducing it randomly into any of the 29 spatial units. We resampled the simulation data proportional to the population size in each region and get the global summaries.

## Parameter Estimation Procedures and Computational Details

To optimize the parameters (fitting) for model $M_0$, we employed two primary search strategies:

1. **Profiling** involves identifying the highest likelihood value for each parameter while keeping other parameters fixed. This step is repeated to find a combination of parameter values that yields a higher overall likelihood. While this method doesn't guarantee the maximum likelihood estimate, it is particularly effective during the initial stages of optimization when parameter values are not well tuned. The log-likelihood for SpatPOMP models was primarily estimated using the Unadapted Bagged Filter (UBF) and the Block Particle Filter (BPF). Shared parameters were evaluated using UBF, while unit-specific parameters were assessed using both UBF and BPF.

2. **Local searching** maximizes the log-likelihood using the iterated unadapted bagged filter (IUBF) or the iterated block particle filter (IBPF)[5,10,11] methods. IBPF is considered more effective when optimizing models consist mostly unit-specific parameters; however, our results indicated that IUBF provided more stable estimates for shared parameters on our models. These iterative methods refine parameter estimates by progressively decreasing perturbation magnitudes, theoretically guiding the parameters towards the neighborhood of the maximum likelihood estimate[6,12].

During profiling, confidence intervals were constructed using profile likelihood, adjusted for Monte Carlo variability[13,14]. These intervals informed decisions on whether a parameter value should be retained for subsequent rounds of optimization. Each search strategy can be applied using three parameter set levels: low, mid, and high (detailed in the source code). Higher levels were utilized in later stages of fitting to balance between computational time and fitting accuracy.

Initially, all parameter values were set to 0.5 (logit-transformed to range from 0 to 1), resulting in an initial log-likelihood of -8,544,184 (estimated using UBF). A series of 41 rounds of parameter tuning, termed the *burn-in* period, led to a parameter set with a log-likelihood of -6,854,808 (estimated using UBF). After the burn-in period, formal parameter optimization began, primarily using UBF and IUBF, with occasional use of BPF or IBPF for unit-specific parameters. Formal optimization involved 11 rounds with low search parameters, 13 rounds with mid search parameters, and 8 rounds with high search parameters. The final parameter set for $M_0$ achieved a log-likelihood of -3,137,774.241 (estimated using UBF).

The fitting process was terminated when further searches using high search parameters and 640 replications on 20 computational nodes failed to produce higher log-likelihood values. The majority of the computational work was performed on the High Performance Computing service provided by the University of Hong Kong, with additional data curation and summarization conducted on a local server.

The burn-in period required 96,618 CPU hours (15 calendar days, dynamically scheduled with 1,143 individual nodes). The formal optimization phase consumed 245,102 CPU hours (20 calendar days, dynamically scheduled with 1,995 individual nodes). The computational hardware included modern CPUs such as AMD EPYC 7542, 7742, 9654, and Intel GOLD 6626R.

### *Model Comparisons and Statistical Analyses*

## Model Comparisons and Statistical Analyses

To evaluate the performance and adequacy of our mechanistic model ($M_0$), we compared it against two alternative models: a simplified mechanistic model ($M_1$) and a basic statistical model ($M_2$). This benchmarking approach allows us to assess whether the added complexity of $M_0$ provides a tangible improvement in capturing the dynamics of COVID-19 transmission.

## Comparison between $M_0$ and $M_1$

Model $M_1$ is a simplified version of $M_0$. While $M_0$ incorporates phylogenetically inferred introduction events and models detailed transmission dynamics across spatial units, $M_1$ excludes these introduction events and only models daily aggregated data, resembling conventional metapopulation compartmental models without phylogenetic calibration.

To compare the fitting efficiency and validate our parameter optimization methods, we used the same starting point and applied an identical fitting process during the first ten rounds of formal parameter optimization for both $M_0$ and $M_1$. We observed that both models exhibited similar behavior under these fitting strategies, validating the robustness of our parameter optimization algorithm and model settings.

During the first ten rounds of fitting with identical parameters, the log-likelihood for $M_0$ improved from $-6,854,808.115$ to $-5,297,706.367$, while for $M_1$ it improved from $-7,516,558.394$ to $-5,221,950.285$ (Supplementary Data 1, access via https://github.com/Leo-Poon-Lab/global_genomic_surveillance_spatpomp_article/raw/refs/heads/main/results/model_data/Supplementary%20Data%201.xlsx). Notably, $M_1$ runs significantly faster than $M_0$—approximately 25 times faster—due to the reduced number of observational variables (Table S1). For instance, in a single round of local searching using the unadapted bagged filter (UBF) with the "low" searching profile on a computation node equipped with a 32-core Intel GOLD 6226R CPU, $M_0$ took 9,259.5 seconds, whereas $M_1$ took only 362.9 seconds.

One challenge we encountered during the fitting process was optimizing shared parameters across spatial units, particularly when using the block particle filter (BPF), which is more suited for unit-specific parameters. We found that the UBF performed better for our models. For the 20 shared parameters in both $M_0$ and $M_1$, we observed that after the first round of fitting with identical starting values, the direction of parameter changes remained consistent between the two models (Table S5). This consistency suggests that the phylogenetically inferred data used in $M_0$ does not conflict with the aggregated data in our model specification.

Although $M_1$ lacks the capacity to study introduction and exportation dynamics due to the absence of origin-specific compartments, our results indicate that it could be a useful tool for studying metapopulation systems when the focus is primarily on aggregated counts.

## Benchmarking $M_0$ against a Simple Statistical Model ($M_2$)

To further assess the effectiveness of our mechanistic model $M_0$, we benchmarked its performance against a basic statistical model $M_2$. This comparison helps determine whether the added complexity of $M_0$ offers a significant improvement over a simpler model in capturing the observed data.

Model $M_2$ is an autoregressive (AR) model that predicts current case numbers based on previous observations. It estimates the expected number of cases or deaths at time $t_n$ using a weighted combination of a baseline mean and the observed value at time $t_{n-1}$:

$$\mu_{Y_u^{cases}} = \frac{1}{N} \sum_{n=1}^{N} Y_u^{cases}(t_n),$$

$$\mu_{Y_u^{deaths}} = \frac{1}{N}\sum_{n=1}^{N} Y_u^{deaths}(t_n),$$

$$\mu_{Y_{u,k}^{seq\_all}} = \frac{1}{N}\sum_{n=1}^{N} Y_{u,k}^{seq\_all}(t_n),$$

$$\mu_{Y_{u,k,o}^{seq\_unit}} = \frac{1}{N}\sum_{n=1}^{N} Y_{u,k,o}^{seq\_unit}(t_n),$$

where $N$ is the total number of time points, and $n$ indexes the time points from 1 to $N$. The predictions are then made as:

$$Y_u^{cases}(t_n) \sim \text{Normal}\left(\frac{1}{2}(\mu_{Y_u^{cases}} + Y_u^{cases}(t_{n-1})), \theta_{cases}\right),$$

$$Y_u^{deaths}(t_n) \sim \text{Normal}\left(\frac{1}{2}(\mu_{Y_u^{deaths}} + Y_u^{deaths}(t_{n-1})), \theta_{deaths}\right),$$

$$Y_{u,k}^{seq\_all}(t_n) \sim \text{Normal}\left(\frac{1}{2}(\mu_{Y_{u,k}^{seq\_all}} + Y_{u,k}^{seq\_all}(t_{n-1})), \theta_{seq\_all}\right),$$

$$Y_{u,k,o}^{seq\_unit}(t_n) \sim \text{Normal}\left(\frac{1}{2}(\mu_{Y_{u,k,o}^{seq\_unit}} + Y_{u,k,o}^{seq\_unit}(t_{n-1})), \theta_{seq\_unit}\right).$$

We calculated the log-likelihoods for models $M_0$ and $M_2$, ensuring that all likelihoods were computed with respect to the same base measure for a valid comparison (details are specified in source code). The results show that $M_0$ outperforms $M_2$ by achieving a higher log-likelihood value ($-2,964,449$ vs. $-4,542,715$) despite using fewer parameters (Table S1). The superior performance indicates that $M_0$ not only provides a statistically adequate representation of the data but also effectively captures the underlying mechanisms of COVID-19 transmission[15].

## Conclusion

The comparisons between $M_0$, $M_1$, and $M_2$ demonstrate that our mechanistic model $M_0$ offers a significant improvement over simpler models in capturing the dynamics of COVID-19 transmission. While $M_1$ serves as a faster, simplified alternative useful for modeling aggregated counts, $M_0$'s incorporation of phylogenetic data and detailed transmission dynamics provides a more accurate and mechanistically plausible explanation of the observed data.

# Supplemental Figures

**Figure S1: The schematic diagram showing the SIEIRD compartmental model of $M_0$. (a)** The top panel shows the overall configuration of the SIEIRD model in each spatial unit. **(b)** The bottom left panel expanded the travelling population in **(a)**, illustrating the compartments for the direct and transit travelers, respectively. **(c)** Notations for major parameters and covariates used in the model.

a

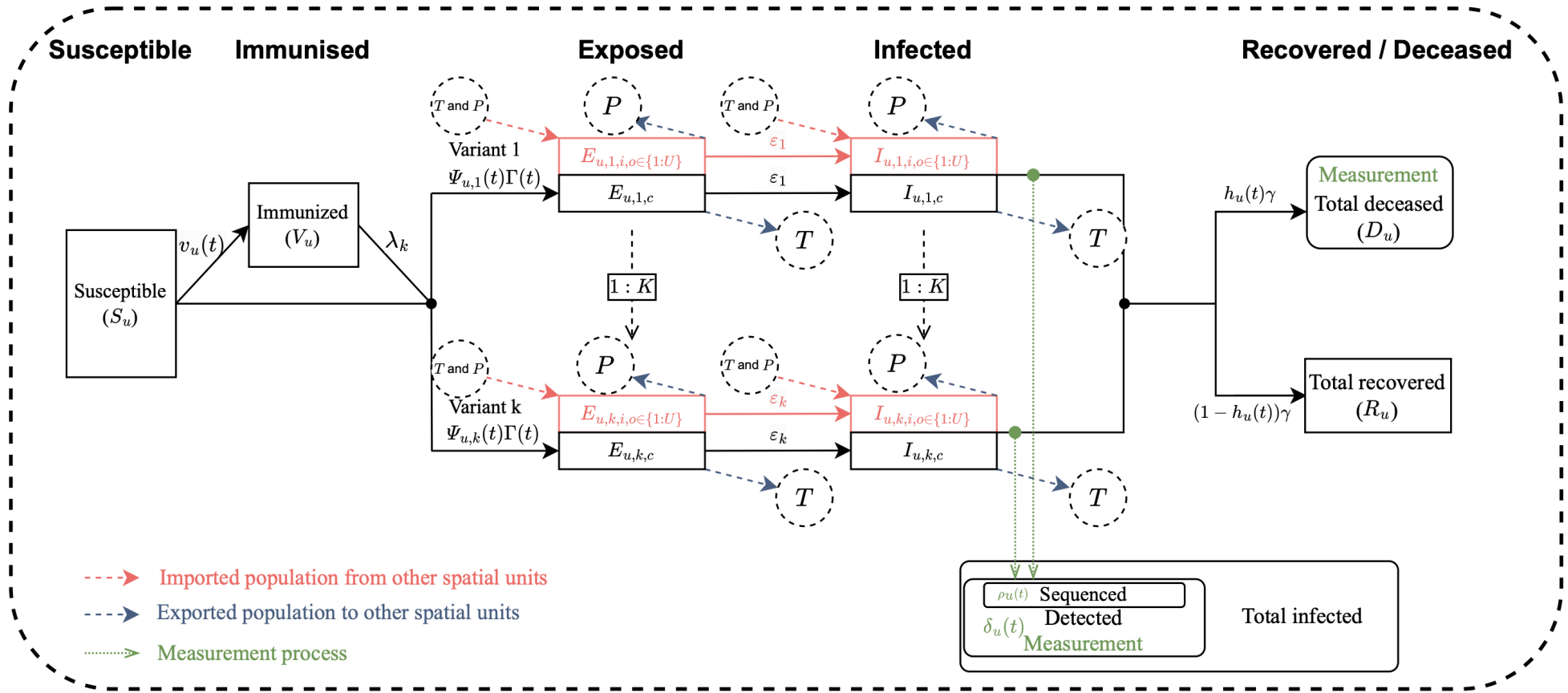

**b**

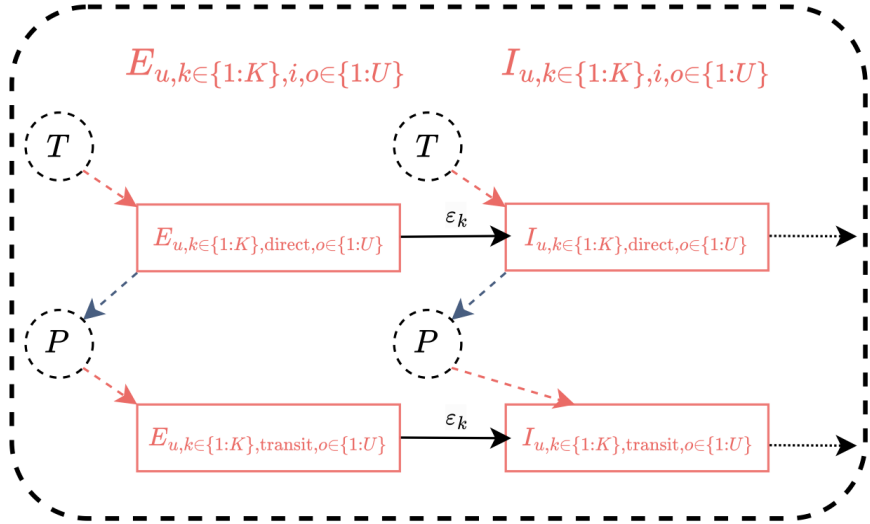

$E_{u,k\in\{1:K\},i,o\in\{1:U\}}$      $I_{u,k\in\{1:K\},i,o\in\{1:U\}}$

$T$        $T$

$E_{u,k\in\{1:K\},\text{direct},o\in\{1:U\}}$   $\xrightarrow{\varepsilon_k}$   $I_{u,k\in\{1:K\},\text{direct},o\in\{1:U\}}$

$P$        $P$

$E_{u,k\in\{1:K\},\text{transit},o\in\{1:U\}}$   $\xrightarrow{\varepsilon_k}$   $I_{u,k\in\{1:K\},\text{transit},o\in\{1:U\}}$

**c**

$$\Psi_{u,k}(t) = \eta_k \beta_u \frac{\widetilde{I}_{u,k}(t)}{N_u(t)}$$

$$\widetilde{I}_{u,k}(t) = \zeta_u(t) I_{u,k,i}(t) + I_{u,k,c}(t)$$

$\beta_u$ : Natural transmissibility in $u$

$N_u(t)$: Population size at time $t$ for unit $u$

$v_u(t)$ : Vaccination rate at time $t$ for unit $u$      $\Gamma(t)$ : Gamma white noise with intensity $\sigma$

$\eta_k$ : Relative increase/decrease in infection rate due to strain $k$

$\lambda_k$ : Relative decrease in infection rate due to immunity for strain $k$

$\zeta_u(t)$: Change in transmissibility due to level $n$ travel control policy at time $t$ for unit $u$

$\varepsilon_k$ : Inverse of the latent/incubation period for strain $k$

$\gamma$ : Rate at which infected individuals become recovered or deceased

$h_u(t)$ : Infection-to-fatality    $\delta_u(t)$: Infection-to-diagnostic ratio at    $\rho_u(t)$: Diagnostic-to-sequencing
ratio at time $t$ in unit $u$      time $t$ in unit $u$      ratio at time $t$ in unit $u$

**Figure S2: The schematic diagram showing the two-step measurement model in $M_0$.**
**(a)** *Step 1*: A fraction of the total infected population is identified through diagnostic testing, determined by the infection-to-diagnostic ratio (IDR), to estimate the total diagnostic population (while not exceeding the diagnostic capacity). *Step 2*: A subset of the diagnostic population is sequenced, determined by diagnostic-to-sequencing ratio (DSR), to estimate the total sequenced population (while not exceeding the sequencing capacity). **(b)** In the models, imported and community populations are treated separately, following the same two-step process described above. In $M_0$, both imported, and community cases undergo Steps 1 and 2 with the same IDR and DSR values. However, in $M_3$ and $M_4$, the models introduce traveler weights, where diagnostic and sequencing capacities are first estimated as totals, then reallocated based on these weights. Notably, we perform checks on the numbers of such reallocated imported populations, ensuring the actual diagnostic and sequencing population size will not exceed the diagnostic/sequencing capacities, or their upper level population sizes (i.e., infected/diagnostic population sizes).

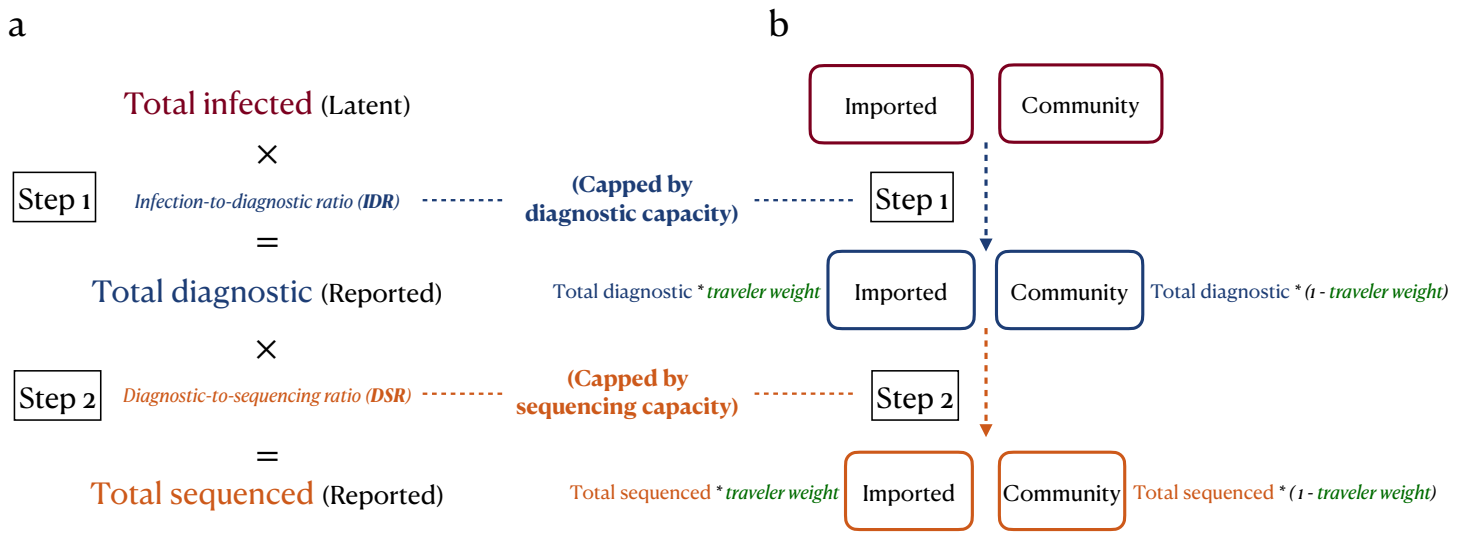

**Figure S3: The temporal changes of infection-to-diagnostic ratio (IDR), diagnostic-to-sequencing ratio (DSR), and travel control level in different spatial units.** The values are shown in percentage. For the travel control level, 100%, 75%, 50%, and 25% refer to the levels 1 to 4 respectively, as specified in Oxford COVID-19 Government Response Tracker data.

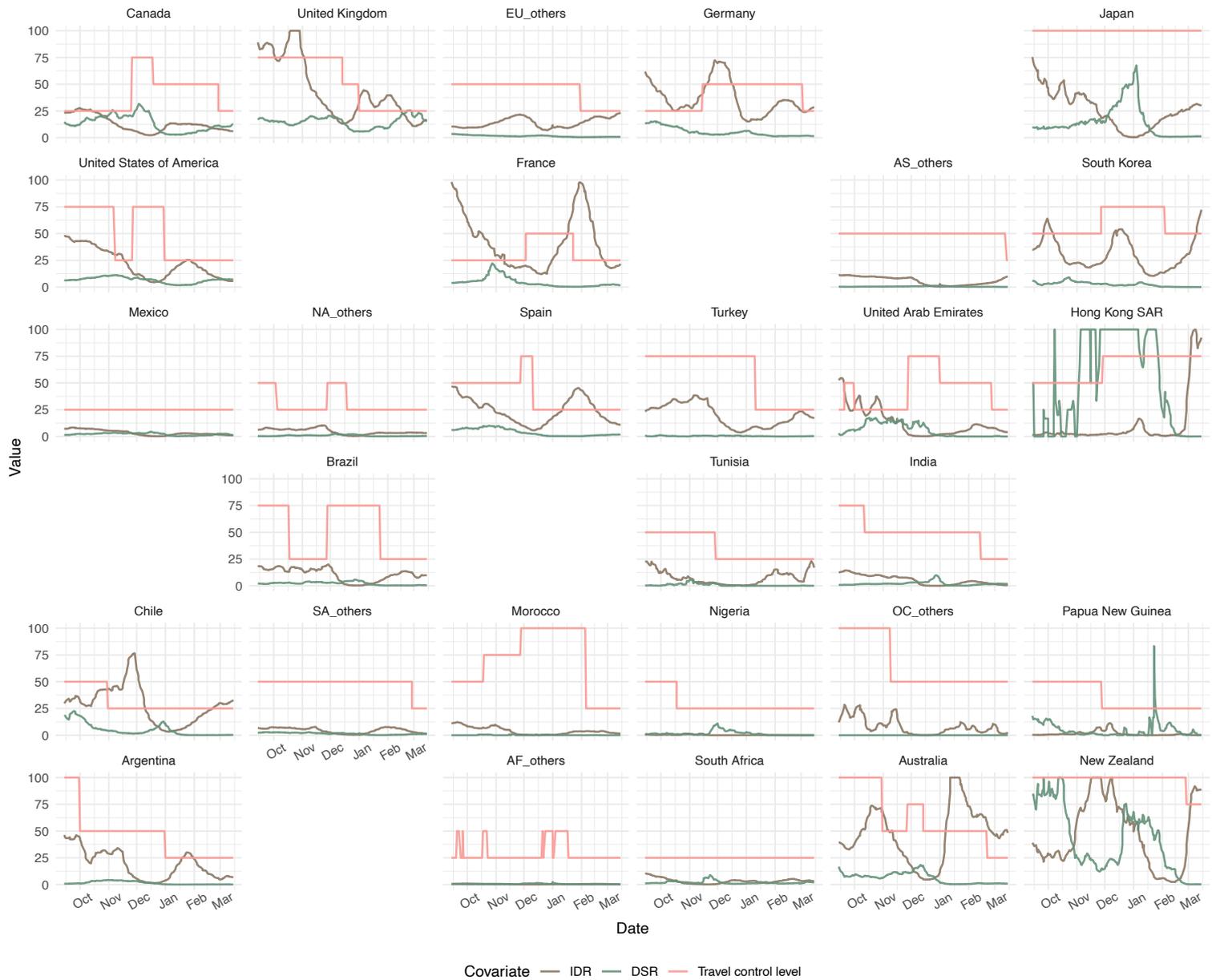

**Figure S4: Association between traveler volume and surveillance lags for Omicron BA.1 and BA.2 variants.** This figure illustrates the relationship between total travel volume and the time lags associated with infection arrival, arrival to diagnostic detection, and diagnostic to sequencing detection for Omicron BA.1 and Omicron BA.2 variants across various regions (stratified by continents in different colors). The panels are divided into different lag types: emergence-arrival lag, arrival-diagnostic lag, and diagnostic-sequencing lag. Each point represents a region, labeled with its respective 2-letter ISO code. The continent-specific lines are linear regression fits generated using the `geom_smooth(method="lm")` function in R. The shaded areas around the regression lines indicate the 95 percent confidence intervals for the linear regression fits.

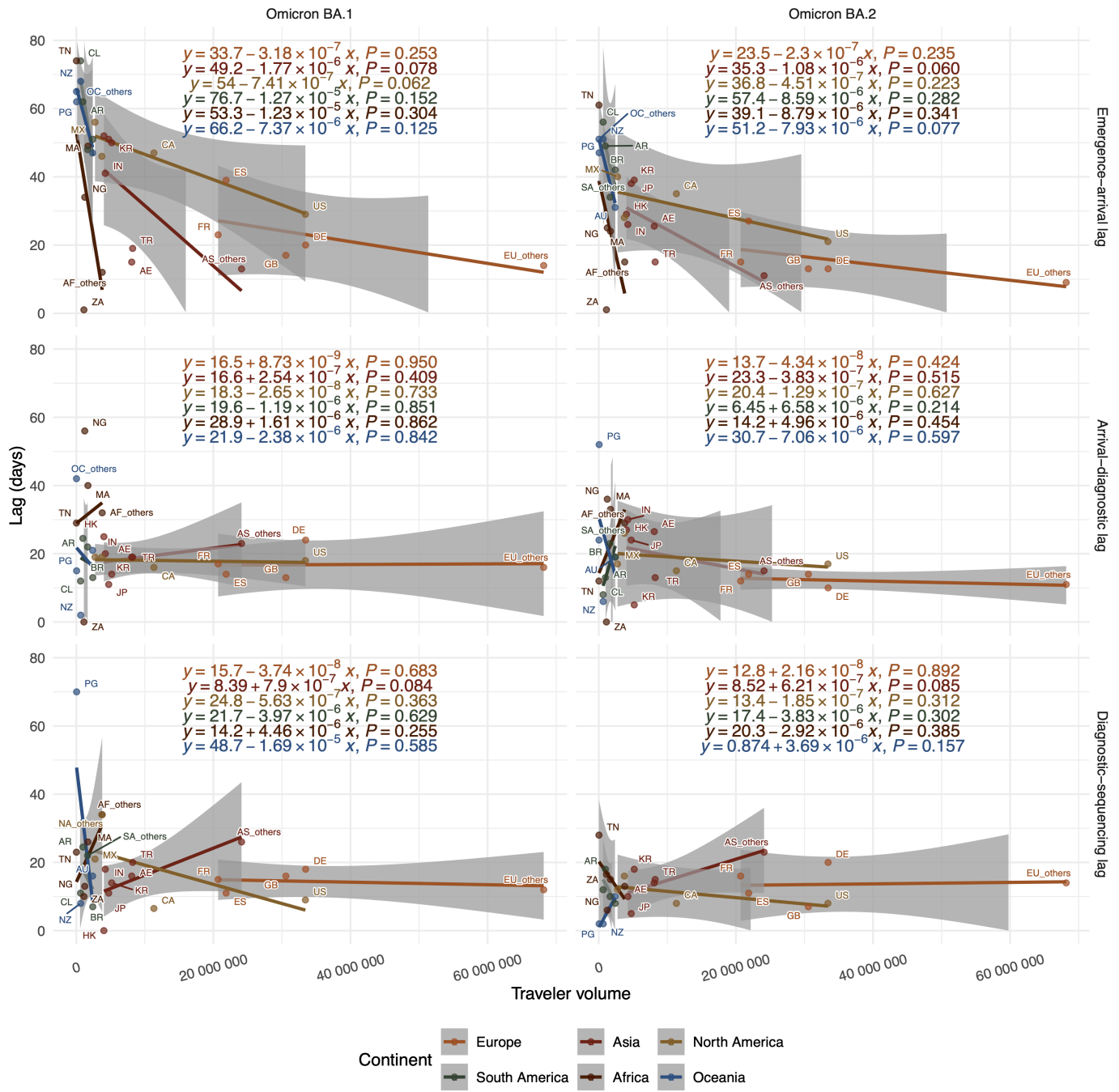

**Figure S5: Distribution of IDR, DSR, traveler weight and travel hubs between two groups of regions.** The data points inside the boxplot are colored by origin regions. Each data point corresponds to one of the lowest identification lag achievements (those texted/labelled simulations in Figure 4a, bootstrapped resampled of 1024 times for each region of emergence).

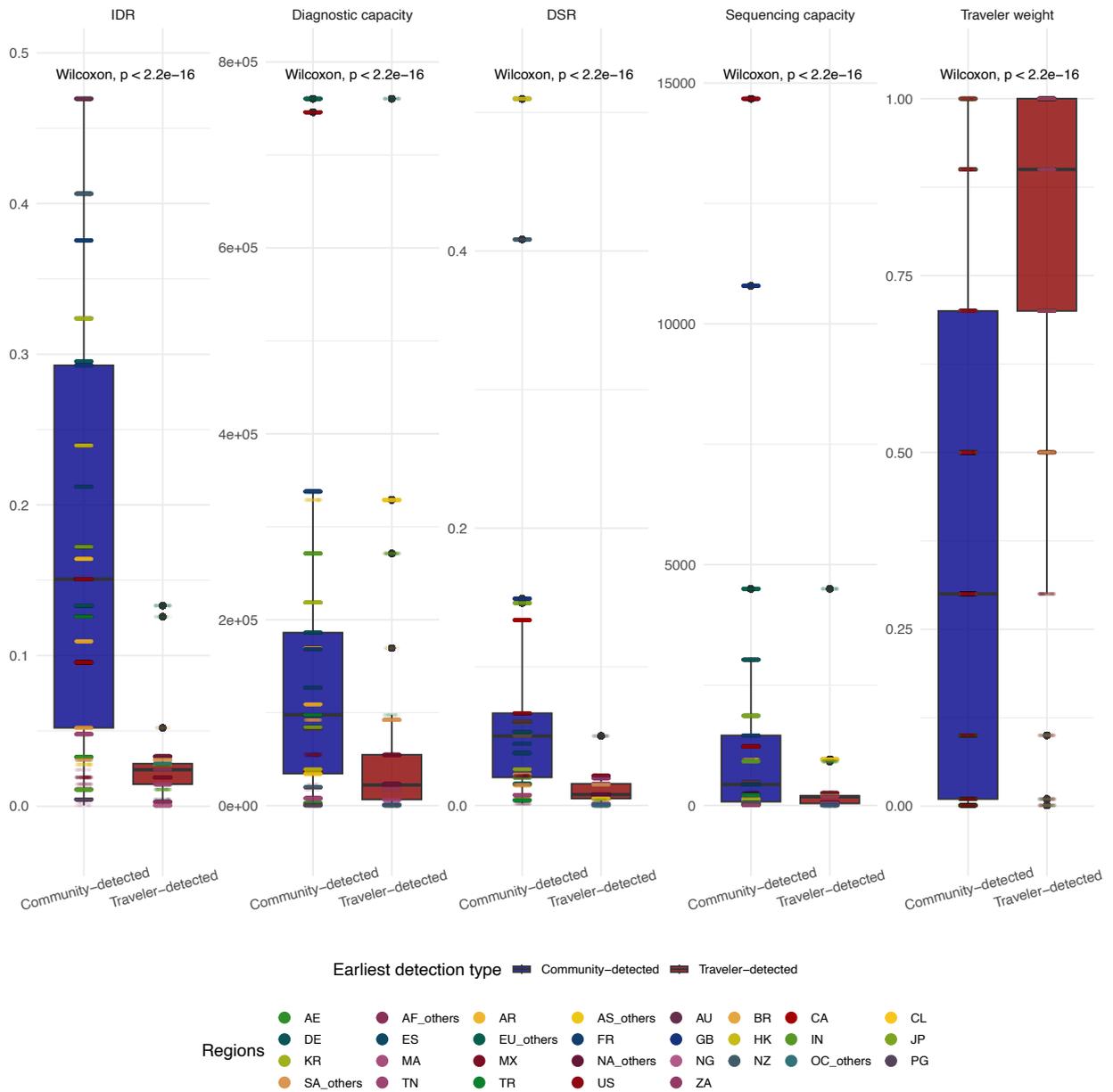

**Figure S6: Identification Lags for the Emergence of a Novel Variant Under Non-Pandemic Conditions, focusing on variations of the P2-H travel hub strategy.** This figure is analogous to Figure 4. The new regions adding to P2 as the variation strategies were annotated with two-letter codes as shown in Table S2. Results are based on 512 simulation

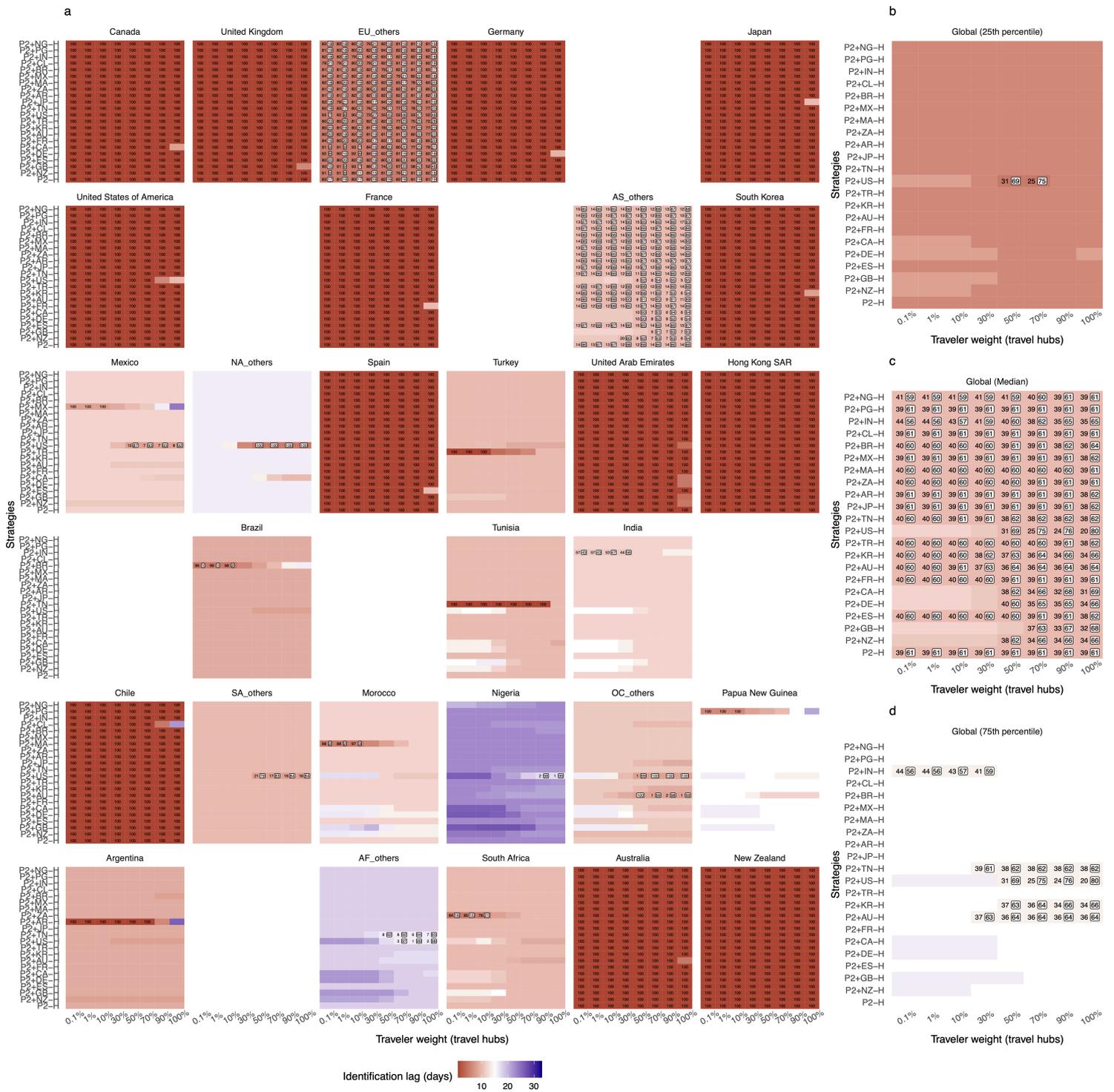

replications.

**Figure S7: The correlation between different sources of air travel data. (a)** The correlation between ADS-B and WTO data; **(b)** The correlation between ADS-B and ICAO data. The black solid diagonal line shows the reference with slope=1 and intercept=0.

**a.**

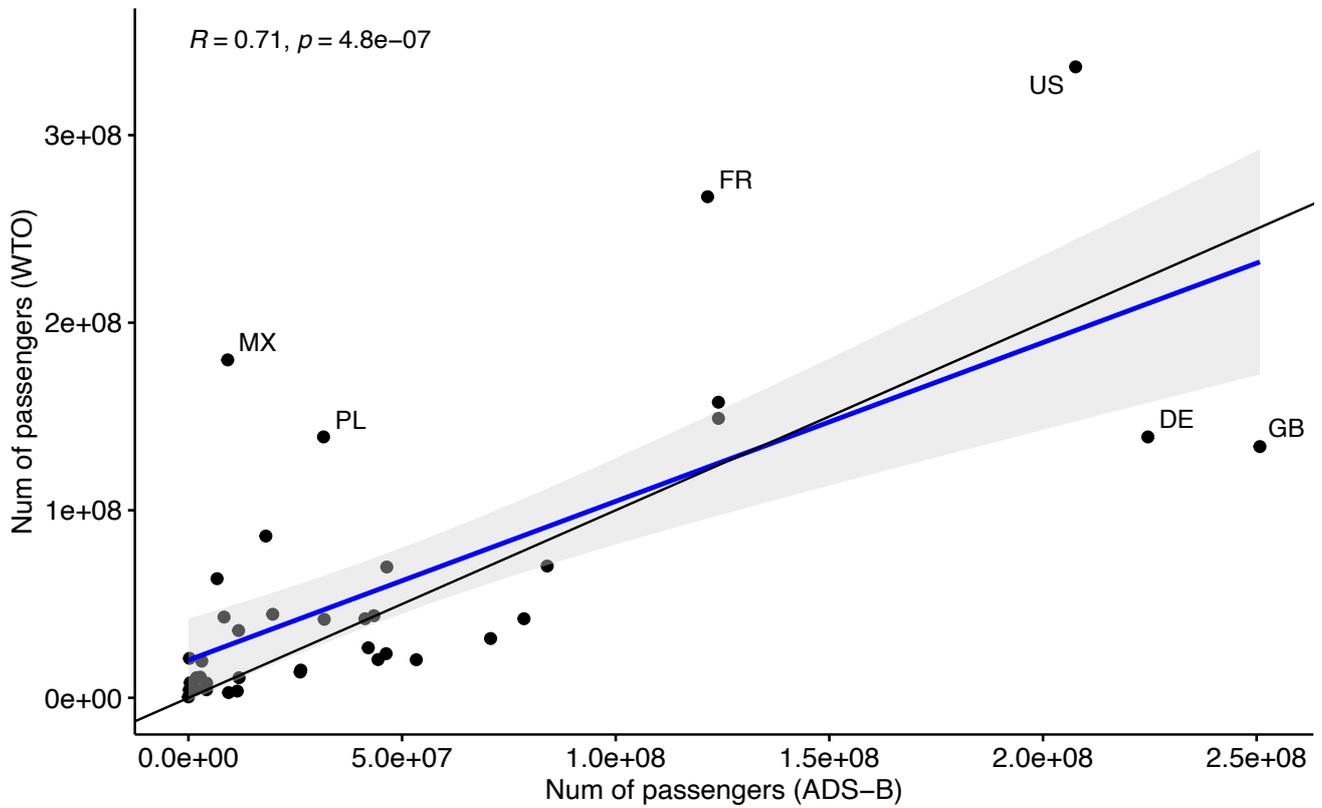

**b.**

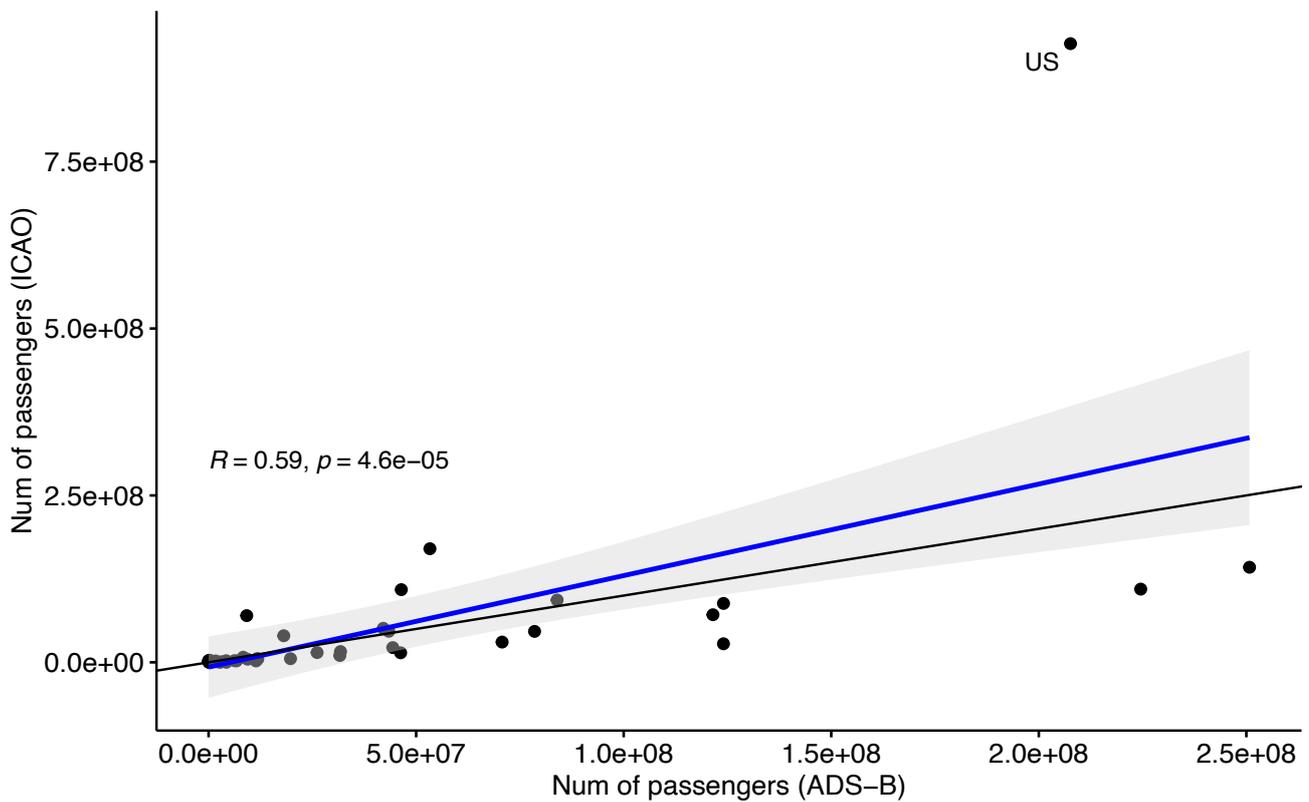

# Supplemental Tables

## Table S1: Description and summary of the models used in the study.

| Model | Description | Framework | Number of observational variables used for fitting | Number of parameters in model | New parameters compared to $M_0$ |
|---|---|---|---|---|---|
| $M_0$ (Base model) | The base multi-strain metapopulation model with origin tracing | Spatpomp | $643916 = 122(\text{observation}) \times 29(\text{regions}) \times 182(\text{timepoints})$, $122(\text{observation}) = 1(\text{Cases}) + 1(\text{Deaths}) + 1(\text{Cumulative sequenced}) * 4(\text{strains}) + 1(\text{Origin inferred sequenced}) * 4(\text{strains}) * 29(\text{origins})$. | $49 = 1(\text{unit\_specific}) * 29(\text{regions}) + 20(\text{shared})$ | NA |
| $M_1$ (For comparison) | A multi-strain metapopulation model without origin tracing | Spatpomp | $31668 = 6(\text{parameters}) \times 29(\text{regions}) \times 182(\text{timepoints})$; $6(\text{parameters}) = 1(\text{Cases}) + 1(\text{Deaths}) + 1(\text{Cumulative sequenced}) * 4(\text{strains})$. | $49 = 1(\text{unit\_specific}) * 29(\text{regions}) + 20(\text{shared})$ | None |
| $M_2$ (For comparison) | A set of autoregressive normal models | Autoregression | $643916 = 122(\text{observation}) \times 29(\text{regions}) \times 182(\text{timepoints})$, $122(\text{observation}) = 1(\text{Cases}) + 1(\text{Deaths}) + 1(\text{Cumulative sequenced}) * 4(\text{strains}) + 1(\text{Origin inferred sequenced}) * 4(\text{strains}) * 29(\text{origins})$. | $3538 = 122(\text{observation}) \times 29(\text{regions})$ | None |
| $M_3$ (For simulation) | $M_0$ with an adjustable sequencing propensity | Spatpomp | Fitting not needed | $52 = 1(\text{unit\_specific}) * 29(\text{regions}) + 23(\text{shared})$ | $\kappa_{IDR}, \kappa_{DSR},$ and $\pi$ |
| $M_4$ (For simulation) | $M_3$ with a neutral movement matrix and a random unit of the novel variant introduction | Spatpomp | Fitting not needed | $49 = 1(\text{unit\_specific}) * 29(\text{regions}) + 20(\text{shared})$ | $\kappa_{IDR}, \kappa_{DSR}, \pi$ and $o \in 1:29$ |

**Table S2: Travel hubs related information for different spatial units.** For continents, AS: Asia, EU: Europe, NA: North America, SA: South America, AF: Africa, OC: Oceania. In the columns for travel hubs ranking, * denotes continental representative travel hubs. The baseline IDR/DSR were averaged for per region daily values over the study period; The baseline diagnostic/sequencing capacities were determined by the 95th percentile of the daily sequences per region during the study period. The Omicron period refer to the study period (2021-09-15 to 2022-03-15). For earliest detection type, C and T denote community-detected and traveler-detected origins respectively.

| Administrative name | Two-letter code | Continent | Population (million) | Travel volume (Omicron period, yearly, million) | Travel volume (2019 period, yearly, million) | Travel hubs ranking (T-ranked, 2019 period) | Travel hubs ranking (T-ranked, Omicron period) | Travel hubs ranking (P-ranked, 2019 period) | Travel hubs ranking (P-ranked, Omicron period) | Proportion of transit passengers | Baseline IDR (daily averaged) | Baseline DSR (daily averaged) | Baseline diagnostic capacity (daily 95th percentile) | Baseline sequencing capacity (daily 95th percentile) | Earliest detection type |
|---|---|---|---|---|---|---|---|---|---|---|---|---|---|---|---|
| Germany | DE | EU | 84.91 | 33.46 | 112.48 | 1* | 3 | 5 | 6 | 47.50% | 29.52% | 5.30% | 186174 | 3026 | C |
| United States of America | US | NA | 327.98 | 33.42 | 115.72 | 2* | 2* | 9 | 12 | 41.50% | 15.07% | 6.65% | 745944 | 14675 | C |
| United Kingdom | GB | EU | 67.22 | 30.57 | 125.54 | 3 | 1* | 4 | 4* | 31% | 29.26% | 14.92% | 168313 | 10789 | C |
| Spain | ES | EU | 46.02 | 21.86 | 64.78 | 4 | 5 | 3* | 5 | 33% | 21.20% | 3.79% | 126813 | 434 | C |
| France | FR | EU | 66.2 | 20.69 | 67.49 | 5 | 4 | 6 | 8 | 34% | 37.56% | 4.44% | 337910 | 1453 | C |
| Canada | CA | NA | 36.52 | 11.32 | 47.56 | 6 | 6 | 7* | 7* | 14% | 9.55% | 13.37% | 37194 | 1220 | C |
| Turkey | TR | AS | 81.36 | 8.22 | 32.84 | 7* | 8 | 10 | 11 | 47% | 12.58% | 0.37% | 97608 | 219 | C |
| United Arab Emirates | AE | AS | 9.24 | 8.09 | 38.1 | 8 | 7* | 1* | 1* | 62.50% | 3.27% | 6.01% | 2801 | 20 | C |
| Korea | KR | AS | 53.4 | 5.17 | 25.29 | 9 | 12 | 12 | 10 | 14% | 32.38% | 2.61% | 218423 | 138 | C |
| Japan | JP | AS | 127.79 | 4.73 | 26.86 | 10 | 11 | 14 | 14 | 6% | 17.22% | 14.61% | 84452 | 1862 | C |
| India | IN | AS | 1390.71 | 4.24 | 26.96 | 11 | 10 | 22 | 21 | 12% | 1.11% | 2.02% | 271292 | 912 | C |
| Hong Kong | HK | AS | 7.5 | 4.03 | 29.74 | 12 | 9 | 2 | 2 | 29% | 23.94% | 50.99% | 39046 | 77 | C |
| Mexico | MX | NA | 124.94 | 2.73 | 9.37 | 13 | 15 | 16 | 18 | 14% | 1.92% | 2.13% | 54752 | 257 | T |
| Brazil | BR | SA | 216.66 | 2.41 | 12.35 | 14* | 14* | 19 | 19 | 14% | 5.20% | 2.23% | 169516 | 914 | C |
| Australia | AU | OC | 24.57 | 2.38 | 19.47 | 15* | 13* | 11 | 9 | 14% | 46.96% | 6.06% | 82692 | 479 | C |
| Morocco | MA | AF | 35.95 | 1.7 | 4 | 16* | 19 | 13* | 17 | 14% | 1.46% | 0.12% | 6649 | 3 | T |
| Nigeria | NG | AF | 214.82 | 1.24 | 3.38 | 17 | 20 | 20 | 23 | 14% | 0.06% | 1.56% | 1232 | 41 | T |
| South Africa | ZA | AF | 55.59 | 1.12 | 6.64 | 18 | 17* | 18 | 16 | 14% | 2.42% | 1.97% | 17413 | 178 | T |
| Argentina | AR | SA | 45.12 | 0.97 | 5.84 | 19 | 18 | 17 | 15* | 14% | 10.94% | 1.47% | 108993 | 66 | C |
| New Zealand | NZ | OC | 4.5 | 0.65 | 8.59 | 20 | 16 | 8* | 3* | 14% | 40.65% | 40.84% | 19822 | 58 | C |
| Chile | CL | SA | 18.2 | 0.63 | 0.42 | 21 | 22 | 15* | 20 | 14% | 16.41% | 5.04% | 34458 | 132 | C |
| Papua New Guinea | PG | OC | 9.87 | 0.05 | 0.17 | 22 | 23 | 21 | 22 | 14% | 0.46% | 5.01% | 344 | 26 | C |
| Tunisia | TN | AF | 11.57 | 0.03 | 2.63 | 23 | 21 | 23 | 13* | 14% | 4.78% | 0.75% | 7984 | 5 | C |
| Europe (others) | EU_others | EU | 467.07 | 68.18 | 282.59 | 24 | 24 | 24 | 24 | 16.57% | 13.31% | 1.55% | 760417 | 4497 | C |
| Asia (others) | AS_others | AS | 2886.53 | 24.11 | 170.94 | 25 | 25 | 28 | 28 | 17.15% | 2.76% | 0.49% | 328900 | 958 | T |
| Africa (others) | AF_others | AF | 798.8 | 3.78 | 19.18 | 26 | 26 | 29 | 29 | 15.79% | 0.29% | 0.52% | 22277 | 189 | T |
| North America (others) | NA_others | NA | 68.18 | 3.73 | 13.66 | 27 | 27 | 25 | 25 | 20.30% | 3.30% | 0.79% | 23130 | 62 | T |
| South America (others) | SA_others | SA | 147.72 | 1.65 | 12.43 | 28 | 28 | 27 | 27 | 14% | 3.07% | 1.49% | 92230 | 198 | T |
| Oceania (others) | OC_others | OC | 0.91 | 0.03 | 0.15 | 29 | 29 | 26 | 26 | 14% | 2.82% | 0% | 330 | 0 | T |

**Table S3: List of parameters, respective notations and fitted values.**

| State | Mechanism | Notation | Description | Type | Fitted values (with 95% confidence intervals) or fixed values |
|---|---|---|---|---|---|
| All | Compartments | $S, V, E, I, R, D$ | The model comprises susceptible, immunised, exposed, Infected, recovered, and deceased compartments. | Symbol | NA |
| All | Travel | $j, u$ | Spatial units $j$ and $u$ | Symbol | NA |
| All | Strains | $k \in 1:4$ | Four different infecting strains under study. | Symbol | NA |
| Immunised | Vaccination | $v_u(t)$ | Vaccination rate at time $t$ in unit $u$ | Covariates | Curated from IHME data, see Methods. |
| Exposed | Infection | $\eta_k$ | Relative increase/decrease in infection rate due to different strain $k$, specifically we set $\eta_1(Delta) = 1$. | Parameter (Estimated) | $\eta_2(BA.1) = 2.93\,(2.91, 2.95)$, $\eta_3(BA.2) = 4.85\,(4.79, 4.92)$, $\eta_4(Others) = 1.35\,(1.12, 1.41)$ |
| Exposed | Infection | $\beta_u$ | Natural transmissibility (infection rate) of the virus in each unit $u$ | Parameter (Estimated) | $\beta_1(AE) = 218.97\,(154.85, 266)$, $\beta_2(AF\_others) = 293.08\,(268.14, 327.28)$, $\beta_3(AR) = 238.92\,(196.89, 277.4)$, $\beta_4(AS\_others) = 226.1\,(201.87, 253.89)$, $\beta_5(AU) = 180.5\,(144.87, 209.71)$, $\beta_6(BR) = 372.88\,(348.66, 402.1)$, $\beta_7(CA) = 338.68\,(308.04, 365.04)$, $\beta_8(CL) = 452.69\,(427.03, 475.49)$, $\beta_9(DE) = 10.91\,(10.91, 722.74)$, $\beta_{10}(ES) = 365.04\,(335.12, 407.08)$, $\beta_{11}(EU\_others) = 556\,(531.78, 580.23)$, $\beta_{12}(FR) = 206.86\,(179.07, 238.92)$, $\beta_{13}(GB) = 200.45\,(170.52, 242.49)$, $\beta_{14}(HK) = 20.89\,(10.91, 722.74)$, $\beta_{15}(IN) = 267.43\,(251.04, 280.25)$, $\beta_{16}(JP) = 77.18\,(10.91, 206.15)$, $\beta_{17}(KR) = 181.92\,(152.71, 210.42)$, $\beta_{18}(MA) = 402.1\,(-Inf, Inf)$, $\beta_{19}(MX) = 434.16\,(382.14, 484.04)$, $\beta_{20}(NA\_others) = 267.43\,(245.34, 285.24)$, $\beta_{21}(NG) = 226.81\,(196.89, 265.29)$, $\beta_{22}(NZ) = 126.34\,(48.68, 206.15)$, $\beta_{23}(OC\_others) = 595.91\,(505.41, 679.27)$, $\beta_{24}(PG) = 210.42\,(186.91, 240.35)$, $\beta_{25}(SA\_others) = 407.08\,(391.41, 424.18)$, $\beta_{26}(TN) = 442.71\,(399.96, 478.34)$, $\beta_{27}(TR) = 455.54\,(383.57, 532.49)$, $\beta_{28}(US) = 318.73\,(284.53, 364.33)$, $\beta_{29}(ZA) = 143.5842$. (Full name of the abbreviations in Table S2.) |
| Exposed | Infection | $R_0$ | Basic reproduction number for reference, calculated as $\beta_u / \gamma$ | Parameter (Estimated) | $R_0(AE) = 2.56(1.79, 3.15)$, $R_0(AF\_others) = 3.43(3.09, 3.88)$, $R_0(AR) = 2.80(2.27, 3.29)$, $R_0(AS\_others) = 2.65(2.33, 3.01)$, $R_0(AU) = 2.11(1.67, 2.49)$, $R_0(BR) = 4.36(4.02, 4.76)$, $R_0(CA) = 3.96(3.56, 4.33)$, $R_0(CL) = 5.30(4.93, 5.63)$, $R_0(DE) = 0.13(0.13, 8.56)$, $R_0(ES) = 4.27(3.87, 4.82)$, $R_0(EU\_others) = 6.51(6.14, 6.88)$, $R_0(FR) = 2.42(2.07, 2.83)$, $R_0(GB) = 2.35(1.97, 2.87)$, $R_0(HK) = 0.24(0.13, 8.56)$, $R_0(IN) = 3.13(2.90, 3.32)$, $R_0(JP) = 0.90\,(0.13, 2.44)$, |

| | | | | | $R_0(KR) = 2.13\ (1.76, 2.49),$ |
| --- | --- | --- | --- | --- | --- |
| | | | | | $R_0(MA) = 4.70\ (-Inf, Inf),$ |
| | | | | | $R_0(MX) = 5.08\ (4.41, 5.74),$ |
| | | | | | $R_0(NA\_others) = 3.13\ (2.83, 3.38),$ |
| | | | | | $R_0(NG) = 2.65\ (2.27, 3.14),$ |
| | | | | | $R_0(NZ) = 1.48\ (0.56, 2.44),$ |
| | | | | | $R_0(OC\_others) = 6.97\ (5.83, 8.05),$ |
| | | | | | $R_0(PG) = 2.46\ (2.16, 2.85),$ |
| | | | | | $R_0(SA\_others) = 4.76\ (4.52, 5.03),$ |
| | | | | | $R_0(TN) = 5.18\ (4.62, 5.67),$ |
| | | | | | $R_0(TR) = 5.33\ (4.43, 6.31),$ |
| | | | | | $R_0(US) = 3.73\ (3.28, 4.32),$ |
| | | | | | $R_0(ZA) = 1.68\ (1.66, 1.70).$ |
| | | | | | (Full name of the abbreviations in Table S2.) |
| Exposed | Infection | $\lambda_k$ | Relative infection rate due to immunity for strain $k$, acting on the Immunised population | Parameter (Estimated) | $\lambda_1(Delta) = 0.22\ (0.21, 0.22),$ $\lambda_2(BA.1) = 0.39\ (0.37, 0.41),$ $\lambda_3(BA.2) = 0.45\ (0.44, 0.45),$ $\lambda_4(Others) = 0.06\ (0.01, 0.11).$ |
| Exposed | Infection | $\zeta_u(t)$ | International travel control policies levels at time $t$ in unit $u$ | Covariates | Curated from Oxford COVID-19 Government Response Tracker data, see Methods and Figure S3. |
| Exposed | Infection | $\iota_n$ | Relative decrease in transmissibility due to international travel control policies (such as quarantine) at level $n$. | Parameter (Estimated) | $\iota_1 = 0.41\ (-Inf, Inf),$ $\iota_2 = 0.31\ (0.18, 0.46),$ $\iota_3 = 0.22\ (0.11, 0.39),$ $\iota_4 = 0.23\ (0.02, 0.3)$ |
| Exposed | Infection | $o$ | The random origin of the novel variant. Only used in $M_4$. | Parameter (Fixed) | $1:29$ |
| Exposed | Stochastic effects | $\sigma$ | Gamma white noise with intensity $\sigma$ | Parameter (Estimated) | $85.47\ (84.39, 86.64)$ |
| Exposed/Infected | Infection | $\tau_{BA.1}$ | The day since $t_0$ when 2000 Omicron BA.1 cases was introduced into both E and I compartments of South Africa | Parameter (Estimated) | $7.33\ (6.1, 8.53)$ |
| Exposed/Infected | Infection | $\tau_{BA.2}$ | The day since $t_0$ when 2000 Omicron BA.2 cases was introduced into both E and I compartments of South Africa | Parameter (Estimated) | $55.73\ (55.22, 56.24)$ |
| Exposed/Infected | Travel | $M_{ju}(t)$ | The number of individuals moving from region $u$ to region $j$ at time $t$. | Parameter (Covariates) | Air travel data, see Methods. |
| Exposed/Infected | Travel | $q_u$ | Fraction of the arrival travelers that are transit passengers in unit $u$. | Parameter (Fixed) | See Table S4. |
| Infected | Incubation | $\varepsilon_k$ | Inverse of the incubation period for strain $k$ | Parameter (Fixed or estimated) | Fixed: $\varepsilon_1(Delta) = 82.95$ equivalent to $4.4\ days$ $\varepsilon_2(BA.1) = 101.39$ equivalent to $3.6\ days$ $\varepsilon_3(BA.2) = 107.35$ equivalent to $3.4\ days$ Estimated: $\varepsilon_4(Others) = 83.14\ (-Inf, Inf),$ equivalent to $4.4\ days.$ |
| Recovered | Recovery | $\gamma$ | Rate at which infected individuals become recovered or deceased (the rate of movement from the $I$ compartment to $R$ and $D$ compartments). | Parameter (Estimated) | $85.47\ (84.39, 86.64)$ equivalent to $4.27\ (4.21, 4.33) days.$ |
| Deceased | Deceased | $h_u(t)$ | Infection-to-fatality ratio (fraction of infected cases that are fatal at time $t$ in unit $u$. | Parameter (Covariates) | Curated from IHME data, see Methods. |
| Infected | Measurement | $\delta_u(t)$ | Infection-to-diagnostic ratio (fraction of diagnostic/reported cases among the infection population at time $t$ in unit $u$). | Parameter (Covariates) | Curated from IHME data, see Methods and Figure S3. |
| Infected | Measurement | $\rho_u(t)$ | Diagnostic-to-sequencing ratio (fraction of sequenced cases among the diagnostic/reported cases at time $t$ in unit $u$). | Parameter (Covariates) | Curated from IHME and GISAID data, see Methods and Figure S3. |

| Infected | Measurement | $\theta$ | Variance parameters in the normal distribution observation model for detected imported and community cases. | Parameter (Estimated) | $\theta_{cases} = 0.99\ (0.99, 0.99)$, $\theta_{deaths} = 0.99\ (0.98, 0.99)$, $\theta_{seq\_all} = 0.99\ (0.99, 0.99)$, $\theta_{seq\_unit} = 0.99\ (0.99, 0.99)$. |
|---|---|---|---|---|---|
| Infected | Measurement | $\kappa_{IDR}$ | Relative change in IDR. | Parameter (Fixed) | 0.1%, 1%, 10%, 20%, 40%, 50%, 60%, 80%, 100%, 150%, 200% |
| Infected | Measurement | $\kappa_{DSR}$ | Relative change in DSR. | Parameter (Fixed) | 0.1%, 1%, 10%, 20%, 40%, 50%, 60%, 80%, 100%, 150%, 200% |
| Infected | Measurement | $\pi$ | The proportion of diagnostic and sequencing capacity allocated to the imported cases. | Parameter (Fixed) | 10%, 20%, 30%, 40%, 50%, 60%, 70%, 80%, 90% |

**Table S4: Overview of Travel Hub Configurations Tested in Figure 3, 4 and 5.**

| Strategies | Number of selected travel hubs | Selected hubs and relative change in diagnostic and sequencing efforts (Omicron period, for Figure 3) | Selected hubs and relative change in diagnostic and sequencing efforts (2019 period, for Figures 4 and 5) |
|---|---|---|---|
| P2-R | Top 2 (P-ranked) travel hubs | NA | Diagnostic efforts: All unchanged. Sequencing efforts: All unchanged. |
| P2a-R | Top 2 (P-ranked) travel hubs with 5 additional continental hubs | NA | Diagnostic efforts: All unchanged. Sequencing efforts: All unchanged. |
| P3-R | Top 3 (P-ranked) travel hubs | Diagnostic efforts: All unchanged. Sequencing efforts: All unchanged. | Diagnostic efforts: All unchanged. Sequencing efforts: All unchanged. |
| P3a-R | Top 3 (P-ranked) travel hubs with 4 additional continental hubs | Diagnostic efforts: All unchanged. Sequencing efforts: All unchanged. | Diagnostic efforts: All unchanged. Sequencing efforts: All unchanged. |
| P7-R | Top 7 (P-ranked) travel hubs | Diagnostic efforts: All unchanged. Sequencing efforts: All unchanged. | Diagnostic efforts: All unchanged. Sequencing efforts: All unchanged. |
| P7a-R | Top 7 (P-ranked) travel hubs with 3/2 additional continental hubs | Diagnostic efforts: All unchanged. Sequencing efforts: All unchanged. | Diagnostic efforts: All unchanged. Sequencing efforts: All unchanged. |
| P11-R | Top 11 (P-ranked) travel hubs | Diagnostic efforts: All unchanged. Sequencing efforts: All unchanged. | Diagnostic efforts: All unchanged. Sequencing efforts: All unchanged. |
| P2-H | Top 2 (P-ranked) travel hubs | NA | Diagnostic efforts: AE:37020%, HK:6675%, and others scaled down to 50% of the original level. Sequencing efforts: AE:60706%, HK:36049%, OC_others:100%, and others scaled down to 50% of the original level. |
| P2a-H | Top 2 (P-ranked) travel hubs with 5 additional continental hubs | NA | Diagnostic efforts: AE:20400%, AR:127%, CA:658%, GB:220%, HK:3715%, NZ:4396%, TN:619%, and others scaled down to 50% of the original level. Sequencing efforts: AE:24415%, AR:329%, CA:220%, GB:115%, HK:14522%, NZ:3599%, OC_others:100%, TN:4627%, and others scaled down to 50% of the original level. |
| P3-H | Top 3 (P-ranked) travel hubs | Diagnostic efforts: AE:32746%, ES:537%, HK:3809%, and others scaled down to 50% of the original level. Sequencing efforts: AE:54642%, ES:1259%, HK:20739%, OC_others:100%, and others scaled down to 50% of the original level. | Diagnostic efforts: AE:29916%, HK:5410%, NZ:6411%, and others scaled down to 50% of the original level. Sequencing efforts: AE:49052%, HK:29136%, NZ:7145%, OC_others:100%, and others scaled down to 50% of the original level. |
| P3a-H | Top 3 (P-ranked) travel hubs with 4 additional continental hubs | Diagnostic efforts: AE:25062%, CA:869%, CL:194%, ES:434%, HK:2936%, MA:853%, NZ:1968%, and others scaled down to 50% of the original level. Sequencing efforts: AE:41252%, CA:327%, CL:358%, ES:974%, HK:15673%, MA:14601%, NZ:2194%, OC_others:100%, and others scaled down to 50% of the original level. | Diagnostic efforts: AE:20400%, AR:127%, CA:658%, GB:220%, HK:3715%, NZ:4396%, TN:619%, and others scaled down to 50% of the original level. Sequencing efforts: AE:24415%, AR:329%, CA:220%, GB:115%, HK:14522%, NZ:3599%, OC_others:100%, TN:4627%, and others scaled down to 50% of the original level. |
| P7-H | Top 7 (P-ranked) travel hubs | Diagnostic efforts: AE:14854%, CA:555%, DE:181%, ES:297%, FR:150%, GB:200%, HK:1776%, and others scaled down to 50% of the original level. | Diagnostic efforts: AE:16120%, CA:540%, DE:163%, ES:235%, GB:195%, HK:2953%, NZ:3491%, and others scaled down to 50% of the original |
| | | Sequencing efforts: AE:17429%, CA:196%, DE:149%, ES:468%, FR:170%, GB:113%, HK:6657%, OC_others:100%, and others scaled down to 50% of the original level. | level. Sequencing efforts: AE:18442%, CA:190%, DE:137%, ES:345%, GB:112%, HK:10980%, NZ:2740%, OC_others:100%, and others scaled down to 50% of the original level. |
| P7a-H | Top 7 (P-ranked) travel hubs with 3/2 additional continental hubs | Diagnostic efforts: AE:13750%, CA:521%, CL:152%, DE:175%, ES:283%, FR:146%, GB:192%, HK:1651%, MA:512%, NZ:1121%, and others scaled down to 50% of the original level. Sequencing efforts: AE:16221%, CA:189%, CL:201%, DE:146%, ES:442%, FR:165%, GB:112%, HK:6200%, MA:5780%, NZ:920%, OC_others:100%, and others scaled down to 50% of the original level. | Diagnostic efforts: AE:15439%, AR:120%, CA:521%, DE:160%, ES:229%, GB:191%, HK:2832%, NZ:3346%, TN:492%, and others scaled down to 50% of the original level. Sequencing efforts: AE:17994%, AR:269%, CA:188%, DE:136%, ES:339%, GB:111%, HK:10714%, NZ:2675%, OC_others:100%, TN:3432%, and others scaled down to 50% of the original level. |
| P11-H | Top 11 (P-ranked) travel hubs | Diagnostic efforts: AE:9463%, AU:147%, CA:389%, DE:152%, ES:225%, FR:132%, GB:163%, HK:1164%, NZ:801%, TR:121%, US:104%, and others scaled down to 50% of the original level. Sequencing efforts: AE:5858%, AU:129%, CA:132%, DE:116%, ES:222%, FR:123%, GB:104%, HK:2279%, NZ:393%, OC_others:100%, TR:159%, US:101%, and others scaled down to 50% of the original level. | Diagnostic efforts: AE:11384%, AU:198%, CA:410%, DE:144%, ES:195%, FR:126%, GB:167%, HK:2110%, KR:133%, NZ:2488%, TR:122%, and others scaled down to 50% of the original level. Sequencing efforts: AE:14338%, AU:223%, CA:170%, DE:129%, ES:290%, FR:140%, GB:109%, HK:8545%, KR:277%, NZ:2149%, OC_others:100%, TR:224%, and others scaled down to 50% of the original level. |
| P2-M | Top 2 (P-ranked) travel hubs | NA | Diagnostic efforts: AE:66556%, HK:11936%, and others scaled down to 10% of the original level. Sequencing efforts: AE:109191%, HK:64808%, OC_others:100%, and others scaled down to 10% of the original level. |
| P2a-M | Top 2 (P-ranked) travel hubs with 5 additional continental hubs | NA | Diagnostic efforts: AE:36640%, AR:148%, CA:1104%, GB:316%, HK:6608%, NZ:7834%, TN:1034%, and others scaled down to 10% of the original level. Sequencing efforts: AE:43866%, AR:513%, CA:315%, GB:128%, HK:26060%, NZ:6398%, OC_others:100%, TN:8249%, and others scaled down to 10% of the original level. |
| P3-M | Top 3 (P-ranked) travel hubs | Diagnostic efforts: AE:58863%, ES:886%, HK:6777%, and others scaled down to 10% of the original level. Sequencing efforts: AE:98275%, ES:2186%, HK:37251%, OC_others:100%, and others scaled down to 10% of the original level. | Diagnostic efforts: AE:53769%, HK:9658%, NZ:11459%, and others scaled down to 10% of the original level. Sequencing efforts: AE:88214%, HK:52365%, NZ:12781%, OC_others:100%, and others scaled down to 10% of the original level. |
| P3a-M | Top 3 (P-ranked) travel hubs with 4 additional continental hubs | Diagnostic efforts: AE:45031%, CA:1485%, CL:270%, ES:701%, HK:5205%, MA:1455%, NZ:3462%, and others scaled down to 10% of the original level. Sequencing efforts: AE:74174%, CA:509%, CL:565%, ES:1674%, | Diagnostic efforts: AE:36640%, AR:148%, CA:1104%, GB:316%, HK:6608%, NZ:7834%, TN:1034%, and others scaled down to 10% of the original level. Sequencing efforts: AE:43866%, AR:513%, CA:315%, GB:128%, |
| | | HK:28131%, MA:26201%, NZ:3869%, OC_others:100%, and others scaled down to 10% of the original level. | HK:26060%, NZ:6398%, OC_others:100%, TN:8249%, and others scaled down to 10% of the original level. |
| P7-M | Top 7 (P-ranked) travel hubs | Diagnostic efforts: AE:26657%, CA:918%, DE:246%, ES:455%, FR:190%, GB:280%, HK:3117%, and others scaled down to 10% of the original level. Sequencing efforts: AE:31291%, CA:272%, DE:189%, ES:763%, FR:226%, GB:123%, HK:11903%, OC_others:100%, and others scaled down to 10% of the original level. | Diagnostic efforts: AE:28936%, CA:892%, DE:214%, ES:343%, GB:270%, HK:5236%, NZ:6203%, and others scaled down to 10% of the original level. Sequencing efforts: AE:33115%, CA:262%, DE:167%, ES:541%, GB:121%, HK:19683%, NZ:4851%, OC_others:100%, and others scaled down to 10% of the original level. |
| P7a-M | Top 7 (P-ranked) travel hubs with 3/2 additional continental hubs | Diagnostic efforts: AE:24670%, CA:857%, CL:193%, DE:236%, ES:429%, FR:183%, GB:266%, HK:2892%, MA:841%, NZ:1939%, and others scaled down to 10% of the original level. Sequencing efforts: AE:29118%, CA:260%, CL:282%, DE:182%, ES:716%, FR:218%, GB:121%, HK:11081%, MA:10325%, NZ:1577%, OC_others:100%, and others scaled down to 10% of the original level. | Diagnostic efforts: AE:27710%, AR:136%, CA:859%, DE:209%, ES:332%, GB:263%, HK:5017%, NZ:5944%, TN:806%, and others scaled down to 10% of the original level. Sequencing efforts: AE:32308%, AR:404%, CA:258%, DE:165%, ES:530%, GB:121%, HK:19205%, NZ:4735%, OC_others:100%, TN:6097%, and others scaled down to 10% of the original level. |
| P11-M | Top 11 (P-ranked) travel hubs | Diagnostic efforts: AE:16953%, AU:184%, CA:619%, DE:193%, ES:325%, FR:157%, GB:214%, HK:2015%, NZ:1361%, TR:139%, US:108%, and others scaled down to 10% of the original level. Sequencing efforts: AE:10465%, AU:152%, CA:157%, DE:129%, ES:320%, FR:142%, GB:108%, HK:4022%, NZ:627%, OC_others:100%, TR:207%, US:102%, and others scaled down to 10% of the original level. | Diagnostic efforts: AE:20411%, AU:276%, CA:658%, DE:180%, ES:271%, FR:148%, GB:220%, HK:3717%, KR:159%, NZ:4399%, TR:139%, and others scaled down to 10% of the original level. Sequencing efforts: AE:25728%, AU:321%, CA:226%, DE:152%, ES:443%, FR:172%, GB:116%, HK:15301%, KR:418%, NZ:3788%, OC_others:100%, TR:324%, and others scaled down to 10% of the original level. |
| T2-R | Top 2 (T-ranked) travel hubs | NA | Diagnostic efforts: All unchanged. Sequencing efforts: All unchanged. |
| T2a-R | Top 2 (T-ranked) travel hubs with 4 additional continental hubs | NA | Diagnostic efforts: All unchanged. Sequencing efforts: All unchanged. |
| T3-R | Top 3 (T-ranked) travel hubs | Diagnostic efforts: All unchanged. Sequencing efforts: All unchanged. | Diagnostic efforts: All unchanged. Sequencing efforts: All unchanged. |
| T3a-R | Top 3 (T-ranked) travel hubs with 4 additional continental hubs | Diagnostic efforts: All unchanged. Sequencing efforts: All unchanged. | Diagnostic efforts: All unchanged. Sequencing efforts: All unchanged. |
| T7-R | Top 7 (T-ranked) travel hubs | Diagnostic efforts: All unchanged. Sequencing efforts: All unchanged. | Diagnostic efforts: All unchanged. Sequencing efforts: All unchanged. |
| T7a-R | Top 7 (T-ranked) travel hubs with 3 additional continental hubs | Diagnostic efforts: All unchanged. Sequencing efforts: All unchanged. | Diagnostic efforts: All unchanged. Sequencing efforts: All unchanged. |
| T11-R | Top 11 (T-ranked) travel hubs | Diagnostic efforts: All unchanged. Sequencing efforts: All unchanged. | Diagnostic efforts: All unchanged. Sequencing efforts: All unchanged. |
| T2-H | Top 2 (T-ranked) travel hubs | NA | Diagnostic efforts: GB:484%, US:211%, and others scaled down to 50% of the original level. Sequencing efforts: GB:132%, OC_others:100%, US:126%, and others scaled down to 50% of the original level. |
| T2a-H | Top 2 (T-ranked) travel hubs with 4 additional continental hubs | NA | Diagnostic efforts: AE:6409%, AU:245%, BR:139%, GB:371%, US:179%, ZA:305%, and others scaled down to 50% of the original level.<br>Sequencing efforts: AE:4958%, AU:211%, BR:143%, GB:123%, OC_others:100%, US:119%, ZA:250%, and others scaled down to 50% of the original level. |
| T3-H | Top 3 (T-ranked) travel hubs | Diagnostic efforts: DE:321%, GB:315%, US:174%, and others scaled down to 50% of the original level.<br>Sequencing efforts: DE:178%, GB:116%, OC_others:100%, US:116%, and others scaled down to 50% of the original level. | Diagnostic efforts: DE:305%, GB:343%, US:171%, and others scaled down to 50% of the original level.<br>Sequencing efforts: DE:172%, GB:118%, OC_others:100%, US:115%, and others scaled down to 50% of the original level. |
| T3a-H | Top 3 (T-ranked) travel hubs with 4 additional continental hubs | Diagnostic efforts: AU:144%, BR:119%, DE:269%, GB:264%, MA:491%, TR:167%, US:156%, and others scaled down to 50% of the original level.<br>Sequencing efforts: AU:131%, BR:119%, DE:161%, GB:112%, MA:3299%, OC_others:100%, TR:312%, US:112%, and others scaled down to 50% of the original level. | Diagnostic efforts: AE:4398%, AU:199%, BR:126%, DE:255%, GB:285%, US:154%, ZA:239%, and others scaled down to 50% of the original level.<br>Sequencing efforts: AE:2986%, AU:166%, BR:126%, DE:154%, GB:113%, OC_others:100%, US:111%, ZA:189%, and others scaled down to 50% of the original level. |
| T7-H | Top 7 (T-ranked) travel hubs | Diagnostic efforts: CA:361%, DE:208%, ES:243%, FR:152%, GB:205%, TR:143%, US:136%, and others scaled down to 50% of the original level.<br>Sequencing efforts: CA:129%, DE:135%, ES:242%, FR:139%, GB:107%, OC_others:100%, TR:222%, US:107%, and others scaled down to 50% of the original level. | Diagnostic efforts: AE:3078%, CA:423%, DE:208%, ES:225%, FR:150%, GB:228%, US:137%, and others scaled down to 50% of the original level.<br>Sequencing efforts: AE:1895%, CA:135%, DE:133%, ES:219%, FR:136%, GB:108%, OC_others:100%, US:107%, and others scaled down to 50% of the original level. |
| T7a-H | Top 7 (T-ranked) travel hubs with 3 additional continental hubs | Diagnostic efforts: AU:125%, BR:111%, CA:328%, DE:195%, ES:225%, FR:145%, GB:192%, MA:319%, TR:138%, US:132%, and others scaled down to 50% of the original level.<br>Sequencing efforts: AU:115%, BR:109%, CA:125%, DE:130%, ES:221%, FR:133%, GB:106%, MA:1667%, OC_others:100%, TR:204%, US:106%, and others scaled down to 50% of the original level. | Diagnostic efforts: AE:2639%, AU:159%, BR:116%, CA:376%, DE:192%, ES:206%, FR:143%, GB:209%, US:132%, ZA:182%, and others scaled down to 50% of the original level.<br>Sequencing efforts: AE:1580%, AU:134%, BR:113%, CA:129%, DE:128%, ES:199%, FR:130%, GB:107%, OC_others:100%, US:106%, ZA:146%, and others scaled down to 50% of the original level. |
| T11-H | Top 11 (T-ranked) travel hubs | Diagnostic efforts: AE:1689%, CA:294%, DE:181%, ES:206%, FR:139%, GB:178%, IN:113%, JP:137%, KR:125%, TR:132%, US:127%, and others scaled down to 50% of the original level.<br>Sequencing efforts: AE:1077%, CA:121%, DE:125%, ES:203%, FR:128%, GB:105%, IN:114%, JP:113%, KR:167%, OC_others:100%, TR:189%, US:105%, and others scaled down to 50% of the original level. | Diagnostic efforts: AE:2173%, CA:325%, DE:175%, ES:187%, FR:135%, GB:189%, HK:400%, IN:123%, JP:158%, TR:135%, US:126%, and others scaled down to 50% of the original level.<br>Sequencing efforts: AE:1337%, CA:124%, DE:123%, ES:182%, FR:125%, GB:106%, HK:696%, IN:124%, JP:120%, OC_others:100%, TR:195%, US:105%, and others scaled down to 50% of the original level. |
| T2-M | Top 2 (T-ranked) travel hubs | NA | Diagnostic efforts: GB:792%, US:301%, and others scaled down to 10% of the original level.<br>Sequencing efforts: GB:158%, OC_others:100%, US:148%, and others scaled down to 10% of the original level. |
| T2a-M | Top 2 (T-ranked) travel hubs with 4 additional continental hubs | NA | Diagnostic efforts: AE:11456%, AU:362%, BR:170%, GB:587%, US:241%, ZA:468%, and others scaled down to 10% of the original level. Sequencing efforts: AE:8845%, AU:301%, BR:178%, GB:141%, OC_others:100%, US:133%, ZA:369%, and others scaled down to 10% of the original level. |
| T3-M | Top 3 (T-ranked) travel hubs | Diagnostic efforts: DE:498%, GB:487%, US:233%, and others scaled down to 10% of the original level. Sequencing efforts: DE:240%, GB:128%, OC_others:100%, US:128%, and others scaled down to 10% of the original level. | Diagnostic efforts: DE:469%, GB:538%, US:227%, and others scaled down to 10% of the original level. Sequencing efforts: DE:229%, GB:132%, OC_others:100%, US:127%, and others scaled down to 10% of the original level. |
| T3a-M | Top 3 (T-ranked) travel hubs with 4 additional continental hubs | Diagnostic efforts: AU:180%, BR:134%, DE:404%, GB:396%, MA:804%, TR:221%, US:202%, and others scaled down to 10% of the original level. Sequencing efforts: AU:156%, BR:134%, DE:210%, GB:122%, MA:5858%, OC_others:100%, TR:481%, US:122%, and others scaled down to 10% of the original level. | Diagnostic efforts: AE:7837%, AU:278%, BR:147%, DE:380%, GB:432%, US:196%, ZA:351%, and others scaled down to 10% of the original level. Sequencing efforts: AE:5295%, AU:219%, BR:146%, DE:197%, GB:124%, OC_others:100%, US:120%, ZA:260%, and others scaled down to 10% of the original level. |
| T7-M | Top 7 (T-ranked) travel hubs | Diagnostic efforts: CA:569%, DE:295%, ES:357%, FR:193%, GB:290%, TR:178%, US:165%, and others scaled down to 10% of the original level. Sequencing efforts: CA:153%, DE:163%, ES:355%, FR:170%, GB:113%, OC_others:100%, TR:319%, US:113%, and others scaled down to 10% of the original level. | Diagnostic efforts: AE:5461%, CA:682%, DE:294%, ES:325%, FR:190%, GB:330%, US:167%, and others scaled down to 10% of the original level. Sequencing efforts: AE:3331%, CA:163%, DE:160%, ES:315%, FR:165%, GB:115%, OC_others:100%, US:112%, and others scaled down to 10% of the original level. |
| T7a-M | Top 7 (T-ranked) travel hubs with 3 additional continental hubs | Diagnostic efforts: AU:145%, BR:119%, CA:510%, DE:270%, ES:324%, FR:182%, GB:266%, MA:495%, TR:168%, US:157%, and others scaled down to 10% of the original level. Sequencing efforts: AU:127%, BR:117%, CA:145%, DE:154%, ES:318%, FR:160%, GB:111%, MA:2921%, OC_others:100%, TR:287%, US:111%, and others scaled down to 10% of the original level. | Diagnostic efforts: AE:4671%, AU:205%, BR:128%, CA:596%, DE:265%, ES:292%, FR:177%, GB:296%, US:157%, ZA:248%, and others scaled down to 10% of the original level. Sequencing efforts: AE:2765%, AU:161%, BR:124%, CA:152%, DE:150%, ES:277%, FR:154%, GB:112%, OC_others:100%, US:110%, ZA:182%, and others scaled down to 10% of the original level. |
| T11-M | Top 11 (T-ranked) travel hubs | Diagnostic efforts: AE:2961%, CA:448%, DE:245%, ES:291%, FR:169%, GB:241%, IN:124%, JP:166%, KR:146%, TR:158%, US:148%, and others scaled down to 10% of the original level. Sequencing efforts: AE:1858%, CA:138%, DE:146%, ES:286%, FR:151%, GB:109%, IN:126%, JP:124%, KR:221%, OC_others:100%, TR:260%, US:109%, and others scaled down to 10% of the original level. | Diagnostic efforts: AE:3832%, CA:505%, DE:235%, ES:256%, FR:163%, GB:260%, HK:640%, IN:142%, JP:204%, TR:164%, US:146%, and others scaled down to 10% of the original level. Sequencing efforts: AE:2326%, CA:143%, DE:141%, ES:248%, FR:145%, GB:110%, HK:1172%, IN:144%, JP:136%, OC_others:100%, TR:271%, US:109%, and others scaled down to 10% of the original level. |
| P11a-R | Top 11 (P-ranked) travel hubs with 2 additional continental hubs | Diagnostic efforts: All unchanged. Sequencing efforts: All unchanged. | NA |
| P11a-H | Top 11 (P-ranked) travel hubs with 2 | Diagnostic efforts: AE:9128%, AU:145%, CA:378%, CL:134%, DE:150%, | NA |
| | additional continental hubs | ES:221%, FR:131%, GB:161%, HK:1126%, MA:372%, NZ:776%, TR:121%, US:104%, and others scaled down to 50% of the original level. Sequencing efforts: AE:5664%, AU:128%, CA:131%, CL:135%, DE:116%, ES:218%, FR:123%, GB:104%, HK:2206%, MA:2061%, NZ:383%, OC_others:100%, TR:157%, US:101%, and others scaled down to 50% of the original level. | |
| P11a-M | Top 11 (P-ranked) travel hubs with 2 additional continental hubs | Diagnostic efforts: AE:16350%, AU:181%, CA:601%, CL:161%, DE:190%, ES:317%, FR:155%, GB:210%, HK:1946%, MA:590%, NZ:1316%, TR:137%, US:108%, and others scaled down to 10% of the original level. Sequencing efforts: AE:10116%, AU:150%, CA:155%, CL:163%, DE:128%, ES:313%, FR:141%, GB:107%, HK:3890%, MA:3629%, NZ:610%, OC_others:100%, TR:203%, US:101%, and others scaled down to 10% of the original level. | NA |
| T11a-R | Top 11 (T-ranked) travel hubs with 3 additional continental hubs | Diagnostic efforts: All unchanged. Sequencing efforts: All unchanged. | NA |
| T11a-H | Top 11 (T-ranked) travel hubs with 3 additional continental hubs | Diagnostic efforts: AE:1470%, AU:118%, BR:108%, CA:267%, DE:169%, ES:191%, FR:133%, GB:167%, IN:111%, JP:132%, KR:122%, MA:261%, TR:128%, US:123%, and others scaled down to 50% of the original level. Sequencing efforts: AE:916%, AU:111%, BR:107%, CA:118%, DE:121%, ES:186%, FR:124%, GB:104%, IN:112%, JP:111%, KR:156%, MA:1219%, OC_others:100%, TR:174%, US:104%, and others scaled down to 50% of the original level. | NA |
| T11a-M | Top 11 (T-ranked) travel hubs with 3 additional continental hubs | Diagnostic efforts: AE:2565%, AU:133%, BR:114%, CA:400%, DE:225%, ES:264%, FR:160%, GB:221%, IN:120%, JP:157%, KR:139%, MA:389%, TR:150%, US:142%, and others scaled down to 10% of the original level. Sequencing efforts: AE:1570%, AU:119%, BR:112%, CA:132%, DE:138%, ES:255%, FR:143%, GB:108%, IN:121%, JP:120%, KR:201%, MA:2114%, OC_others:100%, TR:233%, US:108%, and others scaled down to 10% of the original level. | NA |

**Table S5: Initial parameter values and maximum likelihood estimates (MLE) after the first round of the fitting process for $M_0$ and $M_1$.** NA in the MLE column indicates that estimates are not applicable due to high uncertainty, which occurs when the Monte Carlo-adjusted profiles result in confidence intervals approaching infinity. Note that these values are log-transformed such that the value range is from 0 to 1 during the fitting process.

| Parameter | Initial value | $M_0$ MLE | $M_1$ MLE | Changes in the same direction between $M_0$ and $M_1$? |
|---|---|---|---|---|
| $\eta_2$ | 0.6096 | 0.8066 | 0.7703 | Yes |
| $\eta_3$ | 0.6252 | 0.4799 | 0.4622 | Yes |
| $\eta_4$ | 0.7281 | 0.2307 | 0.5397 | Yes |
| $\varepsilon_4$ | 0.2004 | NA | 0.2955 | NA |
| $\lambda_1$ | 0.7796 | 0.8340 | 0.9900 | Yes |
| $\lambda_2$ | 0.3724 | 0.6319 | 0.5486 | Yes |
| $\lambda_3$ | 0.4772 | NA | 0.0100 | NA |
| $\lambda_4$ | 0.2463 | NA | 0.6839 | NA |
| $\iota_1$ | 0.3807 | 0.3807 | NA | NA |
| $\iota_2$ | 0.6163 | 0.6163 | 0.6319 | Yes |
| $\iota_3$ | 0.7057 | 0.4387 | 0.1836 | Yes |
| $\iota_4$ | 0.4448 | 0.8772 | NA | Yes |
| $\gamma$ | 0.9900 | 0.3092 | 0.1571 | Yes |
| $\sigma$ | 0.9714 | NA | NA | NA |
| $\theta_{cases}$ | 0.5000 | 0.9900 | 0.9900 | Yes |
| $\theta_{deaths}$ | 0.5000 | 0.9900 | 0.9900 | Yes |
| $\theta_{seq\_all}$ | 0.5000 | 0.9900 | 0.9900 | Yes |
| $\theta_{seq\_unit}$ | 0.5000 | 0.9900 | 0.8664 | Yes |
| $\tau_{BA.1}$ | 0.4110 | 0.8066 | 0.7703 | Yes |
| $\tau_{BA.2}$ | 0.6277 | 0.4799 | 0.4622 | Yes |